# Molecular Electronics by Chemical Modification of Semiconductor Surfaces

Ayelet Vilan* and David Cahen

Weizmann Institute of Science,

Rehovot, Israel 76100

* author for correspondence: ayelet.vilan@weizmann.ac.il

## Outline

Inserting molecular monolayers within metal / semiconductor interfaces provides one of the most powerful expressions of how minute chemical modifications can affect electronic devices. This topic also has direct importance for technology as it can help improve the efficiency of a variety of electronic devices such as solar cells, LEDs, sensors and possible future bioelectronic devices, which are based mostly on non-classical semiconducting materials (section 1). The review covers the main aspects of using chemistry to

- control alignment of energy levels at interfaces (section 2):

- passivate interface states (section 3),

- insert molecular dipoles at interfaces (section 4),

- induce charge rearrangement at and around interfaces (section 5).

After setting the stage, we consider the unique current-voltage characteristics that result from transport across *metal / molecular monolayer / semiconductor* interfaces. Here we focus on the interplay between the monolayer as tunneling barrier on the one hand, and the electrostatic barrier within the semiconductor, due to its space-charge region (section 6), on the other hand, as well as how different monolayer chemistries control each of the these barriers. Section 7 provides practical tools to experimentally identify these two barriers, and distinguish between them, after which section 8 concludes the story with a summary and a view to the future. While this review is concerned with hybrid semiconductor / molecular effects (see Refs. 1,2 for earlier reviews on this topic), issues related to formation of monolayers and contacts, as well as charge transport that is solely dominated by molecules, have been reviewed elsewhere,[3-6] including by us recently.[7]

## List of Abbreviations

a     Distance between molecules [Å];

A     Contact (junction) area [$cm^2$];

A*     Richardson coefficient [120 A/$cm^2$/$K^2$, for n-Si];

BB     Band bending [eV]; $BB_0$ at 0V;

CB     Conduction band (bottom of) of a semiconductor;

CNL     Charge neutrality level [eV];

CR     Charge rearrangement;

d     Dipole length or distance between point charges [Å];

$D_{IS}$     Density of interface states [$eV^{-1}cm^{-2}$];

$E_F$     Fermi level;

$E_g$     Forbidden energy gap;

$E_T$     Tunneling barrier height [eV];

$E_{vac}$     (local) vacuum level;

FB     Flat band potential;

HD     Highly-doped;

HOMO     Highest occupied molecular orbital [eV];

i     Insulator (subscript);

I     Current [A];

IS     Interface states (subscript);

IP     Ionization potential [eV], = $E_{vac}$ − HOMO;

ISR     Interface specific region;

J     Current Density [A/$cm^2$];

k     Boltzmann coefficient [$8.62 \cdot 10^{-5}$ eV/K];

K     Geometric correction factor (dimensionless);

L     Monolayer (insulator) thickness or tunneling distance [Å];

LUMO     Lowest unoccupied molecular orbital [eV];

M     Metal (subscript);

MD     Moderately-doped;

mol     Molecular (subscript);

n     Diode's ideality factor [≥1, dimensionless];

$n_e$     Number of electrons;

N     Number of molecules;

$N_D$     Doping level [1/$cm^3$];

p     Molecular dipole [Debey] (1 Debey = 3.336E-30 C·m);

$p_0$     Dipole of isolated molecule;

$p_\perp$     Surface-normal projection of molecular dipole

q     Electron charge [1.6E19 C];

Q     Accumulated charge [C];

S     Index of interface behavior [0-1, dimensionless];

Sat     Saturation (subscript);

SBH     Schottky barrier height [eV];

SC     Semiconductor (subscript);

SCR     Space charge region;

T     Temperature [K];

V     Voltage [V];

VB     Valence band (top of) of a semiconductor;

$W_D$     Width of depletion region [cm];

WF     Work function [eV] (mostly of metal);

α     Polarizability of the monolayer (the surface-perpendicular component of the tensor) [$Å^3$];

Δ     Potential energy step on the monolayer [eV]; $Δ_0$ at 0V;

$ε_0$     Vacuum permittivity or Electric constant [8.854E-14 F/cm];[8]

$ε_i$, $ε_{sc}$     Relative permittivity of the insulator / semiconductor /;

φ     Electrical potential [V];

η     Chemical hardness [eV];

μ     Chemical potential [eV];

$μ_0$     Standard chemical potential [eV];

$\tilde{μ}$     Electrochemical potential [eV];

σ     Hammett parameter;

θ     Molecular / dipole tilt angle (relative to normal);

χ     Electron affinity [eV], = $E_{vac}$ − LUMO;

ξ     Distance of Fermi level from the band edge in semiconductor [eV];

## 1. Introduction: the interface is the device

Electronics is based on asymmetries induced by interfaces, in contrast to electrical properties such as resistivity or electric permittivity (dielectric constant), which are bulk properties. Herbert Kroemer opened his Nobel lecture by stating: "Often, it may be said that the interface is the device."[9] Non-linear functionalities, such as current rectification or gain, lasing, photo-voltage or thermopower are all induced by breaking the symmetry between electrons and holes at a chemical interface.[9] The ongoing miniaturization via micro- to nano-electronics drives home the point that action occurs at very small dimensions. Both academy and industry are still waiting to witness where this journey will halt: what will be the ultimate minimal size of a "bulk" required to maintain an "interface". A different perspective is that of nanotechnology, where new physical properties emerge by breaking a "bulk" into very small objects, or creating huge areas of interface. Indeed, at nm dimensions the distinction between bulk and surface is rather vague. Still, the common concept of nano-technology focuses on the "intrinsic nature" of the nano-objects. In contrast we wish in this review to promote the idea that the inherent functionality is contained at borders, in spatial gradients that result from equilibration between adjacent phases, rather than in isolated objects. Thus, we view the interface not as a perturbation, but as the essence of the function. This is challenging because the best friend of a (well-defined) surface is ultra-high vacuum: with the exception of (layered) compounds with van der Waals surfaces, clean, inorganic surfaces are normally unstable and reactive. This review thus focuses on understanding and manipulating the chemical and electronic properties of semiconductor surfaces and interfaces that involve a semiconductor, toward control of electrical functionality of junctions, containing them.

Most of standard Si technology is based on so-called homo-junctions, where the Si crystal lattice is continuous, and it is only the nature of the foreign dopants that changes across the interface. As argued above, the chemical change is the key for the build-up of an electrical potential difference across the interface, which can induce asymmetric charge transport.[9] A metal / semiconductor interface belongs to the more generic group of 'hetero-interfaces', which includes also interfaces between two semiconductors made of different materials.

In a hetero-junction, the profile of the electrochemical potential of the electron across the interface includes both a bulk (standard chemical potential) contribution and a doping contribution. This double contribution complicates predicting the built-in potential or the carrier injection / blocking characteristics associated with the junction. Nonetheless, understanding electrical potential barriers of hetero-junctions is obviously vital for electronics in general and specifically for novel electronic materials,[10-12] and material combinations, which are mostly hetero-junctions. This review focuses on how molecules can intervene with and affect the interface electrostatic balance, and control it to a large extent. We will focus specifically on a generic metal / molecular / semiconductor interface, and examine how the molecular properties can be related to the charge transport characteristics across the molecular modified junction. However, monolayer preparation aspects are largely ignored, in view of excellent existing reviews: monolayer formation on oxidized Si,[13] oxidized surfaces [14] in general and transparent conducting electrodes[15] in particular, and in contrast on oxide-free Si[16-18] and Ge.[19,20] Issues related to top-contact formation can be found in Refs. 7,21,22.

## 2. The Schottky Barrier Height and Fundamental insights into the Interface energy balance

The rules governing a molecular modified, or hybrid interface are actually the same as those used to describe any electronic interface, regardless of if the modification is by an organic or inorganic foreign layer, or even an atomically abrupt interface between two solid materials. In this sense, molecular modification of interfaces offers a convenient test-bed for elucidating the (electronic) energy level alignment mechanism, irrespective of actual applications. This introductory section defines the basic physical relations required to interpret the experimental evidence, described in later sections.

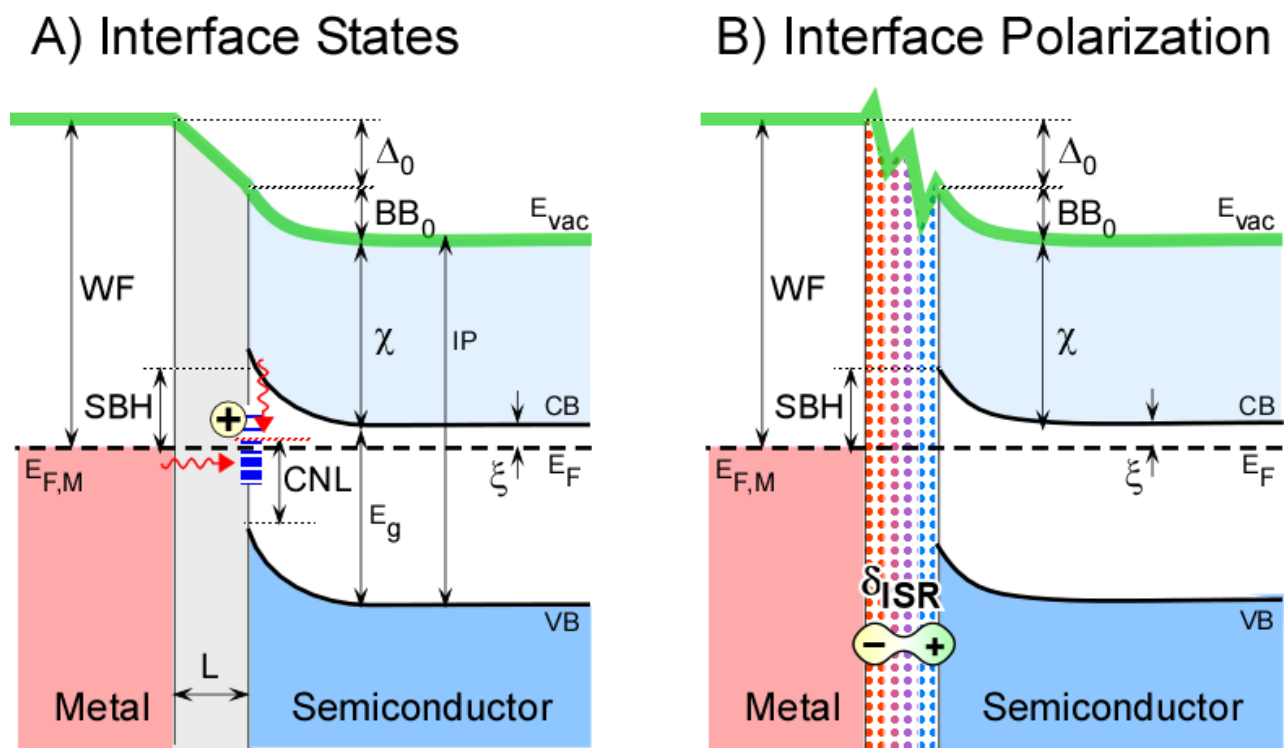

**Figure 1:** Schematic illustration of interface energy alignment and creation of the Schottky Barrier Height (SBH), according to: **A)** Charging of interface states or **B)** Interface polarization.

Both panels show energy per electron (Y-axis) across an interface between a metal (left, red) and an n-type semiconductor (right, blue), separated by either a foreign layer (gray in **A**) or an intrinsic transition region (dotted in **B**), known as ISR (interface specific region), the thickness of which is grossly exaggerated for clarity. The interface charging model (**A**) assumes a density of interface states, marked by blue stripes, centered at a distance CNL (charge neutrality level) from the top of the valence band at the surface. The Fermi level is pinned nearly at the CNL, where the difference between them accounts for the net interface charge, $Q_{IS}$. In this case $E_F$ is below the CNL and therefore $Q_{IS}$ is positive, with a counter negative charge on the metal edge, dictating $\Delta>0$. The red arrows in (**A**) mark possible electronic processes (see text). In the interface polarization model (**B**) a net change in potential, $\delta_{ISR}$, emerges from the bond-polarization between the intrinsic atoms that make up the interface. In both panels the electronic equilibrium state (0 V) is shown with a constant Fermi level (dashed line) across the interface, while the local vacuum level (green top line) varies across the interface. See list of symbols for other symbols.

### a. Work-function balance – The pure Schottky -Mott case

In order to understand the chemical modification of interfaces, we first shortly review the difficulty in predicting energy level alignment at generic metal-semiconductor junctions without molecules, also known as Schottky junctions. Such junctions have technical importance by themselves, as injecting contacts but also for low-cost active junctions in photovoltaics.[23] From a fundamental point of view, Schottky junctions can serve as model system to understand the much wider case of any hetero-junction (i.e., semiconductor / semiconductor).[24] The basic energy diagram as a function of distance perpendicular to the interface is shown in Figure 1. The prime characteristic of such junction is the 'Schottky barrier height' (SBH), which is the energy an electron should gain to transfer from the metal side (Fermi level) to the semiconductor side, which for an n-type semiconductor is the conduction band minimum, CB. The need to predict the extent of the SBH, based on characteristic energy levels of the constituting materials, was recognized already in the very early days of solid-state electronics.[25,26] Nevertheless, the nature of energy level alignment at interfaces remained hotly debated, as detailed elsewhere.[24,27] A 'chemist-friendly' tutorial, in the context of surface catalysis is given in Ref. 28.

These "solid-state" difficulties are somewhat similar to the difficulties to define the potential gradient within an electrochemical cell.[29] Within our motivation to demonstrate the close relation between chemical and electrical characteristics, we would like to begin with fundamental definitions, all referring to an electron (i.e., single negative charge, not molar quantities).

The *chemical potential of the electron*, $\mu$ [30] is composed of a standard component, $\mu_0$ plus a temperature-dependent occupancy term, where kT is the thermal energy and $n_e$ is the number of electrons:

$$\mu = \mu_0 + kT \ln n_e \qquad (1)$$

The chemical potential, $\mu$ combined with the electrical potential, $\phi$,[31] gives the electrochemical potential, $\tilde{\mu}$:

$$\tilde{\mu} = \mu + q\phi \qquad (2)$$

where q is the electron charge. The solid-state equivalent of the electrochemical potential of the electron, $\tilde{\mu}$ is the Fermi level, $E_F$, the energy where the probability to find an electron is 50% (horizontal dashed line in Fig. 1). This relation ($\tilde{\mu} \simeq E_F$) is elaborated on in Refs. 32-35. Thus at equilibrium (i.e., without an externally applied bias) the Fermi level must be the same across the interface.

The second requirement is conservation of energy, namely that the energy to remove an electron from the interface to the vacuum is identical at both sides of the interface. The energy of an electron at rest, and outside the influence of the crystal potential is known as the 'vacuum level', $E_{vac}$, while the difference between $E_{vac}$ and $E_F$ is known as the work function, WF. Note that $E_{vac}$ refers to the vacuum level immediately next to the surface, while the absolute vacuum level, at infinity, cannot be measured.[29,36]

Formally speaking, no vacuum exists at an interface, undermining the relevance of 'vacuum level' to describe solid interfaces (in contrast to surfaces). Nevertheless, it is a simple way to grasp the requirement for energy conservation even within the solid interior. Such virtual level is known as the 'local vacuum level', defined as the energy of an electron at rest and *free from the influence of the crystal potential*[36,37] (i.e., as if virtually cutting a solid and removing an electron out while freezing all its nuclei and electrons). As the gradient in the local vacuum level is proportional to the electric field,[37] the local vacuum level can be thought to be fixed to the actual vacuum level at the real (far-end) surface and just follow the averaged electrostatic potential, to any internal position in a solid. This level is shown by the green thick fluctuating line at the top of Fig. 1, and from now on will be termed just "vacuum level" and denoted by $E_{vac}$.

Comparing again between solid-state and molecular terminologies, the work function (or vacuum level, if the Fermi level is the reference zero energy) relates to the chemical potential plus charging contributions (no charging for metals, $\phi \sim 0$). The WF can be measured using photoelectron spectroscopy[36] or with a Kelvin probe, as the contact potential difference with respect to a reference surface, CPD.[38] It can also be estimated from electrochemistry,[36] an approach that has been shown to be quite effective for organic electronics.[39] Notice that the above definition of local vacuum level excludes the periodically fluctuating potential of the atomic nuclei. However, the breaking of the lattice at the surface adds an abrupt surface component to the WF.[24,36,40] Thus the surface contribution of a (111) plane differs from that of a (100) plane; that of a vacuum reconstructed surface is irrelevant for a chemically-bonded, structurally-relaxed surface. In other words, at the surface we can no longer ignore the contribution of atomic-scale charging, by e.g., a hetero-atomic unit cell or the spill-over of the 'free-electron' wave functions beyond the surface as defined by the nuclei to slightly (1 to a few tenths of nm) outside the crystal. This contribution is illustrated by the fluctuation in the vacuum level at the interface of Fig. 1.B.

Despite these fundamental difficulties, the vacuum level is the accepted way to describe the variation in the _electrical potential_ within a solid, in contrast to the Fermi level, which represents the _electrochemical potential_. Thus Fermi level equilibrium at a contact between material A and B ($\tilde{\mu}_A = \tilde{\mu}_B$ in Eq. 2) of different WFs (or standard chemical potentials, $\mu_{0,A} \neq \mu_{0,B}$, in Eq. 1) can be established by either flow of electrons from the low- to the high WF material (change $n_e$ from high to low $\mu$, in Eqs. 1,2) or by a build-up of an electrical potential difference ($\phi$ in Eq. 2) immediately at the contact between the two phases. Most often these two contributions co-exist. An interfacial potential step (change in $\phi$) is illustrated in Fig. 1 by $\Delta_0$ (where the subscript '0' indicates no applied bias), while the electron flowing from the low to the high WF side (semiconductor to metal in Fig. 1) creates a charged region in the semiconductor, known as space charge region (SCR), which is positive in the case depicted (for an n-type semiconductor). The net accumulated charge creates a gradual potential change known as the built-in potential, or band bending at 0 V: ($BB_0 = E_{C,Surf} - E_{C,bulk}$).[28] The built-in potential and the Schottky barrier height (SBH) are directly related:

$$SBH = BB_0 + \xi \qquad (3)$$

where $\xi$ is the energy difference between the Fermi level and the conduction band minimum in the bulk semiconductor, CB-$E_F$.

As noted above, from the charge transport perspective, this complicated energy balance boils down to one parameter: the Schottky barrier height, SBH (Fig. 1). Following energy conservation, the WF of the metal (left side of the interface in Fig. 1) equals to the sum of (from top to bottom) the interfacial potential drop ($\Delta_0$), the electron affinity of the semiconductor ($\chi$=$E_{vac}$-CB) and the SBH. This provides a straightforward prediction of the SBH based on the 'intrinsic' properties of the two contacting materials (WF and $\chi$, see Fig. 1) for an n-type semiconductor (for brevity, p-type energy diagrams and formal equations are skipped; see Refs. 24,41,42):

$$SBH(n) = WF - \chi - \Delta \qquad (4)$$

We arbitrary define the sign convention of $\Delta > 0$ if the vacuum level increases away from (rather than into) the semiconductor. Eq. 4 is generally known as Schottky-Mott rule, as both scientists recognized it in 1939.[25,26] The hetero-junction analogue to the Schottky-Mott rule is called the Anderson rule (1962).[24] The interface potential, $\Delta$, was originally missing from Eq. 4 and $\Delta$ remains a topic of discussion and research, also today. Indeed, this review is mostly concerned with describing the immense power that interfacial species, extrinsic (to the clean semiconductor(s) and metal), have on dictating Δ (up to 1 eV), thus granting such species, molecules and ions, a gating-like effect on the SBH and, therefore, exponentially on the net current across this interface.

### b. Vacuum level alignment, Fermi-level pinning and the index of interface behavior

Starting in the late 1990's the critical importance of interface potential in dictating the SBH became recognized.[43-47] While adsorbed molecules are one way to control the interface electrostatics, strong effects take place even at pure, abrupt inorganic interfaces.[27,48] Eq. 4 implies that balancing the interface mismatch in electrochemical potential (WF-χ) can be done via creating Δ or SBH (related to the BB). Historically the discussion distinguishes between the two extreme cases where solely one mechanism is active:

*Vacuum level alignment* (Schottky-Mott limit[25,26]) is the case of zero electrical potential discontinuity at the interface: Δφ = Δ$_0$ = 0;

*Fermi level pinning* (Bardeen limit[49]) is the case where the Fermi-level of a given semiconductor is pinned to a fixed energy position at its surface (e.g., the charge neutrality level, CNL in Fig. 1.A). As a result, the SBH is constant for a given semiconductor (often SBH∼E$_g$/2), regardless of the work-function of the metal. In such case, contacts with different metals lead to strong variations in Δ. Thus the limit of Fermi level pinning is also referred to as the case of vacuum level shift, or vacuum level misalignment (*Δ ≠ 0*).

In reality most semiconductors behave somewhere between these two extreme limits. Their interfacial behavior is quantified by a parameter called the index of interface behavior, *S*:[24,50-52]

$$S_{SC} = \frac{d(SBH)}{d(WF)} \qquad (5)$$

As *S* → 1, the vacuum levels align (Eq. 4, with Δ→0), while if *S*→0, there is 'Fermi level pinning', where most of the difference in the vacuum levels is translated into Δ.

There are some variations to the exact definition of *S*, where the denominator is electronegativity,[53] a thermochemical quantity,[52] rather than the work-function,[50] but they all yield a similar picture, one of which is shown in Figure 2.A. *S* is very low for covalent semiconductors, such as silicon and germanium and approaches unity for strongly ionic semiconductors, like oxides.[53] Note that the technologically most important semiconductors, such as silicon, germanium and GaAs have very low *S* values, meaning that device engineers are very limited in tailoring their interface' energetics.

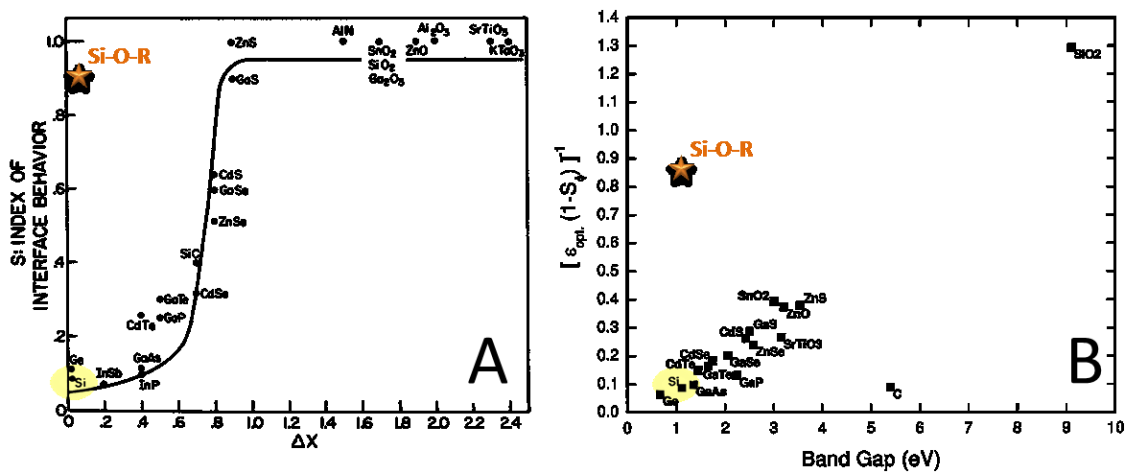

**Figure 2**: Variation of index of interface behavior, *S*, for inorganic materials, showing: A) A plot of S against the difference in electronegativity of the semiconductor component elements; reproduced from Ref. [53] and B) A plot of an the inverse functional, $Y^{-1} = \varepsilon_{SC}(1-S)$ (see Eq. #21) against the semiconductor forbidden energy gap, reproduced from Ref. [47]. The orange stars in (A) and (B) mark the $S_{SAM}$ and $Y^{-1}_{SAM}$ (Eq. &99) for an alkoxy-modified Si(100) compared to the textbook values for a clean metal/Si interface (yellow background).

## c. The Interface Specific Region (ISR) - the interface as a different chemical entity.

Although it was recognized that *S* is a property of a given semiconductor, namely an intrinsic material property,[24,27,52,53] it was also clear that some interface interaction must be involved in order to get the Fermi-level pinning (or the emergence of $\Delta \neq 0$). Over the years a variety of models and terminologies were used to account for the origin of $\Delta$ (see refs. [24,27,48,54] for detailed accounts), and it is beyond the scope of this review to cover them all. Instead, we wish to distinguish between two conceptual scenarios: interface traps (Fig. 1.A) and interface bonds (Fig. 1.B). Although the borders between these views are not always clear, there are a few prominent differences.

Technically, in the interface *traps* view the interface is like a flat plate capacitor, with interface-trapped charges on one plate, and the metal contact as the other plate, and a thin *foreign film* as the dielectric layer (gray area in Fig. 1.A). In contrast, the interface *bond* view suggests that charge rearrangement occurs between the atoms[48] or molecules,[54] directly contacting the metal, leading to polarization of the few unit cells immediately at the interface (illustrated in Fig. 1.B as a red to blue color gradient region). This effect decays exponentially with distance, and therefore predicts Fermi level pinning (large $\Delta$) to be more pronounced in *clean*, abrupt interfaces (e.g., in vacuum-deposited interfaces, in contrast to, e.g., oxidized interfaces in the *traps* view[51]), as is indeed observed for both perfect, MBE-grown inorganic interfaces[24] and thin-film organic electronics (which are composed of closed shell molecules).[55]

A second critical difference is that interface *traps* not only alter $\Delta$ and SBH, they also severely interfere with the charge-transfer process, mostly deteriorating the electrical performance, by trap-ping excited carriers or adding tunneling routes, illustrated in Fig. 1.A by wiggly red downward and right-pointing arrows, respectively. Interface *bond*-polarization can occur without introducing new states in the gap,[24] and therefore can explain large interface potential steps with close to ideal charge transfer characteristics. While the latest understanding is strongly in favor of the interface bond polarization, (Fig. 1.B) the specific context of molecularly modified interfaces actually revives the importance of the traditional, interface states view, simply because the molecular monolayer largely blocks direct bond polarization between metal and semiconductor, and behaves as an interfacial capacitor. However, mechanisms similar to interface-bond polarization can occur between the solid contacts *and the monolayer* and, therefore, they will be briefly explained below.

Considering classical, crystalline metal / semiconductor interfaces, Tung has rationalized the emergence of an interface-specific region (ISR) as follows.[24] The notion that a central part of the work-function originates in the breaking of the periodicity at the surface implies that the work function is not a fundamental material property. The quantum effects that dictate the *surface* contribution to the work-function are different at an *interface*. The electron wave-functions at a *surface* decay into a vacuum while an *interface* requires matching the Bloch wave functions of one periodic solid to those of the adjacent periodic solids. This adjustment occurs over a finite region, which is the ISR.[24] The electronic wave-functions of the ISR or 'system orbitals' differ from those of both bulk phases, yet at each boundary of the ISR they match the Bloch wave-functions of the respective periodic bulk.

The substrate-molecules interface is more similar to that formed between a thin organic film on a solid substrate, as in organic electronics, than to an all-inorganic interface. In organic electronics the prevailing concept is that of induced density of interface states (IDIS), which considers the emergence of new density of states from the interaction between the evanescent wave-function of the metal's electrons with those of the adjacent organic semiconductor. In this view, the interface potential drop ($\Delta$) is attributed to the excess charge localized on the IDIS, up to its charge neutrality level (CNL, see Fig. 1.A) and to the low dielectric permittivity of organic conductors compared to inorganic ones.[54,56-58] The IDIS-CNL view is somewhere in between the two scenarios of Fig. 1, yet it has many fundamental similarities to the ISR view: In both descriptions new, hybrid energy states are formed at an interface, which implies shifts in the spatial location of the density of states due to minute charge rearrangement across the interface, as qualitatively illustrated by the sharp potential spikes at the interface (ISR) of Fig. 1.B. The fact that atoms can hold such huge potential spikes is seemingly against the prevailing concept that the electrostatic potential cannot change abruptly.[37] However, this notion refers to the averaged potential (e.g., the one represented by the vacuum level) while the microscopic electrostatic potential is strongly fluctuating on atomistic length-scales.

Although no specific chemical bonds are necessarily formed at the interface, and the binding strength could be rather weak ("physisorption") the ISR can be viewed as a chemically distinct region, because -a- new energy levels formed and -b- charge shifted between the 'reactants'. Such interactions are evident from DFT computations[24,54,56-60] and can be roughly considered as some degree of coupling between the two phases. The two approaches differ mainly in regard to the energy position of the hybrid states, as further discussed in section 5.a. IDIS-CNL requires a finite density of states within the forbidden energy gap while ISR states in principle emerge along the full energy and momentum scale,[24] with possible donation and back-donation at different energy levels.[61] This is a somewhat controversial issue and the subject of a lively discussion, because gap-states can be detrimental as they interfere with the transport process (section 3).

In summary, balancing the electrochemical potential across an interface leads to two types of charge rearrangements: localized ("molecular-like", $\Delta_0$) and long-range (solid-state, $BB_0$); the in-terplay between them provides an efficient handle to control the electronic performance of the interface. Once we realize the intimate relation between the interface chemistry and its emerging physical properties it becomes very appealing to use chemical tools to modify and control the interface properties.[62,63] In a sense, controlling the interface composition is comparable to introducing dopants or purifying a crystalline structure. In both examples, minute chemical changes are extremely influential on net electrical performance. Because an interface is dimensionless, actually a mono-atomic layer can suffice.[64] Adsorption of organic monolayers adds a slightly thicker modification of 1-2 nm, and in principle inorganic thin layers can also significantly affect the energy alignment.[65,66] While both inorganic thin films and organic monolayers alter the charge rearrangements between the two contacting bulks, organic molecules add their own, 'intrinsic' dipole caused by the closed shell nature of molecules. Thus electron-withdrawing (e.g., halogens, CN, $NO_2$) or -donating groups (e.g., OMe, $NH_2$) can induce huge electric fields within a single molecule. Adsorbing molecules so that they will be at solid interfaces can in principle affect, and thus modify, any of the interfacial mechanisms: interface traps (section 3); molecular dipoles (section 4) and charge rearrangement (section 5). In these sections, the molecules are considered as modifiers of a barrier that is basically located within the semiconductor (SBH). Yet molecules add also their own direct transport blocking effect that is considered in section 6. Section 7 offers technical tools to analyze current-voltage characteristics of molecularly modified metal / semiconductor junctions.

## 3. Electrical passivation: Eliminating interface traps

### a. Interface charge – The Bardeen limit

The traditional explanation for Fermi level pinning, also known as the "Bardeen limit"[49] ascribes it to charging, $Q_{IS}$, of electronic states localized at the interface (see blue stripes in Figure 1.A). The 'charge neutrality level', CNL,[67] is normally located around mid-gap, namely below (above) the bulk Fermi level, for an n-type (p-type) semiconductor. In both cases the surface states (ignoring the metal, for the moment, i.e., before forming a junction) withdraw majority carriers from the bulk, which leaves behind a space charge region depleted from free carriers ($Q_{IS} = -Q_{SCR}$), implying a net potential energy ($BB_0 > 0$, see Fig. 1.A).

Moving from a surface to an interface adds the metal's Fermi level to the picture, which complicates it. Depending on the work function of the metal, the occupation of the interface states changes to keep the CNL near the Fermi level of the system. The interface charge balance has now three contributions: the metal has a net counter charge ($Q_M$) balancing the sum of the charge on surface states ($Q_{IS}$) and the space charge ($Q_{SCR}$): $Q_M = -(Q_{IS} + Q_{SCR})$. The $\Delta > 0$ case depicted in Fig. 1.A, implies a negative $Q_M$, regardless of the exact partition between $Q_{IS}$ and $Q_{SCR}$. Notice also that Fig. 1.A illustrates the interface states to be located at the semiconductor surface, which can be due to e.g., steps or dangling bonds. However, in the general case, the entire interface layer can become charged, as is well-known for oxides and other insulators or some charging of the molecular monolayer, as further discussed in section 5.c.

Often, the surface/interface states have a narrow energy distribution ($D_{IS}$) and, therefore, increasing the net accumulated charge ($Q_{IS}$) will hardly move the Fermi level away from the CNL, i.e. Fermi level pinning to the CNL position:

$$\text{SBH}_{Bardeen} = E_{gap} - CNL \qquad (6)$$

where $E_g$ is the forbidden energy gap of the semiconductor ($E_g$=CB–VB) and CNL is as defined earlier and shown in Fig. 1.A.

A practical way to unify between the two extreme descriptions (Eq. 4 with Δ=0 cf. Eq. 6) uses the earlier mentioned index of interface behavior (*S*, Eq. 5) as weighing factor:[41,42,50,51]

$$SBH = S(WF - \chi) + (1 - S)\left(E_{gap} - CNL\right) \qquad (7)$$

The common representation of *S* in organic electronics is:[56,58]

$$S = \left[E_g - (SBH + CNL)\right]/[WF - (IP - CNL)] \qquad (8)$$

Using the energy diagram of Fig. 1.A reveals that these two definitions are identical. It follows that the interfacial potential drop, Δ, is:[58]

$$\Delta = (1 - S)[WF - (IP - CNL)] \qquad (9)$$

Thus, *S* is related to the ability of the interface to hold / screen charge. For traditional, inorganic interfaces, *Δ* relates directly to the amount of interface charge, $Q_{is}$, and the dielectric properties of the interface (see Eq. 24 below):[41,42,50,51]

$$S^{-1} = 1 + q^2 D_{IS} \cdot L / \varepsilon_0 \varepsilon_i \qquad (10)$$

where $D_{is}$ is the density of interface states, $\varepsilon_i$ and *L* are the relative permittivity and thickness of the interfacial layer. As can be seen from Eq. 10, for interface charging to explain $S \rightarrow 0$ ("Bardeen-limit", Fermi level pinning), high interface charge, $Q_{is}$ by itself is insufficient. The interface must include a low permittivity ($\varepsilon_i$) layer, either as an insulator of finite thickness, *L*,[48] or, as in the case of organic electronics (IDIS-CNL view), the entire film has low dielectric constant and therefore the CNL logic is applicable,[54] as further discussed in section 5. Here we focus on passivation of electronic states that emerge from chemical and structural defects at the surface of the semiconductor, or the traditional CNL case (Fig. 1.A) where the adsorbed monolayer replaces the traditional oxide.[51] Therefore passivation of interface states is critical, especially when working with wet-bench prepared samples, poly-crystalline materials and other cost-effective novel (preparations for) materials, due to electronic states that emerge from chemical and structural defects. Such imperfection-related states must be minimized to better control interface electrostatics. Moreover, a mid-gap population of electronic states is critical beyond energy level alignment, because it presents recombination centers (downward wiggly arrow in Fig. 1.A) that quench excited electron-hole pairs, and slow traps, that affect the system's response to a time-modulated stimulus (e.g., AC voltage). In the context of metal-insulator-semiconductor (MIS) junctions, interface states provide a source or sink for electrons to tunnel across the insulator, within an otherwise forbidden energy gap (right-pointing wiggly arrow in Fig. 1.A), as further considered in section 6.

Here we focus on imperfections in the starting substrate, while interactions with the second contact are discussed in section 5. Therefore, most of this section refers to molecularly modified *surfaces* rather than *interfaces*.

### b. Surface passivation by chemical binding

The electrically active states that we wish to remove (i.e., passivate) are by definition those located energetically within the forbidden energy gap of the semiconductor. Such energy position implies high chemical reactivity, because the stable (un)occupied states are in the (conduction)valence band. Therefore, adsorbing a monolayer onto semiconductor surfaces before creating an interface can serve to minimize such chemical imperfection via chemical reactions with surface structural defects and dangling bonds,[68-74] a phenomenon known as 'electrical passivation'. In ad-dition, an adsorbed monolayer can help preserve the interface by acting as a diffusion barrier to block surface oxidation,[75-78] known as 'chemical passivation'.

The practical evaluation of the degree of so-called 'electrical passivation' varies from measuring the band-bending of the free surface (i.e., the space charge that balances the surface charge, $Q_{IS}$), by surface photo-voltage[38,79,80] or from the shift of XPS core levels[81-83] to measuring the life-time of minority carriers, by methods such as fluorescence decay time[68-70,79] or by the time decay of pho-to-induced Eddy currents, known as the 'Sinton' method.[73,84,85] Life time characterization is generally a more authentic characterization, because it focuses on the direct goal of reducing surface traps, while surface charging can be caused by states outside the forbidden energy gap (which are inactive in charge transfer)[85] (see sections 2.c, 3.e). In principle analysis of the net charge transport after forming a contact can also be indicative of the density of interface traps,[86-88] however, such approach relies heavily on analysis and cannot identify whether the trap originates from surface preparation or from subsequent contact deposition.

### c. Oxidation protection and chemical passivation

The simplest mechanism by which an adsorbed monolayer can protect the interface is by providing chemical stability, i.e., by forming a chemical bond to the surface that is stable throughout manufacturing and operation (mostly a covalent bond).[77] For semiconductors the major degradation process is oxidation by ambient oxygen and humidity. While important pioneering contributions to hybrid molecular / semiconductor junctions used silane binding to thin $SiO_2$,[89,90] avoiding any oxide is considered to be far more efficient in eliminating surface states.[91] Just the covalent binding by itself will generally not withstand ambient conditions, and gradual degradation via defects will occur within days.[92] However, a dense layer of alkyl chains is capable to serve as a diffusion barrier, preventing reactive species such as $O_2$ and $H_2O$ from reaching the surface. Thus, chemical passivation requires both stable binding and a highly dense hydrophobic monolayer;[75] specifically, alkyl-based monolayers directly-bonded to oxide-free Si were shown to withstand harsh temperature and pH conditions.[77] Dense binding is not trivial especially with 'interesting' molecules, like proteins or redox-active molecules. This optimization problem is often addressed by using two consecutive surface reactions, [17,93,94] as discussed elsewhere.[7]

### d. Effect of reaction rate

Considering the passivation of electrically active sites in chemical terms implies that we wish to selectively attack the chemically most reactive sites. A possible strategy would be to use mild reaction conditions or poor binding groups that preferentially bind to the most reactive states. This requirement contradicts the other demand for densely packed monolayer required for long-time surface stability ("chemical passivation", section 3.c). This apparent contradiction can also be solved by a two-step approach. For example, binding to H-Si(111) commonly proceeds via a radical chain reaction, by externally initiating a surface radical, Si•, by means such as UV illumination, heating or chemically (e.g., by hydroquinone).[95] Alkenes (R-C=C) react only via radical formation, while alkyl-alcohols (R-OH) can also react via an SN (nucleophilic substitution) reaction with Si-H (i.e., not a radical). Although the SN reaction generally requires activation, it can occur preferentially on chemically / electrically active sites, such as surface defects. Thus comparing UV-activated binding to H-Si(111) of alkenes to that of alcohols, both give efficient surface coverage (chemical passivation) but only the alcohols can preferentially react with active sites and therefore provide distinctively better electrical passivation.[73]

### e. Charge transfer into the substrate

As noted above there is some confusion between passivation of mid-gap (surface) states and net surface charge, $Q_{IS}$. The latter, $Q_{IS}$, is part of the general quest for controlling the SBH (see Eqs. 7, 10), as further discussed in section 4 and 5. Mid gap states are a constant source of non-ideality because they mask the forbidden energy gap, leading to un-desired processes like reducing the life-time of minority carriers and thus decreasing the photocurrent of solar cells. Therefore, we reserve the term 'electrical passivation' for elimination of gap states, regardless of the amount of $Q_{IS}$. Actually large $Q_{IS}$ can improve the electrical passivation in a mechanism known as 'field-effect passivation'.[85]

Adsorption of hydroquinone / methanol on H-Si(100) leads to outstanding electrical passivation.[96] Remarkably, these surfaces display both long lifetime of minority carriers (i.e., they are well-passivated, electrically) and relatively high band bending (i.e., large $Q_{IS}$).[85] This was attributed to 'field-effect passivation': partial negative charging of the adsorbed hydroquinones (see section 5.e for explaining this charging mechanism), which increases the surface charge, $Q_{IS}$ and its accompanying band-bending. Yet, the energy levels of the hydroquinone are outside the forbidden energy gap of the Si and therefore, do not act as traps or recombination centers. Moreover, the larger is the molecular $Q_{IS}$ the more it repels majority carriers (electrons) from the surface and therefore increases the lifetime of minority-carriers, known as 'field-effect passivation'.[85] A similar reasoning was also used to explain the 10 to 100 enhancement in photo-current decay time for a molecular controlled semiconductor resistor (MOCSER), i.e., with molecularly-modified surface, compared to bare resistor.[97]

Except for organic (semi)conductors, most molecular adsorbates have energy levels outside the forbidden energy gap or are inefficiently coupled to the substrate and, thus, actual capture of charge carriers by adsorbed molecules is not very likely. At the same time, molecules are highly polarizable and can therefore easily accumulate static charge or dipoles. This is in line with the hydroquinone-Si case[85] where the molecular polarizability was sufficient to induce a net BB in the Si, yet none of the molecules was really ionized (i.e., introduced gap-states). The ability of molecules to transfer charge into the substrate and induce band-bending is a recognized adsorption effect, even for 'physisorption' (or temporal adsorption) of small molecules like $O_2$ or $C_4H_8$ on substrate such as $TiO_2$, an issue that is critical for catalysis.[28]

The notion of atomistic electrochemical balance (see section 5.a below, Eqs. 16-20) is critical for semiconductor substrates because the surface adsorbates function as dopants altering the charge concentration in the vicinity of the surface, also known as "surface transfer doping".[98,99] In the above example, hydroquinone acts as a surface acceptor (i.e., p-type dopant) and therefore creates a surface depleted of electrons, which in the case of n-type, implies large band-bending. The ability of adsorbates to charge the surface vicinity of a semiconductor is now studied intensely, mostly for physisorbed molecular "dopants" on low-dimension materials, such as graphene and nano-particles, as well as diamond (see Ref. 99 for a comprehensive review). We will return to the issue of charge rearrangement in section 5, below, in the context of energy-level alignment.

### f. Poly-crystalline and nano-shaped substrates;

An interesting outcome of the above-mentioned 'surface transfer doping' is that molecules can 'gate' the electrical properties of their substrate,[63] without any current actually passing through the molecules. This means that no top contact is needed at all which can greatly decrease the technical challenges in exploiting molecular surface-modifications. This scenario is especially relevant to sensors made by traditional lithography,[100-103] or nano-wires[104-106] as well as to quantum dots.[107-111] The lack of top contact implies that energy level alignment (i.e., the role of interface-induced dipole, discussed in section 4) is largely irrelevant, however, the huge surface to volume ratio, makes these systems extremely sensitive to surface modifications such as density of surface traps. Any charge rearrangement between a nano-particle and surrounding molecules serves practically to dope the entire nano-particle. This was demonstrated and explained in terms of change in bulk doping, several decades ago for adsorption of $O_2$ on µm-sized crystallites of $CuInSe_2$ and other chalcogenide semiconductors,[112,113] and more recently for $O_2$ on organic[114] polycrystalline semiconductors and on PbS quantum dots.[115]

Adsorbing molecules onto multi-faceted substrates (e.g., nano-particles or poly-crystalline films) is challenging because the reaction rates can vary for different crystallic facets, and the defect densities on the different facets may not be the same, either. For example, our attempts to translate the successful passivation of single-crystal Si wafers into multi-crystalline Si substrates used for solar-cells, generally failed.[116] In contrast adsorption of a series of phenyl-phosphonic acids on ZnO was successful, regardless of the huge variation in surface morphology, ranging from ultra-smooth MBE-grown substrates to ALD and chemical-bath deposition of high-aspect ratio pillars.[117] Additives to spin-coated suspensions of ZnO nano-particles provide another means to tune the work-function of novel materials, pointing to the large versatility of using adsorbates for electrostatic modifications.[118]

In summary, odd-shaped, low-dimensional particles are natural candidates for molecular modifications, but the surface chemistry should be carefully verified for each case.

## 4.  Molecular dipoles as a 2D dipolar array

The discussion above on surface passivation is, in a way, a preliminary, enabling step toward the main goal of tailoring the energy level alignment across hetero-junctions. We note though that full electrical passivation is not a strict prerequisite: molecular modification of the energy alignment is possible even for poorly passivated surfaces, with measurable amount of oxides,[119] something that can be understood in terms of the index of interface behavior, $S$ (Eq. 5), being small but not zero. Still, surfaces that are electrically well-passivated ($S \rightarrow 1$), are expected to show stronger dipolar effects than those with poorer passivation. This section considers the molecular effect on the interfacial potential step, $\Delta$, originating in polarization of the molecular back-bone, while the following one (section 5) considers contributions from charge rearrangement with the adjacent substrate(s).

An underlying assumption, common to both types of dipoles, is that the interface dipole is 'rigid' and un-affected by the applied bias. Actually, strongly polarized molecules can isomerize under a high applied bias (30 V, fields of $\sim 10^8$ V/cm!) – an effect that was used to demonstrate reversible switching.[120,121] Still, the assumption of bias-independent molecular dipole might be an over-simplification even at moderate bias ($\sim 1$ V) for junctions where most of the bias falls on the molecule(s), such as metal-molecule-metal junctions[122] or metal/molecule/high-doped semiconductor ones (see section 6). Here we consider cases where the dominant barrier is that of the semiconductor, so that it is reasonable to assume that the far more polarizable semiconductor will respond to the applied bias and therefore, the change in potential drop across the molecular layer is negligible. Indeed, we see that practical Schottky diodes, with thin interfacial dipole modification follow the standard diode equation over considerable bias ranges (± 0.5V), suggesting that the molecular dipole has negligible changes within this bias range. Still, the net potential drop on the interface, $\Delta$, contains a bias-induced charging contribution (Eq. 22 below), which becomes significant at extreme applied bias (see section 6,7).

**a.      A 2D dipolar array: basic electrostatic considerations**

As explained in section 2, an array of microscopic dipoles in a direction perpendicular to the interface occurs either naturally due to the different chemical nature (difference in electronegativity) of the atoms across the interface or by deliberate adsorption of molecules at the interface. We now summarize the basic electrostatics, describing the translation of a 2D array of molecular dipoles into a potential step. The same electrostatics holds regardless of the chemical nature of the individual dipoles (i.e., molecular or atomic bond-polarization, see sections 2.c and 5). At the same time, having a periodically repeating array of dipoles is very different from from an isolated dipole[123,124] (e.g., single molecule). This is illustrated schematically in Fig. 3.B. The dipole field lines of a single dipole (dark red lines in Fig. 3.B) propagate radially and create a slowly decaying electric field. However, neighbor parallel dipoles add field lines (dark blue lines in Fig. 3.B), which are co-aligned (blue marker) within the polar bi-layer but mutually cancelling (yellow marker) outside the polar bilayer. This leads to strong electric field within the dipolar bi-layer, compared to the sharply decaying electric field outside its poles (see Ref. 123 for a formal derivation).

First, for a net molecular dipole, $p_0$, only the component oriented perpendicular to the interface ($p_\perp$) is relevant for the potential step across the interface (see Figure 3.A for illustration):

$$p_\perp = p_0 \cdot \cos\theta \qquad (11)$$

where $\theta$ is the angle between the main molecular axis and the surface normal ($\theta=0°$ for perpendicular orientation). While this requirement is trivial it shows the importance of knowing the geometry of the adsorbed molecules with respect to the substrate. In addition, groups like ether (R-O-R) have a strong dipole pointing to the lone-pair of the O that is transverse to the R-R direction. Thus a strong molecular dipole is not necessarily effective as interface dipole.

An obvious requirement for translating numerous molecular dipoles into a potential step is that these molecules should be aligned parallel to each other (e.g., Fig. 3.A). Naturally, nice conceptual cartoons are not realistic, as entropy works against such a perfect orientation (e.g., Fig. 3.D). Experimental evaluation of layer uniformity and orientation is not trivial, though methods like FTIR[125,126] and NEXAFS,[127-130] backed by computation, can be helpful. Wettability studies show that terminal surface dipoles can be inferred by comparing the contact angles formed by polar-protic liquids with those formed by polar-aprotic liquids (e.g., water vs. acetonitrile).[131] Nonetheless, because the dipolar effect is averaged, disorder is unlikely to completely cancel the dipolar effect, only to reduce it. Therefore, the critical issue is to maintain a net, average orientation.

A second consideration is the 'infinite' size of the array, namely far larger than some characteristic length. In the context of translating the effect of individual dipoles into a potential step, the characteristics length is the distance, $a$, between two adjacent dipoles (molecules, see Fig. 3.B), and thus 'Infinite' means that the molecular array extends over a distance much larger than $a$.[132] While dense binding (small $a$) will make the electric field more uniform, it also leads to stronger depolarization by neighboring molecules.[10,132-134] Each individual molecular dipole exists within an electric field induced by the surrounding dipoles (Fig. 3.B), acting to reduce the actual dipole, $p$, relative to the intrinsic dipole, $p_\perp$:[123,132]

$$p = p_\perp / \left[ 1 + \alpha \left( \frac{K}{a^3} + \frac{U_{Im}(a)}{d^3} \right) \right] \qquad (12)$$

where $\alpha$ is the molecular polarizability, $K \simeq 10\pm1$ is a geometric factor depending on the binding unit cell (e.g., triangular, rectangular and so on), and $a$ is the inter-dipole distance.[123] Therefore, small $a$ values imply strong depolarization of the intrinsic molecular dipole. The molecular polarizability, $\alpha$, is critical: a more polarizable backbone (e.g., phenyl instead of alkyl) increases the net dipole of an isolated molecule ($p_0$), but within a monolayer, high polarizability increases the depolarization by neighboring molecules, such that knowing the net optimization is not trivial.

The intrinsic molecular dipole is screened also by the substrate, known as renormalization of the molecular energy levels.[135] Monti used image dipoles to account for such screening, where $U_{Im}$ is the difference between the sum of all interactions between image dipoles and the direct interaction between the real dipole and its image. The critical length scale here is $d$ – the distance of the dipole from the image charge.[135] The simple classical electrostatic approach of Eq. 12was shown to well reproduce detailed quantum mechanical calculations.[123,135]

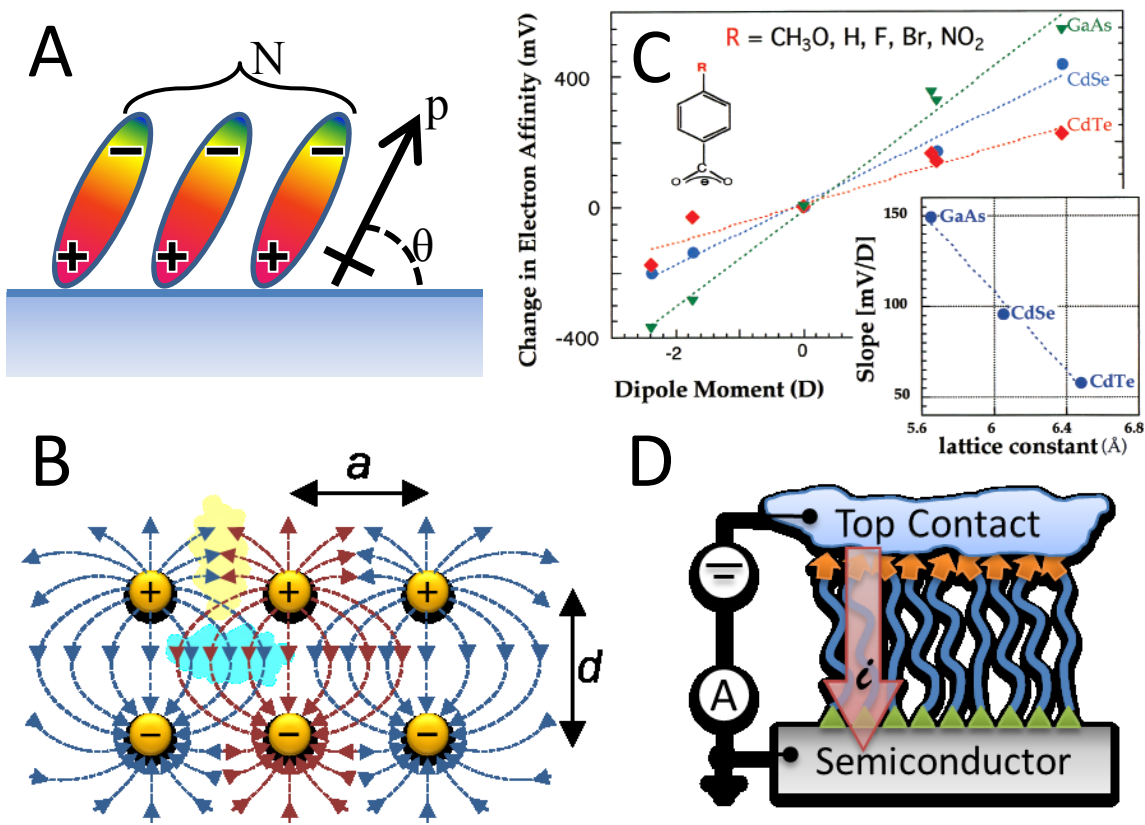

**Figure 3**: Dipole formation by monolayers of polar molecules:

**A)** Diagram illustrating the major quantities controlling the translation of individual dipole into a potential step (cf. Eq. #13);

**B)** Illustration of electric field lines by an array of dipoles: the field is enhanced between the poles (light blue marker) and cancels outside it (yellow marker); also illustrated are the distance between dipoles, *a*, and the dipole length, *d*.

**C)** Variation of the potential step, Δ (relative to non-treated surface), with respect to the nominal dipole, *p*, of individual para-substituted benzoic acids, adsorbed on different semiconductors. Dotted lines are best linear fits, with their slopes plotted in the **inset** against the cation-cation distance, *a*, of each semiconductor. Figure reproduced from Ref. [62].

**D)** A schematic of the complete junction, made of a molecular monolayer adsorbed on a semiconductor (bottom) and covered by a top contact. A voltage source and current meter are connected to the top-contact, while the semiconductor is grounded, driving current (red arrow) across the monolayer. Adsorbed monomers are typically made of a binding group (green triangles), molecular body (wiggly blue lines) and a head-group (orange arrows).

The net potential energy step induced by the individual dipoles (Eq. 12) is proportional to the surface density of dipoles, $N = 1/a^2$ (where for simplicity we assumed a square unit cell for binding, see Fig. 3.A,B):[123,135]

$$\Delta = 37.7 \frac{p_0 \cdot \cos\theta}{\varepsilon_i a^2} \qquad (13)$$

The factor of 37.7 [Å$^2$V/Debye] includes the electric constant, $\varepsilon_0$, and units conversion so that $p_0$ is in Debye and $a$ is in Å. Eq. 13 combines the various depolarization effects into an effective relative permittivity for the monolayer, $\varepsilon_i$. To get a rough scaling between a molecular dipole ($p_0$) and net potential energy step ($\Delta$) we can estimate $\varepsilon_i \simeq 2$-$3$[136] and $a^2 \simeq 20$-$25$ Å$^2$, which yields: $\Delta[eV] \simeq (0.5$-$1) \cdot p_\perp[Debye]$.

Still, we stress that $\varepsilon_i$ is an effective property which may differ significantly from that of a bulk phase made of the same molecules; a more accurate expression for $\varepsilon_i$ is gained by combining Eqs. 12 and 13:[123,135]

$$\varepsilon_i = \frac{1}{4\pi} \left\{ 1 + \alpha \left[ \frac{K}{a^3} + \frac{U_{Im}(a)}{d^3} \right] \right\} \qquad (14)$$

In practice this means that there is a certain optimum binding density to maximize the net potential step induced by an interface dipole, which is often more sparse than that dictated by the available binding sites of a given substrate.[135,137,138]

The binding density is thus a key handle to express the molecular dipole,[139] and can be controlled by molecular design[140] or adsorption conditions.[137,138] Formation of close-packed arrays of small inter-molecular distance and relatively aligned chains is aided by monolayer adsorption, driven by inter-molecular affinity (e.g., van der Waals interactions between alkyl chains) or by weak molecule - substrate interactions (e.g., S on Au).[7] Rigid, non-linear molecular tails and / or substrates with strong inter-atomic bonds (e.g., silicon) can force a larger $a$ (more spacious binding). For example, Fig. 3.C shows the net potential step developed over a series of substituted di-carboxylic molecules adsorbed onto different semiconductors. The sensitivity of the net potential step ($\Delta$) to the molecular dipole ($p_0$), i.e., the slope of the fitted lines in Fig. 3.C varies with the substrate type. The inset to Fig. 3.C shows that the slope is inversely proportional to the surface unit cell,[62] suggesting that for these bulky molecules, the binding distance, $a$ is i) dictated by the substrate and ii) sufficiently large to diminish the depolarization effect. A binding distance, $a$, that is larger than the effective distance for inter-molecular interactions will weaken those interactions, and is expected to increase the entropy, i.e., disorder. Recently it was also noted that very spacious binding favors large tilt angles, $\theta$, which naturally alters the perpendicular dipole component, $p_\perp$ (Eq. 11). While there is a tendency to identify the direction of the molecular dipole with the long molecular axis (as is the case in Fig. 3.A), this is not always the case, which complicates the simple tilt consideration of Eq. 11. Some examples are pendant groups like O-H,[73] and lone-pair electrons in general. Molecules with tilted aromatic groups which are roughly parallel to a metallic substrate induce a stronger push-back effect than perpendicular ones;[120,121,141] in addition their dipole is orientation-dependent.[142,143]

While the electric field outside the dipole-array decays sharply (i.e. , negligible 'far-field') the field within the dipole array ('near-field') is immense.[135] A potential step of 0.1 V across a 1 nm distance equals $10^6$ V/cm and fields as high as $10^8$ V/cm were also reported, leading to measurable Stark shifts in molecular vibrations.[135] If the molecules are sufficiently flexible and the restoring inter-molecular attractive forces are moderate, then the molecules could adopt a non-aligned orientation, in order to minimize the Coulombic repulsion.[144] In certain cases, mutual dipole repulsion can actually hinder the self-assembly process: nitro-phenol-alkyl silane molecules can reach only 80% coverage on oxidized Si, presumably because their dipolar repulsion is too high.[133]

Overall, the translation of an array of molecular dipoles into a significant potential step is well explained theoretically, and simple classical expressions (Eqs. 11-14) can be adequately fit to de-tailed DFT computations.[123,135] Yet, DFT tends to over-estimate the net potential step compared to experiments. The reason is probably that the limited unit cell used for DFT computations, cannot account for molecular dynamics, which acts to minimize the huge internal electric field (i.e., reduce the net potential step).

**b.    The chemical ingredients of a dipole**

While in the previous section we presented the geometrical consideration for efficiently translating a dipole into a potential step, this section focuses on what dictates the dipole itself, in terms of molecular structure. For a molecule to form a relatively ordered monolayer it helps if it has a binding group that dictates its connection to the substrate (e.g., green triangles in the schematic monolayer cartoon shown in Fig. 3.D) and / or another driving force to self-assemble.[7] In this review we consider a 'sandwich' configuration (see Fig. 3.D) where the monolayer is first deposited on a semiconducting substrate, to create a molecularly-modified surface (as discussed here and in section 3 above) and then a metallic contact is deposited on top of the monolayer. Technically, this allows comparing between the surface and interface or junction properties. In this sub-section we are concerned with surface modification, while the next one considers junctions. Issues of monolayer preparation and top-contact formation are reviewed elsewhere.[7] While saturated molecules, like alkyl chains can induce a rather large potential energy step of 0.4 eV,[145-148] each methylene unit adds rather little: about 10 meV/$CH_2$.[131,149] The small dipolar contribution of $CH_2$ despite the rather polarized C-H is because any couple of methylenes mutually cancel each other, and most of the dipole is formed by the terminal methyl and binding group.[150] Still, in a few specific cases shifts of the surface potential by as much as 40 meV/ $CH_2$ with increasing length of adsorbed alkyl chain was reported, for example for alkyl thiols on Ag,[151] or alkyl alcohols on Si.[74] We suggest that this extra effect has to do with charge-rearrangement with the substrate as discussed in the section 5.

Adding aromatic moieties to the adsorbed molecules enhances their polarizability compared to saturated alkanes; combined with substituent groups that push or pull electrons, this can lead to a dipole per molecule on the order of a few Debyes.[132] The net dipole per molecule can be computed,[132,152] but can also be inferred from physical-organic empirical parameters like the Hammett parameter.[68,153,154] Figure 4 shows an example of such a correlation. Styrene reacts with H-Si(111) via a radical chain reaction initiated by UV-irradiation. The minute amount of matter in monolayers presents a challenge to get clear FTIR spectra. However, recent advances in FTIR accessories, such as the Ge hemispherical ATR accessory (Harrick) enhances the FTIR sensitivity down to a 1 nm thick coating, as shown in Fig. 4.A. For metallic substrates, the method of polarization-modulation IR reflectance absorbance spectroscopy (PMIRRAS) provides excellent signal to noise ratio even for a monolayer.[125]

Fig. 4.A demonstrates the sensitivity of the adsorbed monolayers to the nature of the substituent. The strongest peak near 1500 $cm^{-1}$ is that of the ring vibrations, which clearly shifts with the substituent. An electron-withdrawing substituent, like methyl (Me, $CH_3$) increases the vibration's energy while an electron-donating group, like Br, reduces it. This effect can be roughly quantitated by the Hammett parameter,[155] as shown by the red-circles and dotted line in Fig. 4.B. The net potential developed on these monolayers was measured by a Kelvin probe and is also plotted in Fig. 4.B, as black squares. The contact potential difference values in the figure are relative to that of an H-Si(111) reference sample. Although the substituent affects both the vibration energy and the dipole, different mechanisms are involved. The dipole (or CPD) correlates with the inductive component of the Hammett parameter, while the ring-vibration energy shifts according to the standard, para-position, parameter, $\sigma_P$. These differences are expected, because they originate in different interactions of the substituent with the ring. In any case, Fig. 4 demonstrates that a combi-

nation of aromatic ring with substituent strongly affects the vibrational modes of the ring, and it follows known rules of organic chemistry.

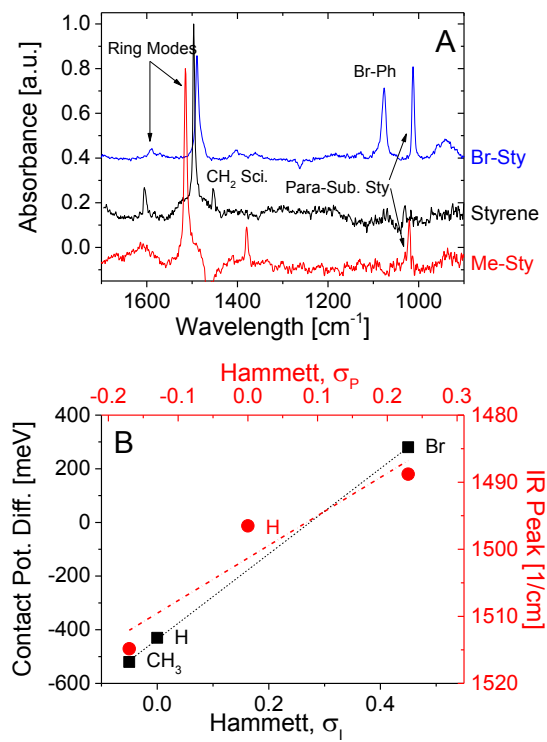

**Figure 4:** Electrical effect of styrene-based monolayers on H-Si(111), using un-substituted styrene ('H'), or para-substituted styrene with bromine ('Br') or methyl (CH₃, 'Me').

A) FTIR spectra of substituted styrene monolayers on H-Si(111);

B) Correlation of surface dipole (black squares, left Y- axis) to inductive Hammett parameter (bottom. X-axis) and the position of IR ring modes (red circles, right Y-axis) against Hammett p-parameter (top X-axis).

Source: Ref. [163] and its SI. Hammett parameters taken from Ref. [155].

Para-substituted phenyls were also used with phosphonic acids,[15,117] benzoic acids,[156-160] dicarboxylic acids[46,70,157,161] and thiols.[162] Binding of substituted phenyls to oxide-free Si was demonstrated by various binding groups.[60,82,163] Substituted aromatics apparently are the major focus for self-assembled monolayer-induced surface dipoles and, indeed, the possibilities are practically endless (see e.g., Refs. 15,164). Major recent efforts are to embed the polarizable groups within the molecular skeleton[164,165] to minimize counter-polarization by the contacts (see section 5).

Still, conjugation is not a strict requirement. Amide bonds are also rather polar and therefore peptides and proteins can develop a significant net dipole across them. However, Ref. 144 shows that, flexible molecules can rearrange to minimize the Coulomb repulsion and this possibility should be considered in designing new molecular systems. An amine group at the terminal of an alkyl chain can also serve to reduce the work function, while a fluoride,[166,167] or other halides,[93,168] increase it. The effect of systematic increase of the degree of fluorination on the work-function was extensively studied.[43,131,145,151] Increasing the number of per-fluorinated methylene (CF₂) groups affects a variety of properties like bulkiness and polarizability, but is not expected to affect the dipole directly: although each C-F bond is strongly polarized, the dipoles of neighboring CF₂ units cancel each other. Therefore, the main dipole is at the bond between the –CF₂-CH₂–,[131] though the number of F atoms may affect the charge rearrangement with the substrate[78] (see section 5).

### c. Correlating the interface dipole with the Schottky barrier of the junction

We now move from a surface dipole to an interface dipole, such as results from inserting a molecular monolayer into a metal / semiconductor ('Schottky') junction (see Fig. 3.D for an illustration). Molecular-dipolar modification of a Schottky barrier was demonstrated on numerous metal / sem-iconductor junctions, made with different semiconductors (Si,[74,93,163] oxidized Si,[162] GaAs,[46,161] SiC,[166] TiO$_2$,[158] ITO[15,157] and ZnO[160,169]) and using different types of monolayers and top contacts. This vast evidence is represented here by two examples in Fig. 5. Panel A shows current as a function of voltage through a molecularly modified n-ZnO(0002)/Au junction.[169] All molecules consist of an identical skeleton of two carboxylic binding groups, which binds via condensation with the ZnO surface hydroxyls. The only difference is the identity of the para-substituted X groups, varying from electron donors (OCH$_3$, CH$_3$) via neutral (H) to electron acceptors (CF$_3$, CN). As can be seen in Fig. 5.A, as the substituent becomes more negatively charged, the onset of the diode current shifts to higher voltage. The quantification of this effect is done using the thermionic emission model (see section 7.a below), where the effective-SBH is proportional to the saturation current in the "off" state of the diode (negative voltage for the n-type semiconductors, shown in Fig. 5).

The saturation current is better visualized using a semi-log plot, as shown in Fig. 5.C. Here a series of alkyl alcohols of varying length was adsorbed on (oxide-free) H-Si(100), with hydroquinone serving as catalyst.[74] Notice that the current increases exponentially (logarithmic Y-scale) with the length of the alkyl chain – opposite to the standard tunneling effect in molecular junctions (see section 6 below). We explain this paradox (longest molecule is best "conductor"), by variation in the induced molecular dipole ($\Delta$) with the length of the alkyl chain, which in turn controls the SBH (Eq. 4) – the dominant barrier of the junction (see further in section 7.d).

Using the (strong) assumption that the molecular-induced dipole does not vary significantly between a *sur*face and an *inter*face, Eq. 4 predicts a linear dependence of the SBH on the molecularly-modified electron affinity ($\chi+\Delta$) of the *surface*. That quantity is measured by the Kelvin probe technique, which yields the above-mentioned contact potential difference, CPD, which is a relative observable. To eliminate the contribution of the surface band-bending to the measured CPD, the sample is irradiated with high-intensity white light so that the photo-generated carriers can practically neutralize the surface, thus virtually eliminating the band-bending (noted as CPD$_L$, X-axis of Fig. 65B). Calibrating the work-function of the reference probe against a known standard (e.g., HOPG) allows to get the absolute value of the effective electron affinity ($\chi+\Delta$ = WF(probe) + CPD$_L$−$\xi$; see Fig. 1 for illustration) as shown on the X-axis of Fig. 65D.

The linear dependence predicted by Eq. 4 is indeed observed in Fig. 5.C-D, but the magnitude and sign of the slope vary between the different systems. The Hg/alcohol-Si junction (Fig. 5.D) has a *negative* slope in agreement with Eq. 4 while the Au/di-carboxylic-ZnO (or GaAs) shows a clear correlation but of an opposite sign. The clear correlation between the diode behavior (SBH) and the molecular property (change in $\chi$) suggests that a given substituent group induces an *opposite* dipole when placed next to air (Kelvin probe) or in a close proximity to Au. The possible cause for such a dipole inversion is further discussed in section 5.d. At this point we focus on the magnitude of the slope, *S*.

We can adapt the concept of index of interface behavior, *S* (Eq. 5) to quantify the dipole effect. Because both the metal (WF) and the semiconductor ($\chi$) are fixed, the *S* factor has to be redefined with respect to the molecular dipole:[161]

$$S_{SAM} = -\frac{d(SBH)}{d(\chi+\Delta)_{Surf}} \qquad (15)$$

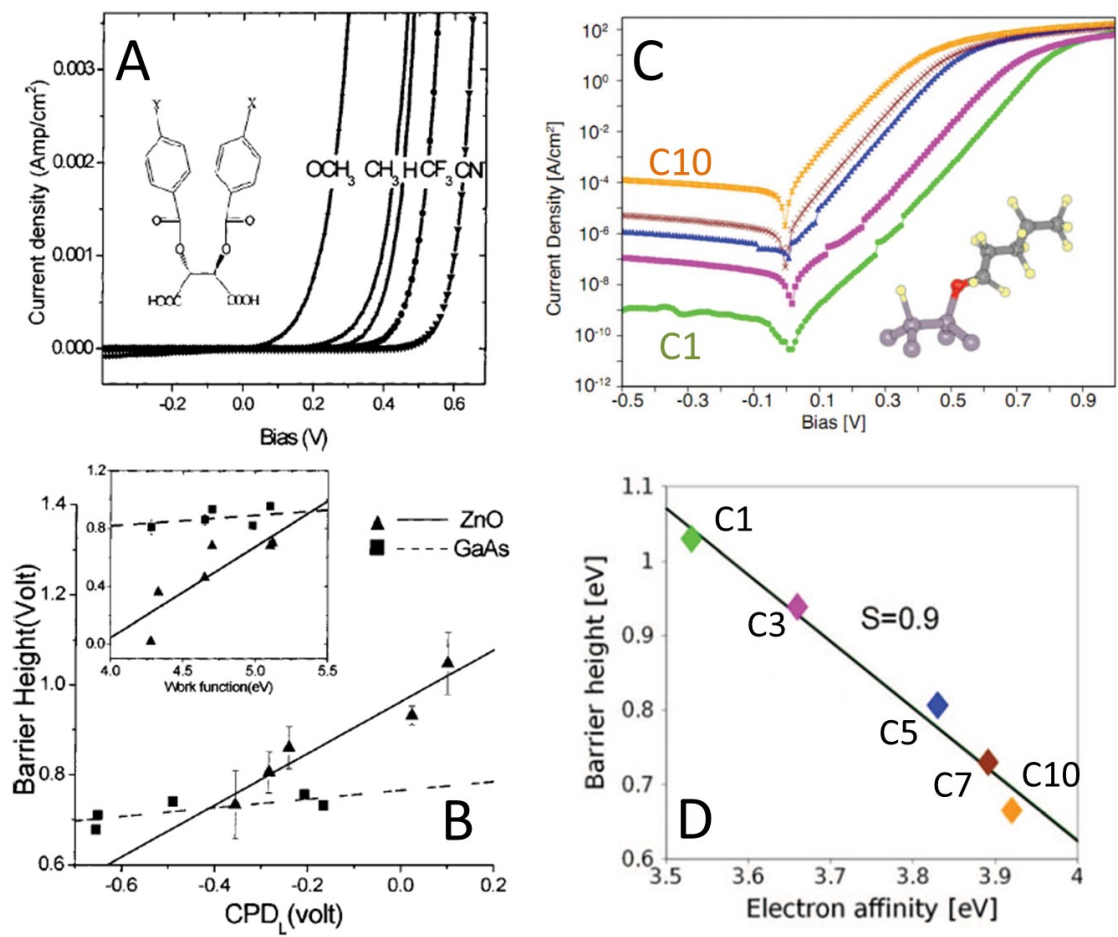

**Figure 5:** Tuning the Schottky barrier height (SBH) by molecular modifications, showing the systematic variation in current-voltage curves (A,C) and the correlation between the SBH and the effective work function (B,D), modified by adsorption of dicarboxylic phenyls with varying para-substituents, on ZnO and contacted by Au (A,B) or varying length alcohols (methanol to decanol) on Si(100) contacted by Hg (C,D). The current-voltage curves are shown on a linear (A) or semi-log scale (C). The effective electron affinity of the semiconductor was measured by Kelvin probe and is shown as raw data of contact potential difference under illumination (CPD$_L$, B) or absolute values (D). Panel B compares the ZnO results to identical molecules adsorbed on GaAs. In all cases, the semiconductor is moderately doped n-type, measured at room temperature. Panels were reproduced from Ref. [169] (A,B ) and Ref. [74] (C,D).

Fitting the SBH, extracted from the I-V curves in Fig. 5.B,D, yields *S*=0.55 for ZnO,[169] 0.09 for n-GaAs and 0.9 for Si.[74] The close to unity *S* value for Si is very impressive, because Si is known to have *S*~0.1 for clean, un-modified interfaces (see orange stars in Fig. 2).[48,52,53] This high *S* value has a fundamental importance because it provides direct evidence that the widely spread idea of Fermi-level pinning in Si ($S{\rightarrow}0$), cannot be justified by an intrinsic property of Si. Rather, it is a property of the way the interface was made and the chemical interactions across it.[24,47,48] For the same reason, the larger $S$ for ZnO compared to n-GaAs is not because the ZnO is more ionic than GaAs (although their extracted $S$ values are indeed very close to literature $S$ values for these semiconductors[169]). Furthermore, the same experiment with p-GaAs yielded $S \cong 0.6$,[161] which is practically the same as for ZnO. Ideally, we wish to get $S{\rightarrow}1$ for any semiconductor / metal combination. The higher $S$ value for the metal/RO-Si (Fig. 5.D) compared to the metal/Mol-ZnO (Fig. 5.B) junctions may be attributed to the excellent electrical passivation of Si by hydroquinone-alkoxy monolayers.[74,85] However, the drastic change in $S$ values for carboxylic acid-modified n-GaAs compared to p-GaAs, is not readily understood. It cannot be due to a doping effect on the monolayer quality, because the span in molecularly-induced surface potential was 0.5 eV, regardless of doping type.[161] A plausible explanation is the fact that these carboxylic acid derivatives have a net effect of acceptors; therefore they a) better passivate the (positively charged) surface states of p- than those of n-type and b) the polarization of the molecular skeleton, common to all derivatives, is to reduce the effective work-function (i.e., opposite to Δ depicted in Fig. 1), which has the effect of reducing the SBH for p-type and increasing it for n-type. Therefore, defective patches have a relatively lower (higher) barrier for n-type (p-type), which will dominate (be insignificant for) the transport.[170] In summary, some compromise on monolayer quality can be made if one aims at reducing the SBH (as in e.g., injecting contacts), but monolayer homogeneity is a strict requirement for blocking interfaces (e.g., active interfaces for solar cells or LEDs). Finally, we note again that the dipole is inverted between the surface and the interface for the combination of Au contact on dicarboxylic acids (the extracted $S$ value is positive, opposite to prediction). Therefore it is possible that the dipole inversion was more efficient on p-GaAs than on n-GaAs, as further discussed in section 5.

### d.     Inversion: interface induced-doping

In section (4.c) we showed that a molecular dipole layer can lead to an almost ideal dependence of the SBH on the effective work function ($S$=0.9). We now ask, what is the extent of the interfacial control – how strong it can be? Traditional solid-state device physics classifies the SBH by the density of carriers in the surface region relative to that in the bulk.[41] The doping of the semiconductor dictates the type of majority carriers (electrons or holes for n- or p-doping, respectively) and their equilibrium concentration (i.e., ξ; see Fig. 6.A), but surface charging leads to bending of the energy bands (red arrows in Fig. 6). Such bending shifts the Fermi level, near the surface, within the gap, i.e., with respect to the conduction band minimum and the valence band maximum. For example, Fig. 6.B depicts an n-type semiconductor with an upward band-bending (BB>0). In this case, the energy difference between the Fermi level and the conduction band minimum is larger at the surface than in the bulk; hence the concentration of electrons at the surface is less than in the bulk. This is the situation, which creates a barrier for transport of majority carriers and allows a meal/semiconductor junction to be rectifying, is known as '*depletion*', namely the majority carriers (e.g., electrons for n-type) are depleted near the surface. Fermi level pinning tends to fix $E_F$ near mid-gap at the surface, which makes depletion the case that is encountered most commonly.

Most often we want to be able to dictate the surface charging; for example injecting (or 'Ohmic contacts') are achieved by increasing the relative concentration of majority carriers at the surface which appears as BB<0, a situation known as '*accumulation*', as depicted in Fig. 6.C. In the other direction, increasing the depletion, and thus the upward band bending will bring the Fermi-level at the surface to a mid-gap position (dash-dot line in Fig. 6.A), where the density of holes and electrons is equal. Pushing the Fermi level below mid-gap implies that at the surface the density of (bulk) minority carriers exceeds that of (bulk) majority carriers. Thus, at the surface of a bulk n-type semiconductor [holes]>[electrons], i.e., at the surface the material is p-type (and *vice versa* for a bulk p-type material). This regime is called '*inversion*' (of type of carriers), a situation that is illustrated in Fig. 6.A. The value of the BB that defines the inversion threshold depends on the intrinsic doping via $\xi$: $BB_{inv} \geq E_g/2 - \xi$.

Crossing this threshold implies that minority carriers dominate at the surface, but their density is still small compared to the bulk density of majority carriers. The two populations become comparable in the '*strong-inversion*' regime, when the BB exceeds twice this energy: $BB_{StrInv} \geq E_g - 2\xi$. Under strong inversion, the semiconductor surface region behaves practically like a p-n junction, which makes it technologically very appealing.[23] The importance of the different SCR regimes is demonstrated in Fig. 8.A, showing current – voltage traces measured over a series of n-Si(111)-Styrene-X / Hg junctions, where the substituent, X, is either Br, $CH_3$ or H (un-modified styrene; same monolayers as in Fig. 4). Depending on the substituent, the trace changes from highly rectifying (as for $CH_3$) to high-current, Ohmic bias dependence for Br. The latter, Ohmic behavior, characterizes a junction under accumulation; depleted junctions show asymmetric, rectifying I-V traces, and their saturation current under reverse bias becomes exponentially smaller as the SCR changes from depletion to strong inversion. How a single substitution of Br for H can induce 7 orders of magnitude change in current is further explained below.

In principle the transition between the different surface charging regimes can be achieved by tailoring the difference in WFs between the metal and the semiconductor (Eq. 4). However, in reality, this hardly occurs for metal-semiconductor junctions due to the low index of interface behavior ($S \ll 1$, see Fig. 2). As argued in section 2, adding a dipole layer at the interface is equivalent to varying the effective WF by the potential drop at the interface, $\Delta$, imposed by that layer. Therefore, minute chemical changes at the interface can be equivalent to altering the metal. This is illustrated in Fig. 6.A-C depicting three metal / semiconductor interfaces at equilibrium (V=0) with identical (WF-$\chi$) difference, but with varying interfacial dipole layer. A negative interface dipole (acting to reduce the apparent WF of the semiconductor, e.g., $CH_3$-Styrene) will drive the SCR toward deeper inversion (Fig. 6.A), while a positive interface dipole will drive the SCR toward accumulation (Fig. 6.C, e.g., Br-Styrene). In principle, applying voltage across the interface (Fig. 6.D,E) also changes the charging of the SCR, but the accepted classification is based on the junction's state at electrical equilibrium or zero voltage (Fig. 6.A-C). Therefore, we can quantify each of the I-V traces of Fig. 8.A to extract their SBH (as done in Fig. 5; see section 7 for further details) and using Eq. 3 verify the equilibrium $BB_0$ with respect to the inversion / depletion threshold.

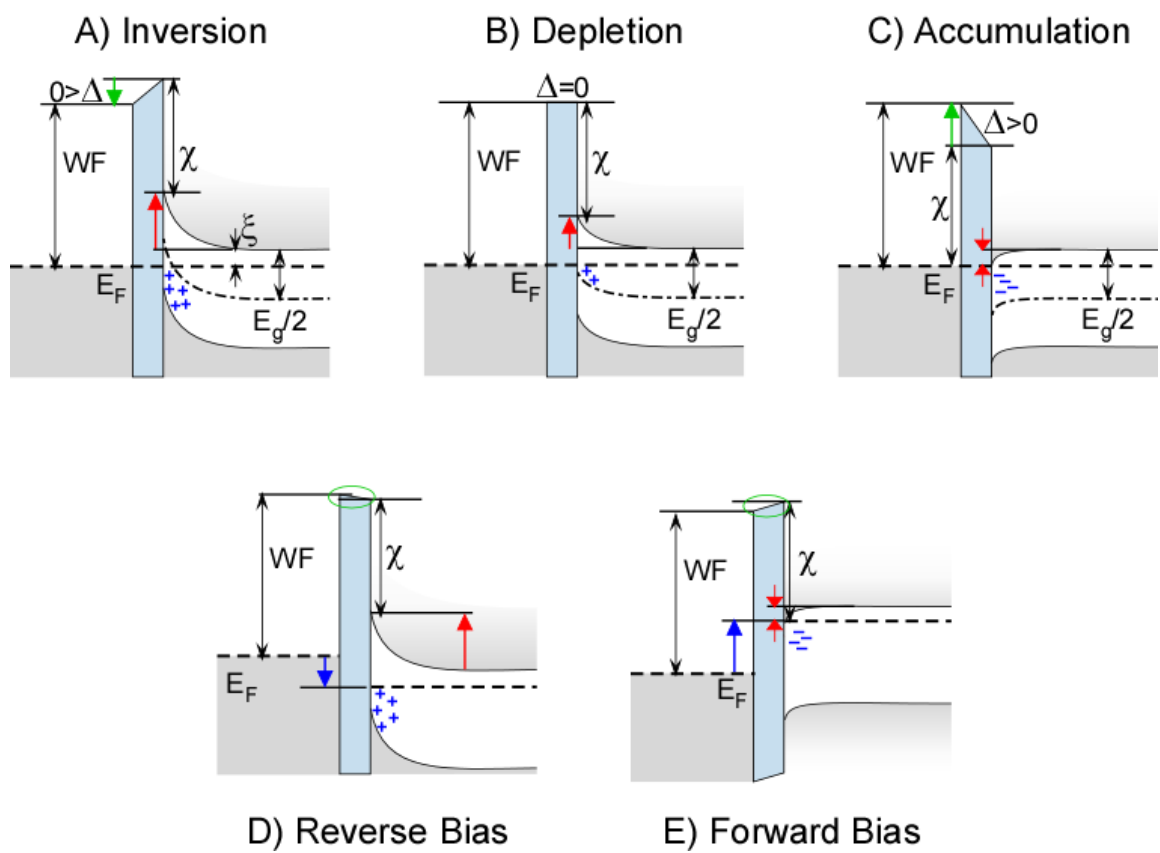

**Figure 6**: Energy schemes illustrating the different types of the semiconductor space charge region; for a fixed combination of metal (left) work-function (WF) and n-type semiconductor (right) electron affinity (χ), a dipolar interfacial layer (light blue) can control the space charge by the potential drop (Δ, green arrow) that it imposes: **A)** Δ<0 → inversion; **B)** Δ=0 → depletion, or **C)** Δ>0 → accumulation, where the (in)equalities refer to an equilibrium state with no applied bias. Red arrows indicate the band bending and blue + , − depict the sign of the space charge, $Q_{SCR}$. The lower panels show the depletion junction (**B**) under **D)** reverse bias or **E)** forward bias, namely V<0 or V> 0 on the metal side, respectively, as marked by a blue arrow. The space charge (stronger depletion in **D** than in **B**; accumulation in **E**) induces a small capacitive polarization of the insulator, marked by a green oval. This effect becomes more significant the closer the majority carrier's quasi-Fermi level approaches the band edge.

The extracted SBH (left Y-axis) and corresponding $BB_0$ (right Y-axis) of the Si-Styrene-X/Hg junctions are plotted in Fig. 8.B against the surface potential of the molecularly modified surface. The horizontal grid of Fig. 8.B shows the threshold BB values for transition between the different charging regimes for this specific Si. The $V_{FB}$ value, extracted from C-V measurements (red-triangles) shows that the methyl-styrene is on the verge of strong inversion, while the normal styrene is on the threshold between inversion and depletion. As far as the Br-styrene barrier can be trusted (based on I-V extracted barrier) this junction is nearly at flat band (between depletion and accumulation). This is a fine example that not only the SBH follows the surface dipole (i.e., S→1), the extent of change in SBH can reach extreme states. The I-V extracted SBH is less reliable for deducing if the semiconductor is in depletion or (strong) inversion, as further considered in section 7.

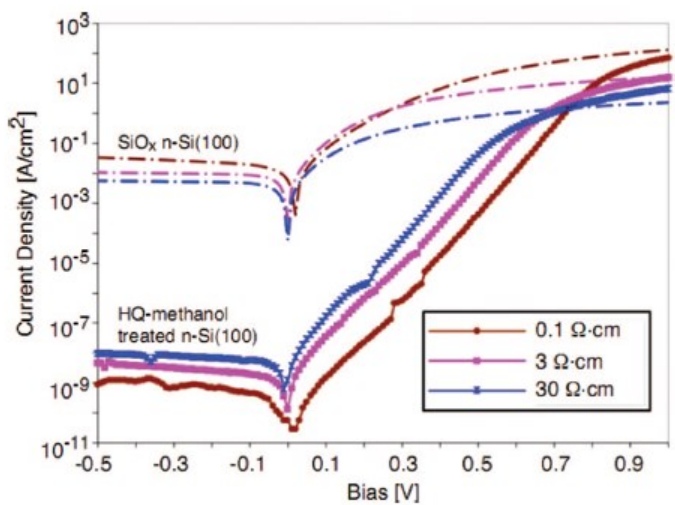

**Figure 7**: The effect of Si doping levels (see in-figure legend) on current-voltage measurements of n-Si (100) treated for 3h with 0.01M hydroquinone-methanol solution (symbols) or including a thin $SiO_2$ film, grown for 30 min in piranha solution (dashed lines). All junctions were with Hg top contact. The Si doping levels, $N_D = 7 \times 10^{16}$, $10^{15}$ and $10^{14}$ cm$^{-3}$, correspond to wafer resistivities of 0.1, 3 or 30 Ω·cm, respectively. The inverse relation between doping level and current is a direct indication that transport is dominated by minority carrier generation-recombination currents. This, rarely demonstrated, behavior for metal-semiconductor junctions is due to the effective molecular passivation of the Si surface, which does not exist in the oxide-modified interfaces (dashed lines). Figure reproduced with permission from Ref. [74].

Further direct evidence that adsorbed molecules are capable to induce a p-n junction close to the interface is shown in Fig. 7. It shows current-voltage traces of a Hg/Si(100) junction, where the Si surface is either naturally oxidized (dashed lines) or modified by hydroxyl-quinone / methanol treatment (solid lines). The three curves present cases of junctions made with three different doping levels of the n-Si substrate, moving from low doping (30 Ωcm) to medium-high doping (0.1 Ωcm). The highest current among the molecular monolayer-modified interfaces is obtained for the most *resistive* wafer. This can be understood only if the current is limited by the availability of *minority* carriers, which is inversely proportional to the doping level, $N_D$. The oxidized interface shows the common case of a junction in depletion (dashed lines): the rectification is much poorer and the current is largest for the least resistive wafer, indicating transport is by majority carriers.

Therefore, both Fig. 8.B and Fig. 7 show that monolayers adsorbed at a Hg/Si interface can override the original doping of the Si surface and induce a p-n junction in the SCR. This complements the high *S* value shown in the previous sub-section, (Fig. 5) which together point to the great power of judicially chosen and applied chemical modifications of interfaces in tailoring the SBH of metal / semiconductor junctions. In addition to their own dipole, molecular monolayers can provide good electrical passivation and block direct metal / semiconductor interactions. Figure 6 exemplifies the interchangeability between the effect of a localized electrostatic effect, that of a molecular dipole layer (Δ) and of a long-range effect, as expressed by band bending (BB) or the SBH. The last critical ingredient in this description is the (unintended) induced dipole which is formed by the close proximity of the organic monolayer to either the metal or the semiconductor as discussed in section 5.

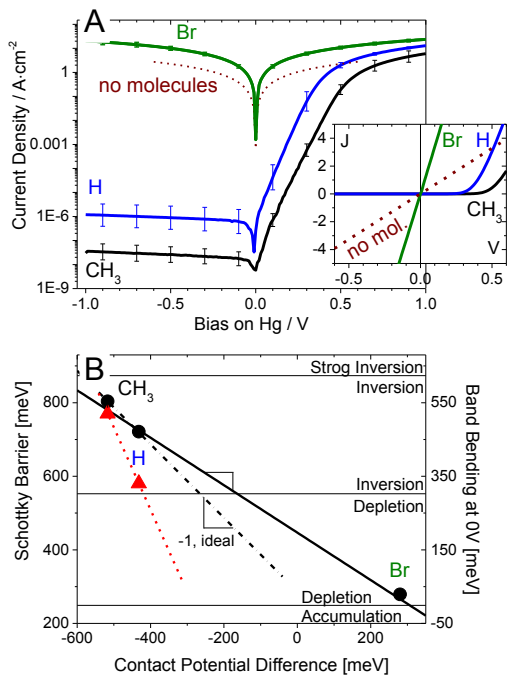

**Figure 8** Change from accumulation to inversion for junctions with differently substituted styrene monolayers at the Hg/Si interface, showing **A)** Current density-Voltage, J-V, characteristics and **B)** Schottky barrier height, SBH, as function of surface potential of molecularly-modified Si. Styrene monomers were para-substituted by -Br, CH₃, or none (H, see legends in plot). J-V traces (**A**) are shown on semi-log scale (main panel) and on linear scale (**inset**). Dotted line in (**A**) shows result for bare junction without molecules (n-Si/Hg). The SBH in (**B**) was extracted from the J-V (black circles) or capacitance-voltage (red-triangles) data (see section 7). Right Y-axis is corresponding band bending at 0 V (BB₀, see Eq. #3) and the horizontal grid refers to this scale (see text). The black solid line is a fit through the three points; the red line and the dotted black line are guides to the eye.

## 5.  Interface doping and charge rearrangement

After describing in the previous section the great power that molecular dipoles can exert over junctions, we now focus our attention on the ways that the contact influences the net interface dipole. That dipole is not only the result of intrinsic polarization of the molecule, but has a considerable contribution due to interaction with the contacts.[10] Such an 'induced dipole' is the focus of this section. The following discussion assumes chemically intact, clean and abrupt interfaces[58] as imperfections were already considered in section 3. Issues of monolayer degradation and defects within it have been described elsewhere.[7] It is common to distinguish between physical and chem-ical effects of the substrate on the induced dipole,[171] though the ISR view (section 2) implies that these effects are highly entangled. The so-called, physical contribution originates in the 'surface dipole' intrinsic to any clean solid surface, defined as the difference between the work function and the chemical potential of the electron in the bulk.[171] It can be visualized as the spill-over of the electron wave-function outside the solid.[48] The net result is a negative pole pointing away from the substrate, which we define as a positive dipole (e.g., $\Delta_0$ in Fig. 1 is positive, viewed from the semiconductor). Any adsorbate will hinder such tunneling-out of the electron wave-function (also known as exchange or Pauli repulsion[172]), and therefore adsorption has a universal component of reducing the work-function compared to the clean surface, known as 'push-back', 'pillow' or 'cushion' effect.[172] Practically, assessing the reduction in the intrinsic surface dipole is challenging, both experimentally and computationally, due to the lack of a clear reference surface. The push-back negative dipole is considered 'physical' because it is generally insensitive to the nature of the molecule,[172] while it increases for metals of higher work-function[171] and more polarizable metals (e.g., Cu).[171,172] The push-back effect has been widely covered elsewhere[45,48,58,172] and therefore will not be dwelled upon here. The 'chemical' effect of the substrate implies some degree of charge exchange with the adsorbate, existing even for very weak interactions (e.g., without formal chemical binding). This 'chemical' contribution of the contact to the molecular polarization is the topic of this section.

### a. Electrochemical balance at a solid surface

While the dipole component induced by the molecule itself is conceptually simple, the component emerging from local charge rearrangement ('induced dipole') is less trivial. A detailed quantum mechanical consideration of abrupt inorganic interfaces can be found in Ref. 24. Here we provide some rough guidelines, based on balance of the electrochemical potential.[40,47,48,171]

Already in the late 1960's it was recognized that the index of interface behavior, $S$ (see Eqs. 5, 7) is related to the ionicity of the semiconductor, the difference in the electronegativity of the anion and cation forming a compound semiconductor.[53] Namely the more covalent is a semiconductor the smaller is its index of interface behavior. For example, clean Si (no adsorbed molecules as in Section 4) has $S \simeq 0.1$ while oxidized Si has an almost ideal $S$ value ($\sim$1). However, this was an empirical observation without a clear physical basis. Tung later suggested to view a metal (M) – semiconductor (SC) interface as a parallel array of M-SC bonds (or "molecule-like", where M and SC represent single atoms).[47] Charge will be exchanged between these two species to equilibrate the local (atomistic[47]) electron chemical potential across the two sides of the interface.

The *driving force* is the difference in the electrons' *chemical potential*, $\mu$ between the atoms constituting the interface, while the *amount* of transferred electronic charge is dictated by their *hardness*, $\eta$ defined as:[171,173,174]

$$\mu = \frac{\partial E}{\partial n_e} \approx -(IP + \chi)/2 \qquad (16)$$

$$\eta = \frac{\partial^2 E}{\partial n_e^2} \approx (IP - \chi)/2 \qquad (17)$$

where E is the energy and $n_e$ is the number of electrons. The exact, derivative-based definitions can be approximated by the frontier orbital energy levels: the ionization potential, IP, and electron affinity, $\chi$. The approximated chemical potential (Eq. 16) is also known as the Mulliken or absolute electronegativity.[174]

For organic conductors, the concept of induced density of interface states (IDIS) enables defining a charge-neutrality level (CNL) as the highest occupied IDIS level, for the electrically neutral system.

This CNL is not necessarily at mid-gap,[58] which is obviously more accurate than the crude definition of Eq. 14. The continuous adjustment of the position of the CNL within the molecular gap is justified by the existence of IDIS, which provides a continuum of finite density of states at any energy within the gap.[58] This is in contrast to the electron chemical potential view which considers $\mu$ as a virtual level, like the position of $E_F$ within a forbidden energy gap of semiconductors[47] or oxides.[175] Experimentally, the CNL is deduced from measurement of the work function and the frontier orbital energy level, before and after contact.

For organics, the existence of IDIS or the broadening of the molecular levels is well documented and explained by interaction of the molecules with the metal,[56] and / or on various other interactions with neighboring molecules.[11] Yet, the metal hardly affects the CNL position, which appears as a characteristic of the organic molecule regardless of the type of metal or interaction strength with it (or level broadening).[56,176] Furthermore, CNL works qualitatively well even without a metal, for organic / organic interfaces.[57] In terms of energy alignment, the 'IDIS-CNL'[54,56-58] and 'chemical potential' concepts [47,171] are very similar; since we focus here on the intimate, direct organic / inorganic interface, there is little doubt that some energy mixing between the organic and the substrate take place, and therefore both views are relevant.

Combining the chemical potential (Eq. 16) and the hardness (Eq. 17) gives an estimate for the amount of charge, $n_e$, shifted across an A-B bond, or across an interface in our context:[40,47,171]

$$n_e = \frac{1}{2} \cdot \frac{\mu_A - \mu_B}{\eta_A + \eta_B} \qquad (18)$$

The net displaced charge, $n_e$ over a distance $d_B$ creates a dipole: $p = d_B n_e$, which is further translated into a net potential step, $\Delta_{CR}$ (see Eq. 13):

$$\Delta_{CR} = qN_B \frac{n_e \cdot d_B}{\varepsilon_0 \cdot \varepsilon_i} \qquad (19)$$

where $N_B$ is the number of interface polarized "bonds" and $\varepsilon_i$ is the relative dielectric constant of the interface (see Eq. 13). In the case of no molecules (or contaminations) at the metal/ semiconductor interface, it can be roughly approximated as the dielectric constant of the bulk semiconductor[47] (i.e., 11.7 for Si). Tung transcribed Eq. 18 to solid state terminology by setting for a metal: $\mu_M = WF$; $\eta_M \propto 1/DOS(E_F) \simeq 0$ and for a semiconductor identifying IP with $E_v$ and $\chi$ with $E_C$ (see Fig. 1), which gives $\mu_{sc} = \chi + E_g/2$; $\eta_{sc} = E_g/2$. Placing these terms into Eqs. 18, 19, yields[47] (sign definition follows the convention that $\Delta$ in Fig. 1 is positive):

$$\Delta_{CR} = qN_B \frac{d_B}{2\varepsilon_0\varepsilon_{SC}} \frac{WF - \chi - E_g/2}{E_g + \kappa} \qquad (20)$$

where $\kappa$ accounts for Coulombic interactions but is small for inorganic interfaces[47] and probably even smaller for covalent binding or physisorption of organic matter[40,171], but might be more important for ionic binding groups like carboxylic or phosphonic acids.

Inserting Eq. 20 to $\Delta$ of Eq. 9 (using CNL=$\chi + E_g/2$) provides a new quantitative definition for the index of interface behavior:[47]

$$S = 1 - \frac{q^2 N_B d_B}{\varepsilon_0 \varepsilon_{SC}(E_g + \kappa)} \qquad (21)$$

Eq. 19 predicts that (1-S) is inversely proportional to the forbidden energy gap of the semiconductor (expressing its "chemical hardness"), which is found to be correct for clean inorganic interfaces[47] as shown in Fig. 2.B. Compared to the "surface states" explanation of $S$ (Eq. 10), the hardness/chemical potential logic (Eqs. 20, 21) does not require any interface insulator and it provides a natural explanation for why SBH values tend to be often pinned to mid-gap of inorganic semiconductors ("CNL" $\approx E_g/2$): it is because the chemical potentials ($\mu_{sc}$) of the semiconductor's constituent atoms approximately coincide with the mid-gap position.[47]

Alternatively, for organic electronics, an expression similar to Eq. 10 was devised where the induced density of states in the organic layer (IDIS) near the Fermi level replaces $D_{IS}$, and the image charge distance replaces $L$. [56,58] The index of interface behavior, $S$, thus emerges as a measure of the interface polarizability or how well the interface screens potential differences. $S \rightarrow 1$ when the interface is chemically 'hard', allowing little charge rearrangement upon contacting ($n_e \rightarrow 0$, Eq. 18; $\Delta_{CR} \rightarrow 0$, Eqs. 9-10 or Eqs. 19-20). Such limited charge rearrangement is also known as a case of 'poor screening'. It follows that interfaces that are made up of materials with wide band-gap and low dielectric constant are less polarizable ($S \rightarrow 1$), while the large density of interface states (Eq. 10), either induced (IDIS) or due to imperfections, yields screening due to polarizable interfaces, with $S \rightarrow 0$. Notice that the same term, 'CNL', is used for two very different scenarios: a situation where there is an interface insulator and traps (Fig. 1.A and Eq. 10), and, for organic electronics, the IDIS-CNL view which does not include any insulator or surface defects. Overall, the IDIS-CNL concept predicts very well the interface dipole formed at metal / organic and organic / organic interfaces.[54]

It is not trivial to reconcile between the 'physical' push-back contribution and the 'chemical' charge-rearrangement one ($\mu$ or CNL). While the magnitude of the pure push-back ($\Delta_{PB}$) effect can be very large ($|\Delta_{PB}| > 0.5$ eV),[172] Vázquez et al., have argued that the index of interface behavior, $S$, is decisive also in this case:[58]

$$\Delta = S \cdot \Delta_{PB} + (1-S) \cdot [WF - (IP - CNL)] \qquad (22)$$

where the second term is taken from Eq. 9 and in principle can be replaced by $\Delta_{CR}$ and $S$ of Eqs. 20, 21. The physical push back effect, $\Delta_{PB}$, dominates only junctions of poor screening ($S \rightarrow 1$) for which vacuum level alignment holds. In contrast, organic electronic thin films are often highly polarizable ($S \rightarrow 0$) so that even a nominally large $\Delta_{PB}$ has limited effect on the final energy alignment.[58] This is in line with the ISR view (section 2) that views the entire interface as a new chemical species,[24] i.e., the quantum-mechanical consideration ('Pauli repulsion') cannot be separated from the chemical identity of the interface (charge rearrangement).

 The electrochemical balance is a very-short range one. Therefore, the specific M-SC bonds considered in Eq. 20 are not expected to take place in an MIS interface where a thin insulator (molecular or inorganic) is inserted at the interface. Instead, two electrochemical balances are maintained between each contact and the molecule, and they may possibly be localized to a part of the molecule (e.g., binding or terminal groups).[177]  Compared to inorganic interfaces, organics have fairly localized nature of energy states and therefore charge reorganization at these interfaces was recognized already at a quite early stage.[45,178] This also means that the difference between an ISR and space charge (short and long range, respectively) is somewhat less defined in organic semiconductors. The hardness and chemical potential of the organic monolayer is expected to greatly vary with the nature of its atoms (e.g., halogen substitution for H on a molecule) and degree of saturation.  However, this fascinating issue is as yet largely unexplored, especially in the context of monolayers.

**b.  Charge rearrangement at metallic contacts to molecules**

The notion of Fermi-level (electron chemical potential) equilibration (Eq. 4, section 2.a) is fundamental to any electrical interface. While the previous sub-section (5.a) borrowed molecular concepts to describe inorganic interface atoms,[47] the opposite analogy – describing the alignment of energy levels of molecular adsorbates by equilibration of their chemical potential / Fermi level is rarely used. This sub-section attempts to apply the above 'bulk' consideration to the extreme limit of a molecular monolayer. The energy-alignment issue is common to fields ranging from thin film organic electronics via monolayers to single molecules. The confusion originates in the lack of formal chemical bond at the interface (in most cases) and the fact that no free carriers are expected for the organic component. Thus, the concept of 'electron flow' from low to high work function side of the interface is inapplicable to most molecular materials. For these reasons, organic / organic or organic / inorganic interfaces were originally often assumed to follow 'vacuum level alignment' (cf. Fig. 9, below) or the ideal Schottky-Mott rule ($\Delta$=0 in Eq. 4).[40,55]

However, there is plenty of recent evidence for localized, fractional (per molecule) charge rearrangement that acts to maintain the electrochemical equilibrium across the interface.[45,97,100] This section is based on analogy between molecular monolayers and organic electronics, or molecular films which are ≥ 50 nm thick.[40,55] Since these are molecular solids, the fundamental processes are assumed to be similar to those in monolayer films. Thus, we can generally say, that although molecules that do not have free electronic charges, formally we cannot associate with them a Fermi-level. There is, though, a mechanism for equilibrating the electrochemical potential of the electron across the interface as detailed in section 2 in general and in the previous sub-section (5.a) specifically. Figure 9 illustrates schematically different possible scenarios of energy alignment across molecule / metal interfaces. Notice that the horizontal scaling in Fig. 9 is misleading: the whole organic film (green shade in Fig. 9) refers to a monolayer (without a "top contact" onto it), and we zoom in on the few atomic layers right at the border between the metal and the organic.

As for inorganic interfaces (Fig. 1), there are two key energy characteristics for the interfaces, which are related: the potential step ($\Delta_{CR}$ or "bond dipole") and the tunneling energy barrier ($E_T$, "injection barrier" in organic electronics terminology). The only difference between a monolayer and a thicker film (section 5.a) is that here we examine the extremely steep and narrow, localized potential step, $\Delta_{CR}$ located within 1-2 atomic layers from the interface, rather than across the length of the molecule (i.e., $\Delta_{mol}$ is ignored). This is a conceptual simplification, because in reality the two contributions of intrinsic ($\Delta_{mol}$) and induced dipole ($\Delta_{CR}$) are intertwined as considered in the following sub-sections. The transport energy barrier, $E_T$ is analogous to the Schottky barrier at inorganic interfaces, and is important when the organic substance is the active transport material, as in e.g., thin-film organic electronics,[40,55] or for molecular junctions with pure metal contacts.[7]

As an illustration, Fig. 9 shows two extremes of low (red shade) and high (blue shade) work-function metals, where the interface is either chemically inert or hard (large η in Eq. 18) which is marked by a vertical gray line (Fig. 9.A,C) or chemically active, with significant IDIS or low η, marked by a gradual color change across the interface in Fig. 9.B. Fig. 9.A depicts a chemically inert interface where localized charge rearrangement is blocked (poor interface screening or $S \to 1$, see Eq. 22) and therefore the *vacuum levels are aligned* and no electrical equilibrium is established across the blocking interface. In such a case, the tunneling barrier, $E_T$ (arbitrarily chosen to be relative to the HOMO) follows closely the changes in the substrate work function, and the interface is said to be in the Schottky-Mott limit (cf. Eq. 5 with $E_T$ instead of SBH, with $S \to 1$). In all the different panels of Fig. 9, the amount of displaced charge, $n_e$ is so small, compared to the huge metal density of states, that the WF is practically fixed, i.e., the metal's Fermi level with respect to $E_{vac}$. Thus, not shown is a possible WF change due to the push-back effect, as described above.[171,172]

The second column of Fig. 9 shows the extreme opposite to vacuum level alignment, bond polarization. Combining Eqs. 2 and 16, and assuming the molecule is at equi-potential (i.e., $\phi$=0 in Eq. 2), places the chemical potential of the electrons in the molecule roughly at the middle of the HOMO-LUMO gap. In such a view, the molecular levels have a fixed alignment with respect to the Fermi level of the substrate, regardless of the metal's work function (compare top and bottom panels of Fig. 9.B). As a result the tunneling barrier, $E_T$ is *independent* of the metal's work-function and coincides more or less with the molecular mid-gap, or $S{\rightarrow}0$. Nonetheless, the variation in work-function should be balanced. The mechanism to maintain this balance is polarization of charge localized in the immediate vicinity of the bonding atoms (Eqs. 16-20), as shown in Fig. 9.B.

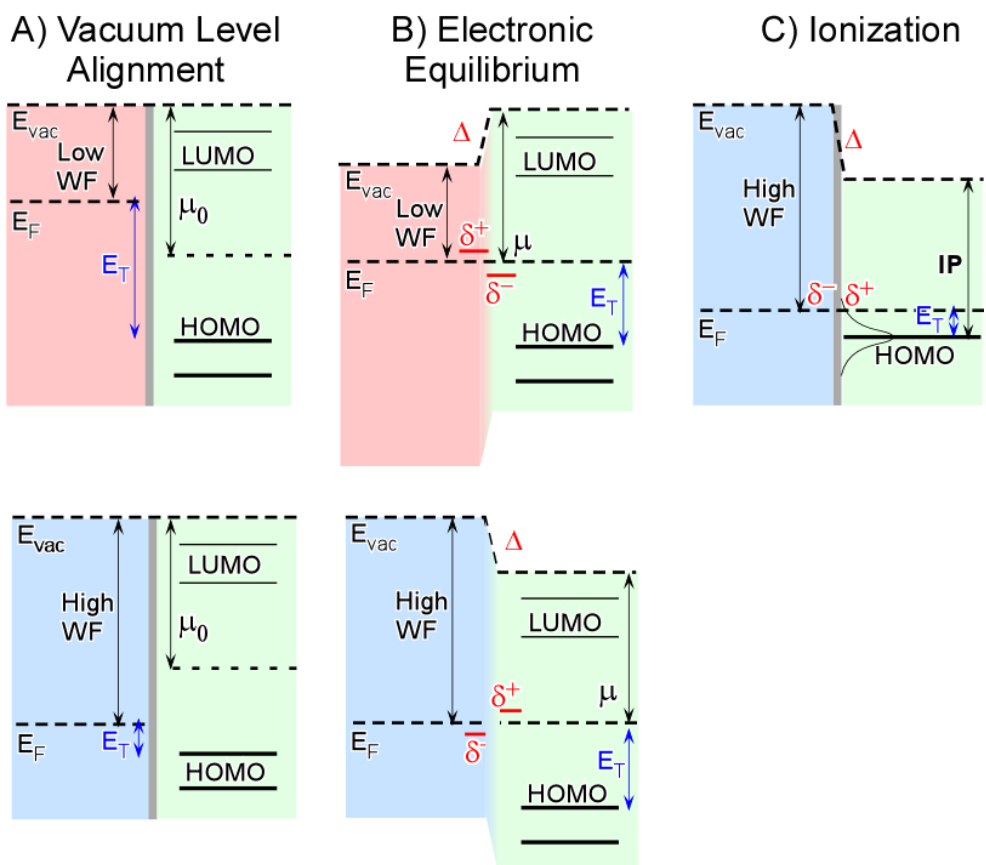

A) Vacuum Level Alignment

B) Electronic Equilibrium

C) Ionization

**Figure 9:** Different scenarios for molecule / substrate energy alignment. The molecule is depicted by a green background and 4 molecular levels (horizontal lines); the solid substrate has red or blue background for low or high work-function, respectively. A gray line at the interface marks lack of interface states. In all panels, $\Delta$ is the interface potential step and $E_T$ is the tunneling barrier or the distance between the metal Fermi level (dashed line) and the molecular HOMO level.

A.  Vacuum level alignment is in this case a state of non-(electrical) equilibrium;

B.  Bonding groups or hybrid states localized at the immediate interface can be polarized so as to maintain a net equilibrium between the electron chemical potential of the molecule ($\mu$) and the Fermi level of the solid, $E_F$; this situation results in an extremely sharp induced potential energy step, $\Delta$;

C.  In cases where the substrate's Fermi level is outside the molecular gap, (IP < WF in our example) the molecule would get charged. However, the low interface permittivity enforces a non-zero $E_T$.

Energy alignment between a molecule and a metal is conceptually similar to a solid-state heterojunction (section 2, Eqs. 4, 5): vacuum level alignment ($\Delta$=0, $S$→1 or 'Schottky-Mott limit') means that the difference in the chemical potential of the isolated phases is balanced by a long-range charging (e.g., space charge that is the origin of BB in semiconductors, or uniform charging of a monolayer), while Fermi-level pinning happens when the balancing charge rearrangement is localized to the bonding atoms ($\Delta$≠ 0, $S$→0 or 'Bardeen limit'). The latter process is termed 'charge rearrangement', [171] (CR subscript) to stress that generally, no actual ionization is involved. The possi-bility that molecules develop an abrupt electrical potential profile within them is often overlooked, possibly because their net length is so short (1-2 nm). But, their localized energy levels suggest that such internal polarization is actually highly possible.

The charge rearrangement described in Figure 9.B involves extremely small amount of charge per molecule (≤ 1% of an electron) because of the very low IDIS. Actual ionization is possible only with respect to the real molecular levels (not the chemical potential), as shown in Fig. 9.D. For example if the molecular HOMO is at higher energy than the substrate's Fermi level (IP < WF) the molecule will donate electrons to (is oxidized by) the substrate as depicted by the δ± signs in Fig. 9.C. The opposite case (not shown) is for WF < χ, where the molecule accepts a negative charge from (is reduced by) the substrate. Seemingly, such scenario should yield a perfect alignment (resonance) between the frontier molecular orbital and the substrate Fermi level. However, extensive studies in organic electronics [40,179] indicate that IP is pinned slightly *below* $E_F$, even if by vacuum alignment it should be above it. This effect was originally attributed to polaron energy,[40,179]; it was later realized that it is another manifestation of the low interface polarizability.[175] The low permittivity of organic matter, and more so if contacted by oxides[179] or air-born contaminations,[40] implies that the interface develops a capacitive potential drop (e.g., Eq. 24below), which limits complete charging of the organic layer. Therefore, Δ in Fig. 9.C is due to capacitive charging of delocalized charges, different from Δ in Fig. 9.B, which is due to localized polarization. In both cases the low permittivity of the organic material allows it to sustain immense electric fields, larger than known breakdown fields in bulk dielectrics.[175] Ley et al. have estimated that even when IP is 2 eV smaller than the WF, only 3% of the molecules in a monolayer are actually ionized and as few as 0.4% when a 6 nm oxide buffers between the monolayer and the metal.[175] This fraction is depicted in Fig. 9.C as a distribution in energy of the HOMO level where only its remote edge is actually aligned with the substrate's $E_F$, and its center is significantly below it. In the absence of IDIS (Fig. 9.A,C), individual molecules can be either neutral or fully ionized (also known as integer charging), while if IDIS or ISR are formed (Fig. 9.B) the molecular energy levels are altered and therefore, compared to their gas-phase state, can appear as if each molecule has accepted / donated a fraction of an electron.[61]

An intuitive explanation why certain organic / inorganic interfaces follow vacuum level alignment while others are governed by bond polarization, is still missing, although there was immense progress in this field.[55,177] Obviously, electrochemical equilibrium requires a "chemically-soft", polarizable interface. Metals are extremely polarizable (η → 0) compared to organic matter and therefore intimate contacts of organics to clean metals often show large charge-rearrangement (Fig. 9.B, S→0), as, for example dye molecules evaporated on a clean metal surface under UHV conditions.[55] On the other hand, a few Å wide interfacial contamination (e.g., gray line in Fig. 9.A) would block the charge rearrangement between the polarizable metal and the electrically active organic layer, leading to an apparent vacuum level alignment. Practical examples for non-intimate contact include: air-borne contaminations underneath spin-coated polymers[40,55] or a thin oxide skin on contact to large-area molecular junctions (e.g., GaOx on EGaIn) as well as a mechanical gap between top contact and substrate due to surface roughness or deposition method[161] or in traditional STM (not break-junction). A few monolayers of salt underneath dye molecules also serve as efficient buffers.[61] Formally, the thin gap (Fig. 9.A) is capacitor-like, but so little charge rearrangement is required to maintain the electrochemical balance that the electrical potential difference across the interface gap is negligible, and vacuum level alignment is maintained.

Formation of IDIS in molecule / Si interfaces was demonstrated by DFT, as shown in Figure 10. It shows DFT-computed density of states (DOS) at the interface between alkyl/alkenyl chain monolayers and oxide-free Si(111), for alkyl(Fig. 10.A) or alkenyl (Fig. 10.B), with a Si-CH₂-R or Si-CH=CH-R link, respectively.[180] The plots show the spatially resolved DOS (gray scale) as a function of energy (Y-axis) and position across the interface (X-axis). The position is given both in Å (bottom axis) and in atomic coordinates (top axis), from 1 (at the interface) to 6 (furthest away from the inter-face). The white region in the middle corresponds to the forbidden energy gap, which is clearly narrower for Si (left side) than for the alkyl monolayer (right side). The interesting part (marked by a red background) is the transition between these two well-defined gaps. Even with the fully satu-rated, closed-shell, chemically 'hard' alkyl chain, the transition is not abrupt and extends from the last Si atoms ('Si1') till beyond the second C atom ('C2'). If the first couple of C atoms are unsatu-rated as in alkene monolayers (Fig. 10.B) the transition region extends well to the 4$^{th}$ C atom. This demonstrates that although the C=C π orbitals do not formally participate in the Si-C binding, they strongly interact with the Si levels, as evident by the penetration of Si states deep into the alkene chain. Such hybridization of π orbitals with Si states was demonstrated also for aromatic monolay-ers on Si.[60] This finding is not trivial because of the large energy difference between the frontier Si levels (i.e., valence band maximum and conduction band minimum) and the alkyl HOMO and LU-MO. Still, there is a clear level-hybridization, or IDIS, even for the fully saturated alkyl (Fig. 10.A). The emergence of unique interface states is the basis of the ISR and IDIS-CNL concepts described above (section 5.a). What Fig. 10 shows is that the extent of these states is very sensitive to the chemical environment, beyond the immediate binding atoms.

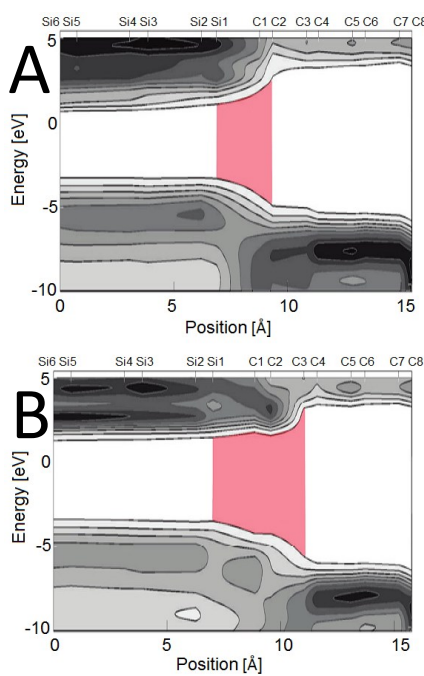

**Figure 10:** Contour maps of the local density of states for **(A)** alkyl and **(B)** alkenyl mono-layers on Si(111). The interfacial transition region is emphasized in red, and is under-stood as some sort of Induced Density of In-terface States, IDIS. Depending on if the bond-ing carbon is saturated (Si-CH$_2$-, as in **A**) or unsaturated (Si-CH=CH-, as in **B**), the narrow Si gap extends up to the 2$^{nd}$ or 4$^{th}$ carbon. Fig-ure reproduced from Ref. [180].

### c. Dipole induced by Charge-rearrangement

The distinction between an adsorbate dipole, which originates in the molecule ($\Delta_{mol}$, section 4) and that due to charge-rearrangement with a substrate ($\Delta_{CR}$) is not very clear. A good example for $\Delta_{CR}$ is spin-coating of a partial monolayer of an amine-rich polymer (polyethylenimine ethoxylated, PEIE or branched polyethylenimine, PEI),[181] which effectively reduces the work-function of many technology-relevant materials (metals and transparent electrodes) by ~1 eV.[181] An equivalent om-ni increase of work-function is achieved by spin-coating of Fluor-rich polymer.[182]

Such a stochastic polymeric structure cannot have aligned segments of internal molecular dipole ($p_{mol}$), as required for a net change in the potential energy (section 4a). Its exceptional ability to reduce the work-function stems from a different source of induced dipole, namely charge rear-rangement between the adsorbate and the substrate. Thus, the amine-rich polymer donates electrons to the substrate, and leaves the polymer partially positively charged, creating a net dipole between the adsorbate and the *substrate* rather than *within the adsorbate* as discussed in section 4. It is equivalent to a sheet of positive charge on the ultra-thin coating, with a parallel sheet of negative charge on the outer-most atoms of the substrate, which act to reduce the effective work-function ($\Delta_{CR} < 0$).[181] Gradual modification of surface potential by fractional coverage of disordered molecules[137] probably also reflects $\Delta_{CR}$ rather than $\Delta_{mol}$ because the latter should be cancelled by the disorder.

Thus, the net interface dipole, $\Delta$, can be roughly considered as the sum of the molecular polarization ($\Delta_{mol}$) where both poles are within the adsorbate (section 4), and the charge rearrangement polarization ($\Delta_{CR}$) where one pole is on the adsorbate and the counter pole is on the solid. Obviously, this is a crude classification that provides a qualitative insight into the possible sources of the net induced interface dipole, while in reality these are mutually related mechanisms.

Notice that Fig. 9 considers only a surface, while in junctions there are two $\Delta_{CR}$: at the monolayer interface with the substrate and with the top contact:

$$\Delta = \left(\Delta_{CR,substrate} + \Delta_{mol}\right) + \Delta_{CR,top\ contact} \qquad (23)$$

In the schematic illustration of Fig. 3.D $\Delta_{CR,substrate}$ refers to the green triangles while $\Delta_{CR,top\text{-}contact}$ refers to the orange arrows. In principle the origin of the charge rearrangement dipoles, $\Delta_{CR}$ is the same at both contacts, regardless of the deposition sequence (i.e., substrate or top contact). However, from experimental perspective, one can only measure the term in the brackets of Eq. 23, namely the adsorbate induced change in work-function, by photo-electron spectroscopy or Kelvin probe measurements. However, a top-contact induced $\Delta_{CR,top\text{-}contact}$ cannot be directly measured because the interface is now buried and inaccessible for direct measurement of the surface electrostatics. This contribution can be inferred indirectly from the Schottky barrier in the semiconductor, deduced from either I-V or C-V measurements of the full junction (see section 7 below). In some cases one can get information on contact-induced $\Delta_{CR}$ by monitoring variation in UPS signals during *in situ* growth of organic films[10,183,184] or by delamination of an interface to analyze its exposed inner faces.[185] The first approach is limited to systems where *in situ vacuum deposition* of metallic films on top of molecular monolayers does not damage the molecules directly. Such damage is actually the only detectable effect when an interface is delaminated,[185] and such analysis will not reveal the much subtler effects of the $\Delta_{CR}$ mechanism.

The extent of $\Delta_{CR}$ varies considerably between treatments: while for PEI / PEIE example it was on the order of 1 eV,[181] adsorption of phosphonic acids on transparent electrodes is argued to yield negligible $\Delta_{CR}$ ("bond-dipole") compared to major $\Delta_{mol}$.[15] Possibly this difference is due to the low polarizability of both oxide electrodes and phosphonate binding groups. Adsorption of redox species is expected to increase $\Delta_{CR}$,[17] (see Fig. 9.C for an illustration) yet in many cases the adsorbate is not ionized, and accepts only a small charge, which is about 1% of an electron charge per monomer, or even less.[100] The cross-talk between molecular states and electronic processes within the semiconductor also depends on the coupling strength between the molecular levels and the semiconductor DOS.[97]

Another key difference between the two possible dipole sources is that $\Delta_{CR}$ affects the tunneling barrier (see $E_T$ in Fig. 63a,b) while $\Delta_{mol}$ is not directly related to transport across the molecule. This consideration is critical for molecular electronics, as discussed elsewhere.[7,122]

### d. Surface to interface: inversion of the dipole direction

As noted in Eq. 23, charge rearrangement can occur with respect to both contacts. Often however, justly or not, one of the interfaces is ignored, and $\Delta_{CR}$ is assumed to occur only with the substrate. For example, the universal work-function reduction by amine-rich PEIE, implies that the polymer is positively charged which is reasonable (its water solution has pH $\sim$ 10), therefore it behaves like a surface donor. Still, using this approach to modify devices (e.g., organic solar-cells or LEDs) means that now the positively-charged PEIE is in an intimate contact with two bulk materials: for example a transparent electrode (i.e., the substrate) and a thin organic film (i.e., the semiconductor). PEIE modification is successful because the balancing *negative* charge is accumulating on the transparent electrode and not on the deposited organic semiconductor.

While this effect is undoubtedly real, there is no simple thumb rule telling where the *counter* charge is formed. Clearly, the location of the counter charge on either side of the interface inverts the net dipole direction. Eq. 18 stresses the importance of hardness, which implies that the balancing charge accumulates on the more polarizable ("softer") contact. This logic provides a rough rational to the PEIE effect: conductive organic materials are chemically-harder than metallic or semi-metallic electrodes. For this reason, the negative charge balancing the protonated PEIE is preferably located on the metallic side rather than on the organic side.

Notice though that the logic of counter-charge does not depend on the assembly process (substrate cf. top) but on polarizability. Thus, if a monolayer is chemically bonded to an inorganic semiconductor substrate (e.g., Si) and a metal is afterwards applied as top contact, the more polarizable side is the top rather than the bottom (substrate) contact. The more oxidized is the substrate (e.g., natural oxide capping) the less polarizable is the substrate. This implies that correlating between the molecularly-induced change in the effective work-function of the substrate (i.e., $\Delta_{CR,sub} + \Delta_{mol}$) and the net Schottky barrier (Eq. 4) should work only for cases where $\Delta_{CR,top}$ is very small, relative to $\Delta_{CR,sub} + \Delta_{mol}$. One example where this assumption holds is alkyl chains: these saturated molecules are highly non-polarizable (large $\eta$ in Eq. 18) and therefore we see a correlation between the change in the work-function by molecularly modified Si and the resulting Schottky barrier.[74,148,186] Notice though that a Hg contact preserves the general direction of the dipole not only for alkyls, but also for substituted phenyls[163] (Fig. 8.B) and substituted alkyls (Fig. 13.C below).[93]

The sign-inversion of $S$ for Au contact to substituted aromatics (Fig. 5.B for ZnO)[169] is attributed to flipping of the counter charge from the semiconductor to the Au side. In addition to ZnO, such an inversion was observed for several contacts to the same set of di-carboxylic acids, such as GaAs, either n- or p-doped, and with either Au or Al top contacts.[161] Out of the five tested substituents in this series, $CF_3$ is a unique case because it induces the largest *surface* potential (change in $\chi$), yet only moderate effect on the SBH of Au/monolayer-GaAs junctions (i.e., of *interface* dipole). We found empirically that the molecular chemical potential (HOMO-LUMO mid-gap by DFT for gas-phase monomers) is a much better predictor for the substituent effect on the SBH, both in terms of the $CF_3$ effect relative to other substituents and in explaining the sign of the effect.[161] A large (deep) $\mu$ (as for CN substituent) withdraws electrons from Au to make $\Delta < 0$ (relative to the semiconductor, e.g., top of Fig. 9.B) and therefore forces a larger SBH(n) to maintain Eq. 4.

The top-metal contact in these junctions was softly-deposited using 'lift-off float-on' technique.[187] Interestingly, the dipole inversion was sensitive to the details of the floatation procedure: the SBH follows the surface dipole (i.e., 'correct', negative slope with respect to modified electron affinity, as in Figs. 8.B and 5.D) in cases of non-intimate contacts, with either residual trapped solvent or deliberately adsorbed thiol spacers on the gold side to form a bilayer, while dipole inversion was characteristic of intimate contacts.[161] Such dependence on the metal to molecule gap is in-line with the electrochemical balance rational detailed in section 5.a (e.g., Eq. 18), as well as with ac-cumulated knowledge on other organic / inorganic interfaces:[55,122] Charge-rearrangement dipole, $\Delta_{CR}$, requires a close proximity between the two phases (Fig. 9.B), while Å-wide separation suffices to block $\Delta_{CR}$ (e.g., Fig. 9.A).

The chemical nature of the contact might also play a role. As mentioned above, charge rearrangement will be largely buffered by residual surface oxide. , Yet, even for (ambient) clean contacts, a difference in interface dipole across an alkyl (i.e., 'rigid molecule') monolayer exists between Au or Hg top contacts. This is evident from the SBH extracted from I-V traces measured across p-doped Si(111)-alkyl / metal. The effect of metal work-function on the SBH for p-doped semiconductor is opposite to that of Eq. 4, thuse the ~0.5 eV higher work function of Au cf. Hg, is expected to give significantly smaller SBH with an Au contact (almost no barrier). In practice however, the SBH of Au was even slightly larger than that of Hg (0.42 cf. 0.36 eV, respectively).[188] Namely, the 0.5 eV difference in the work-function of these two metals disappeared somehow.

As noted above, we cannot exclude that the origin of this effect is a contaminated interface in the case of Au, which was deposited by floatation (despite our extreme efforts to bring this technique into perfection).[189] Still, it is possible that the deeper Fermi level of Au (large WF, e.g., bottom panels of Fig. 9) withdraw electrons from the alkyl monolayer inducing an opposing dipole to the one between the alkyl monolayer and the Si, and therefore altering the monolayer dipole compared to that under Hg contact. A metal-dependent $\Delta_{CR}$ could originate in different surface polarizability,[171] or the orbital character of the metal.[190]

### e.   Rigid cf. polarizable molecular groups

Dipolar modifications have been commonly introduced using substituted phenyls (e.g., Figs. 4, 5.A, 8), where a polar substituent is placed in a para position to the binding group. The reason is a combination of ease of synthesis, avoiding bulky group in inner position to improve dense packing, or simply convention. Nonetheless, this strategy implies a polarizable moiety pointing out of the surface, and accessible for interaction with the top contact. A nice example for embedding dipoles at an inner position is the phenyl-pyrimidine-phenyl chain, with a net effect of up to 1 eV depending on whether the pyrimidine nitrogens point toward the gold substrate or away from it.[165,191]

The sensitivity of the outer group to depolarization by the top contact is illustrated in Fig. 8.B, which is an SBH vs. CPD plot similar to those of Fig. 5.B,D, for a system of n-Si-styrene-X/Hg (see section 4.b above). It compares two methods for extracting the SBH from current-voltage curves (I-V, black circles, see section 7.a) and from capacitance-voltage traces (C-V, red triangles, see section 7.b). Junction made with Br-Styrene were basically Ohmic (not a diode), and therefore their capacitance was below the measurement sensitivity. Despite the admittedly low number of points, the Br-styrene seems to be less effective compared to the other two (a net slope of $S$=0.65 cf. ~1 for only the two highest points), similar to the under-effect of $CF_3$ on GaAs-di-carboxylic-phenyl-X / Au (or Al) junctions[161] mentioned above.

Furthermore, considering the C-V extracted SBH (red-triangles in Fig. 8.B), reveals a huge change in SBH, way beyond the maximal possible slope of $S$=1 (Eqs. 4, 5). Since C-V extracted SBH reflects better the averaged interface potential (see section 7.b), it is unlikely to be an artefact. It probably reflects a different value for $\Delta_{CR,top}$ of methyl-styrene than simple styrene, which affects the actual interface dipole. A plausible explanation is that the methyl buffers the interaction between the Hg and the phenyl ring which is stronger for the un-substituted phenyl. If such interaction withdraws charge from the phenyl ring it will cause $\Delta > 0$ (e.g., Fig. 9.B bottom) against the original dipole direction ($\Delta \simeq -0.4$ eV for H-styrene-Si *surface*, see Fig. 4.B) and therefore reduces the net SBH.

As a thumb rule, conjugation along the molecular skeleton promotes charge rearrangement with the substrate (e.g., the 'resonance' case in Fig. 9.C). For example, the di-carboxylic molecules that showed dipole inversion (see inset to Fig. 5.A) are bonded via a carboxylate and have an ester tether, both expected to reduce the coupling between the polarizable phenyl and the semiconducting substrate.[161,169] On the other hand, a similar carboxylate binding but using a fully conjugated molecular skeleton did not show dipole inversion.[159]

A complete conjugation between a phenyl ring (Ph) and oxide-free Si can be achieved by adsorption of aniline or phenol that yields Si-N-Ph or Si-O-Ph, respectively. Such conjugation is broken in adsorption of styrene or benzyl alcohols that yield Si-$(CH_2)_2$-Ph or Si-O-$CH_2$-Ph, respectively. A combined UPS and DFT work shows that the frontier molecular levels are completely altered in the case of full conjugation, while the phenyl character is largely preserved for molecules with a spacer.[60] Namely, in the fully-conjugated monolayers, new hybridized states are formed (in analogy to either the ISR view of section 2.c or IDIS-CNL), while their creation is effectively blocked by merely two methylenes spacer. The creation of such hybridized states inevitably leads to bond polarization, $\Delta_{CR}$, and alters the molecular 'intrinsic' dipole, $\Delta_{mol}$. Somewhat similarly, Van Dyck et al., computed the DOS of photo-chromic core (dithienylethene) tethered to an Au contacts by different linkers. They found that the FWHM of the HOMO transmission spectrum could change from 200 meV to <3 meV for different linkers of otherwise identical core, suggesting that short saturated segments are extremely effective in blocking the molecule / metal hybridization, and therefore as a consequence the energy alignment of the molecule[122] (or $\Delta_{CR}$ in our terminology).

### f. The effect of substrate's work-function on molecular dipole

We finalize this chapter with a few intriguing experimental examples of substrate-dependent $\Delta_{CR}$, in line with the concept of electrochemical balance across the interface. The concept 'bond dipole', already noted in section 5b above, where we equated it to $\Delta_{CR}$, is not new. It is often taken literally as the result of partial ionization of the atoms constituting the actual chemical bond between the molecule and the substrate. For example, the Au-S bond is known as partially ionic, with the negative pole on the S, as is evident from shifts in the S2p peak in XPS.[92] Such atomic polarization implies $\Delta_{CR}>0$ (e.g., Fig. 9.B top), *which is opposite* to the net negative dipole, induced by adsorption of alkyl thiols on Au.[92,145,192,193] Similarly, the XPS C1s peak indicates that the C that is bound to Si is negatively charged (i.e., positive dipole) while the net dipole of alkyl-Si monolayers is negative.[83,194195]

Regardless if this apparent discrepancy is due to a 'push-back' effect,[172] or due to other charge-rearrangement mechanism (ISR or IDIS-CNL discussed above) it is clear that $\Delta_{CR}$ is not dictated solely by a bond, but it reflects a balance between larger moieties. Adsorption of various saturated[192] and conjugated[196,197] thiols on Ag, Au or Pt reveals an almost fixed work-function of the monolayer-modified metals despite 1.4 eV difference in the work-functions of the clean metals (i.e., strong Fermi level pinning as depicted in Fig. 9.B). Namely the monolayer-induced dipole varies drastically depending on the substrate. Because the molecular skeleton is identical, it must be the charge-rearrangement dipole, $\Delta_{CR}$, that is varying and the use of 'bond dipole' to describe this is misleading.

If $\Delta_{CR}$ is dictated by the polarization of the S-metal bond, it should follow the electronegativity scale (Au>Pt>Ag) but $\Delta_{CR}$ clearly follows the WF scale (Pt>Au>Ag).[196,197] This supports the conclusion that $\Delta_{CR}$ is dictated by the $(E_F - \mu)$ balance described in section 5.a. Such balance breaks the realms of traditional disciplines, because it connects a macroscopic thermodynamic property of a solid, $E_F$, with a localized molecular property, $\mu$.

Semiconductors are excellent candidates to test the $(E_F - \mu)$ balance hypothesis, because doping changes the work-function while the binding chemistry remains identical, as demonstrated in Figure 11, for monolayers of saturated alkyl chains on oxide-free Si, of either n- or p-type with either moderate (MD) or high (HD) doping levels. All together these different substrates provide an identical Si-C bond, for a four different Fermi level positions, varying by ca. 1.1 eV from being almost at CB (for HDn) to almost VB (for HDp). Nonetheless, adsorption of a C10 alkyl monolayer yields an almost constant Si effective work function, varying by less than 0.3 eV,[83] suggesting that the dipole on the monolayer varies with the work function, as reported for thiols on different metals.[192,196,197] Not just that the chemical bond is identical (C-Si cf. different metal in former works), but both the alkyls and the Si are poorly polarizable compared to conjugated molecules and metals, respectively. Still, such varying surface dipole can be only understood by variation in $\Delta_{CR}$ across the alkyl-Si interface.

The down-side of semiconductors for this purpose, is that now the total work-function is modified by both band-bending and surface dipole (consider Fig. 1 without the metal). The BB contribution is isolated by following the Si 2p peaks[81,83,194,198,199] of the C10 alkyl monolayers on the four different Si wafers. Core levels of semiconductors are fixed with respect to the band edge (e.g., VB), but XPS binding energy is measured with respect to the Fermi level, $E_F$ of the substrate. Thus, the Si 2p binding energy shifts with the Fermi level position within the gap, as shown in Fig. 11.A. The few nm detection depth of XPS implies that shifts in Si 2p binding energy reflect the surface position of $E_F$ which varies from the bulk position, in cases of significant BB.[81,194] This trend is quantitatively examined in Fig. 11.B, displaying the Si 2p peak position with respect to the bulk position of $E_F$ within the Si gap, deduced from the nominal wafer resistivity.[83] As can be seen, the experimental binding energy roughly follow a linear dependence on doping level (dashed line), though not strictly. Actually, the Si 2p binding energy is almost the same for both medium-doped (MD) samples, suggesting an apparent pinning of $E_F$ within the Si gap. In contrast, for highly doped (HD) samples, the doping is able to shift $E_F$ within the gap.

We now turn to the C1s peak to gain insight in the potential profile over the monolayer. The peak position of C1s (Fig. 11.C) is almost constant compared to the pronounced doping effect on Si 2p (Fig. 11.A). This is further quantified by plotting the difference in binding energies (C1s − Si 2p) against the Fermi level position in Fig. 11.D. While this difference mostly reflects the intrinsic core-levels, it is modified by any potential developed over the monolayer. The C1s peak center reflects the averaged potential over the 10 carbons along the adsorbed alkyl chain. Fig. 11.D shows a similar potential drop on the two MD samples, compared to ca. 0.4 eV large / smaller potential drop on HD p/n Si substrates. Namely, an alkyl adsorbed to HD-n (HD-p) is more positively (negatively) charged than the same molecule adsorbed onto MD Si.

Both effects can be reconciled using the principal considerations of section 5.a. A monolayer of alkyl chains has a chemical potential somewhere in the middle of the Si gap; therefore electrons are shifted from the Si to the monolayer (from the monolayer to the Si) for n-Si (p-Si). This fractional charging is distributed between the Si space charge and the monolayer, depending on the doping level. For HD samples, the Si is more polarizable (softer, metallic-like) than the monolayer and therefore the position of $E_F$ within the Si gap is fixed (Fig. 11.B), while the molecular dipole varies (Fig. 11.D). For MD samples, the Si polarizability is comparable to that of the monolayer and, therefore, the adsorbates forces the $E_F$ position within the Si gap (Fig. 11.B), while the molecular dipole is roughly fixed. The ca. 0.4 eV higher or lower potential energy as a result of adsorbing the monolayer on HD, instead of MD Si (p or n, respectively, Fig. 11.D) over a distance of ≤ 1 nm is enormous; it translates into an electric field of 4 MV/cm. The arrows in Fig. 11.C point to a shoulder due to negatively charged C in Si-C bond. This shoulder can be observed only for MD Si, and we propose that the large induced electric-field in HD samples obscures this shoulder.[83] Overall, the combination of both contributions leads to a net effective WF, which is almost fixed, regardless of the 1 eV difference in the Si nominal $E_F$.

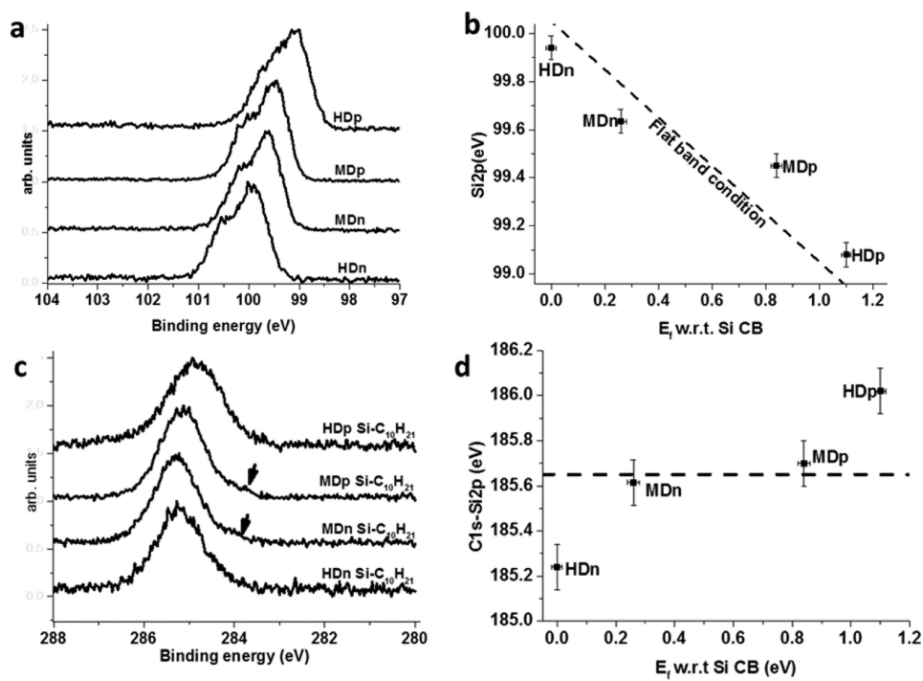

**Figure 11**. **(a)** Si 2p core level from Si–$C_{10}H_{21}$ samples with different Si doping density and type. **(b)** Si 2p peak position as a function of the Fermi level, $E_F$, relative to the Si conduction band (CB) edge, the position of which was extracted from the nominal doping type and density of each Si sample. **(c)** C 1s core level from Si–$C_{10}H_{21}$ samples with different Si doping density and type. Arrows mark the C 1s component from C attached to the Si (see text). **(d)** Difference between the C 1s and Si 2p peak positions as a function of $E_F$, relative to the Si CB edge. The dashed line is a guide to the eye that represents an ideal case, in which the charge rearrangement is unaffected by the doping density of the Si substrate. The error bars for the binding energies represent the measurement accuracy, and those of the $E_F$ reflect the uncertainty in doping density provided by the Si wafer manufacturer. Reprinted with permission from Ref. [83].

A similar effect of Si doping was observed for adsorption of various monolayers on both Si(111) and Si(100)[198] and for much shorter adsorbates, H, $CH_3$ or Br on oxide-free Si.[194] That comprehensive study revealed a characteristic BB and dipole for each adsorbate, yet of roughly opposite trend: the electric potential decreases towards both sides of the monolayer/semiconductor interface, due to the molecular dipole layer on one side and due to the band bending in the semiconductor on the other side. Over-all this leads to at least a partial cancellation of effects which moderates the change in WF due to the molecules.

This odd observation was rationalized by a surface charge ($Q_{IS}$, inducing the BB), which is on the exterior face of the monolayer (with respect to the semiconductor. Thus, $Q_{IS}$ is *not* located on the Si surface, but is due to adventitious organic matter or humidity adsorbed on top of the deliberate chemical modifications.[194]

These two examples of doping-dependent dipole formation[83,194] provide direct support for a significant charge-rearrangement which is independent of the ionicity of the surface bond; similar to organics on oxides[175,179] charge-rearrangement occurs even when $E_F$ is located within the forbidden energy gap of both the substrate and adsorbate, in contrast to the traditional view that Fermi-level pinning is caused by evanescent electron wave-functions at the metal's $E_F$. The non-trivial interplay between surface dipole and BB,[83,194] demonstrates the great power of molecular and other adsorbates in inducing doping[98] that overcomes the intrinsic material doping (section 3.e and 4d ) and is extremely relevant for sensors[97,200] or catalysis[28] applications.

In summary energy level alignment at hetero interfaces controls the barrier for transport. While the net thermodynamic (electrical) equilibrium is dictated by the difference between the work functions (i.e., electrochemical potentials of the electrons) of two contacting electrodes (Eq. 4), this difference is divided into short ($\Delta$) and long-range (BB) charging. The BB is a characteristic semiconducting contacts (i.e., irrelevant for metallic ones), because the wide space charge region pose a far larger transport barrier (SBH) than the narrow interfacial layer. However, the localized charging, $\Delta$, is often comparable in magnitude to the BB, and therefore provides an efficient handle for manipulating the SBH. The interface-localized charging has several sources: 1) trapped interface charge, $Q_{IS}$ due to chemical / structural defects (section 3); 2) foreign modifiers introduced at the interface with additional localized dipoles or energy levels (section 4); and 3) interface charge rearrangement leading to localized dipoles (section 5). The last issue specifically is still debated and it is also highly relevant for molecular electronics between two metallic electrodes, as it dictates the transport barrier by the monolayer itself.[7] The direct transport barrier of the interfacial layer itself has been ignored so far. Such consideration is relevant for cases where the interfacial layer grows thicker, by deliberately adsorbing dense molecular layers (or growing thicker oxides in general context). The interplay between these two barriers is the topic of the next section.

## 6. The concept of tunneling MIS and the dual barrier

### a. **Tunneling** (direct molecular) **vs. SCR** (Space Charge Region)-**limited' transport**

The majority of molecular electronics is focused on manipulating the tunneling probability, often denoted as T, the transmission probability. Nevertheless, the net conductance is the product of the transmission probability times the density of *filled* states in the source electrode with a matching density of *empty* states in the drain electrode. Prominent examples where the surface DOS is critical are scanning tunneling spectroscopy (STS), where there is a huge (vacuum) tunneling barrier, and the conductance-voltage traces are indicative for the surface DOS, or spintronics, which is based on spin mismatch between the DOS of the two contacts. Still, for most of molecular electronics where both electrodes are metallic, the role of the electrodes' DOS is indeed not particularly interesting. This does not hold, though, for a semiconducting electrode, where the DOS of electrons and holes is a complicated function of a variety of factors. Actually, this limited availability of carriers is the fundamental cause of asymmetric transport (rectification) across metal/semiconductor interfaces, and therefore, we designate the "DOS contribution" as semiconductor or Schottky-limited transport. In contrast, if transport is limited by the tunneling probability, we call it insulator-limited or transport in the tunneling regime.

The rate-limiting step generally affects the characteristic behavior: for example, semiconductor-limited transport will show strong temperature activation, while tunneling-limited transport will show exponential attenuation with insulator thickness. The major characteristics of the two transport regimes are summarized in Table 1. However we stress, that this is a gross simplification and in reality it is very rare to have either pure tunneling or pure Schottky transport across a thin-insulator MIS.

**Table 1:**
**Experimental characteristics of tunneling and Schottky-limited transport mechanisms**

| Dependence on / of | Tunneling | Schottky-limited |
|---|---|---|
| Temperature | Very weak | Exponentially activated |
| Insulator thickness | Exponential attenuation | ∼Independent |
| Bias Polarity | ∼Symmetric | Asymmetric |
| Conductance-V relations | Parabolic ($V^2$) | Exponential ($e^{V/kT}$) |

The entangled tunneling and SCR (DOS) contributions lead to rich transport effects, but also prevent a generic simple description, because that requires simultaneous solving of the electrostatic potential balance and steady-state fluxes of two types of carriers. In analogy to chemical reactions, this is maybe similar to describing highly exothermic reactions for which heat and mass balance must be solved simultaneously. Here we describe the major considerations and provide some simple relations, applicable to special cases.

The two panels of Fig. 12 depict two generic cases of insulator- or semiconductor-limited transport, referring to n-type semiconductors only; (see Refs. 24,41,42 for p-type). Any electrical interface in equilibrium or steady state, must obey the following balances:[51,201,202]

<u>Charge balance</u>:

$$\Delta = \frac{L}{\varepsilon_0 \varepsilon_i A}(Q_{IS} + Q_{SCR}) \tag{24}$$

Here Δ is the potential drop on an insulator of thickness $L$ and relative permittivity, $\varepsilon_i$. Notice that the interfacial insulator is the same one as discussed in section 4b. Fig. 6 illustrates the two possible sources of the potential drop, Δ, over the interfacial layer namely the intrinsic molecular dipole Fig. 6 (A-C) and charge balance of charges extrinsic to the molecular dipole layer. The two bottom panels, Fig. 6 (D-E) illustrate the potential difference developed on the insulator (green ovals) due to extensive $Q_{SCR}$ under extreme conditions. These two possible sources for Δ are expected to be additive and for simplicity, this section ignores the dipole (localized) source. The interface insulator serves as a parallel plate capacitor with a net charge on its metallic side ($Q_M$, see Fig. 12) that balances the net charge on the space charge region (SCR) of the semiconductor, $Q_{SCR}$ and any charge localized at the interface, $Q_{IS}$ (either on insulator-defects or interface states). Yet at this stage, for simplicity, we omit the insulator charge from the basic consideration ($Q_{IS}$=0 here; see section 3 for discussion on $Q_{IS} \neq 0$).

Unlike a metal, a semiconductor has a limited charge screening capacity, and, therefore, the charge on the semiconductor side ($Q_{SCR}$) is spread relatively deep into the semiconductor (SCR), inducing band bending, BB, dictated by the Poisson equation. At the same time BB and Δ are also connected by the energy balance.

<u>Energy balance</u>:

In the absence of interface states ($Q_{IS}$ =0, section 3) or interface dipole (see sections 4-5), the electron energy diagram (Fig. 12.B) under applied bias, $V$, implies the following balance:

$$WF = \Delta + BB + \chi + \xi + qV \tag{25}$$

Eq. 25 is very similar to Eq. 4, except that Eq. 4 is at equilibrium (V≡0) and written in terms of SBH (see Eq. 3). A special case is the flat band voltage where $\Delta \equiv BB \equiv 0$ (this condition requires $Q_{iS}$ =0):

$$V_{FB} = WF - \chi - \xi \qquad (26)$$

Thus, $V_{FB}$ contains the essence of the metal / semiconductor energy alignment, regardless of the insulator properties. In this simplistic case of no fixed interface charge or dipole, the sign of $\Delta$ and BB is identical. Both are positive for V< $V_{FB}$, and negative for V> $V_{FB}$. Substituting $V_{FB}$ instead of the fixed energy alignment terms in Eq. 25 we find that the sum of the two interface potentials always equals a constant minus the applied bias: $\Delta + BB = V_{FB} - V$. It is simpler to consider the change in BB and $\Delta$ relative to their value at equilibrium (V=0). So, if we define: $dBB = BB - BB_0$ and $d\Delta = \Delta - \Delta_0$, and note that at zero voltage: $V_{FB} = BB_0 + \Delta_0$, it follows that:[51]

$$V = -d\Delta - dBB \qquad (27)$$

Often this partition is discrete (bipolar): the added voltage falls almost completely on either the insulator or on the space charge layer, corresponding to *tunneling* -limited ($dV \approx -d\Delta$) or *SCR* -limited ($dV \approx -dBB$) transport, respectively.

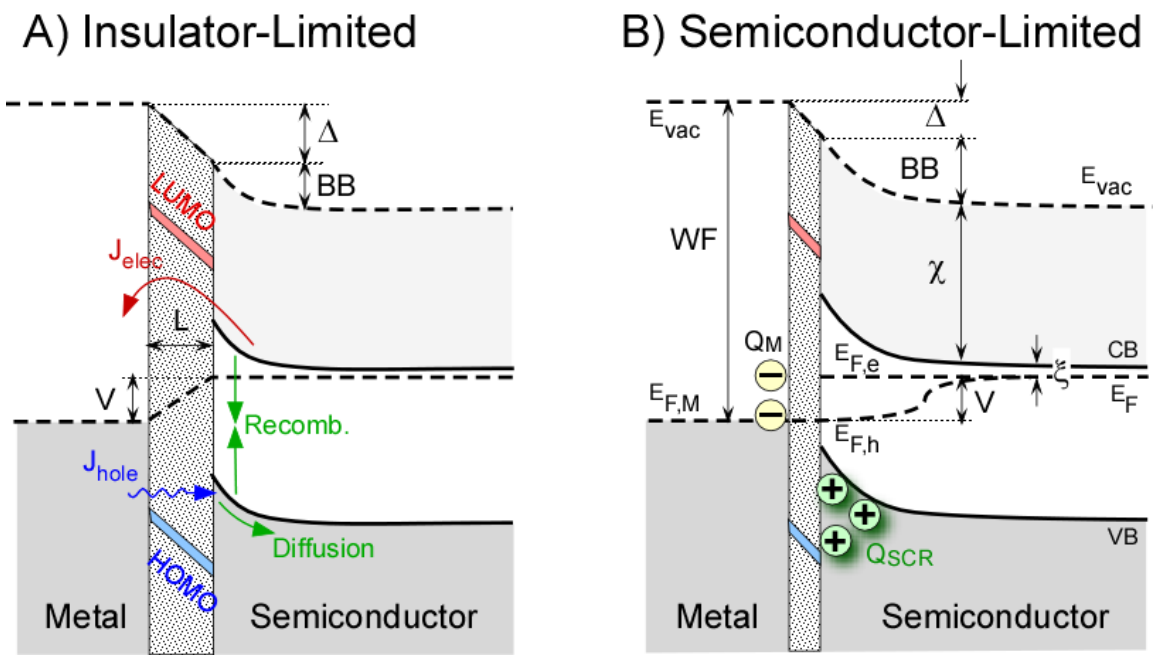

**Figure 12**: One electron energy (Y-axis) schemes across metal-insulator-semiconductor interface (X-axis), showing **A)** thick insulator and **B)** thin-insulator, with an n-type semiconductor under forward bias (positive on the metal), *V*. Each panel shows the metal (left), interfacial layer (dotted bar, with width that is not to scale!), and an n-type semiconductor (right), where dark and light grays are the valence and conductions bands, respectively and the white region between them is the forbidden energy gap. Red and blue stripes on the insulator represent the frontier molecular levels (LUMO and HOMO, respectively). $\chi$ is the electron affinity of the semiconductor and $\xi = |E_C - E_F|$ (for n-type semiconductor).

The large difference between the local vacuum levels, $E_{vac}$, (top dashed line) of the two materials (i.e., the difference between the metal work function, *WF* and ($\chi + \xi$) for an n-type semiconductor) drops partially on the insulator, $\Delta$, and partially in the semiconductor, expressed as the band-bending, BB.

Panel (**A**) depicts the main transport process, including transport of majority carriers ($J_{elec}$, red) by thermionic emission over the barrier, tunneling of minority carriers across the insulator ($J_{hole}$, blue), and its propagation into the semiconductor (green arrows) by either recombination or diffusion.

Panel (**B**) plots the key energy levels and barriers: $E_F$ is the Fermi level (low dashed line) for the metal ($E_{F,M}$) and the quasi-Fermi level for the two types of carriers in the semiconductor side ($E_{F,e}$, $E_{F,h}$). $E_C$ and $E_V$ are the minimum and maximum of the conduction and valence bands, resp.

The charging sources are:

$Q_M$ – Metal side balancing charge; $Q_{SCR}$ –space charge of the semiconductor. Charge trapped in interface states ($Q_{IS}$) is omitted for simplicity's sake (see Fig. 1).

In our early papers we suggested that the tunneling- and semiconductor SCR-limited transport can be imagined viewed as a two-step process, limited by the slower rate (or highest resistance),[203,204] similar to the Lindemann-Hinshelwood chemical reaction kinetics. This view was corrected later,[148] as the dBB / dΔ partition reflects mostly the charge balance (Eq. 24): the higher is the screening capability of the semiconductor (i.e., more excess charge carriers, either minority or majority ones), the smaller is the net BB and its variation with bias, dBB. This interplay is extensively discussed in MOS textbooks,[41,42,205] and the specific conditions for transition between these regimes are explained in section 6.b.

<u>Current balance</u>:

This balance is illustrated in Fig. 12.A. A steady state current implies that the flux of tunneling carriers equals the flux of carriers into the bulk of the semiconductor. As noted by Green and co-workers,[201] this criterion is easily fulfilled for majority carriers in a semiconductor; therefore this process is represented by a single red arrow in Fig. 12.A. In contrast, the supply of minority carriers to the interface (green arrows in Fig. 12.A) is orders of magnitude smaller and is often of the same order of magnitude as that supplied by tunneling (blue arrow).

The property that reflects the density of carriers near the surface is the quasi-Fermi level.[201,202,206] The assumption of practically unlimited supply of majority carriers implies that we can take the Fermi level of the bulk semiconductor as the quasi-Fermi level for electrons ($E_{F,e}$, majority). However, the quasi -Fermi level for minority carriers ($E_{F,h}$, minority) can differ considerably from the real (bulk) Fermi level. Such a separation is illustrated in Fig. 12.B. If the tunneling probability is much smaller than the supply of minority carriers, $E_{F,h}$ will not deviate from the bulk $E_F$ and the potential step will occur across the insulator (see Fig. 12.A). However, if the supply of minority carriers is significantly less than the tunneling probability, $E_{F,h}$ will follow the metal $E_{F,M}$ and deviate considerably from the bulk semiconductor Fermi level value (Fig. 12.B).

The splitting between the quasi Fermi levels ($E_{F,e} − E_{F,h}$) is critical because the minority carrier current (either by diffusion or generation-recombination) increases exponentially with this value. Notably, the minority carrier current is almost independent of the energy alignment (i.e., $V_{FB}$; cf. Eq. 26) or tunneling probability (i.e., barrier width or molecular energy levels), and it scales quite rigidly with various semiconductor materials constants like doping level, carrier life-times and diffusion coefficients.[202] Thus, only the majority carrier current scales directly with interface details, while the minority current imposes a low saturation limit on the current, fixed by materials properties. Minority carrier-dominated transport is often the case for n-Si-alkyl /Hg junctions, as is further discussed in section 6.c below.

### b. Tuning the dominant barrier

The interplay between tunneling and Schottky barrier and types of carriers, although confusing at first glance, is actually quite logical. This section covers the different experimental handles available to manipulate the relative importance of the two barriers, with the help of Figure 13 and Table 1. Fig. 13 is a compilation of results from various reports, relative to a common, 'standard junction' made of a Hg/C16 (16 carbons long alkyl chain) - oxide-free n-doped Si(111), i.e., the C16 is bound covalently to Si, by reacting a terminal alkene with H-Si.[17,95] All junctions have a liquid Hg drop (diameter ca. 0.3 mm) as top contact,[7] except for the one that yielded the dashed curve in Fig. 13.B, which has an evaporated Pb contact.

Fig. 13.A compares the current-voltage characteristics of moderately doped Si ('MD', black) to highly-doped Si ('HD', blue),[180] on a semi-log scale. Although Si cannot be doped to a truly degenerate level, it is clear that a high doping level of 1E19 cm$^{-3}$ suffices to produce more or less symmetric current voltage relations, compared to 6 orders of magnitude rectification with moderately doped (1E15 cm$^{-3}$) Si. As noted in Table 1, symmetric current-voltage curves are characteristic for transport by tunneling. We thus conclude that using highly-doped Si minimizes the (effect of a re-sidual) Schottky barrier and these junctions can be considered as tunneling-dominated ones.

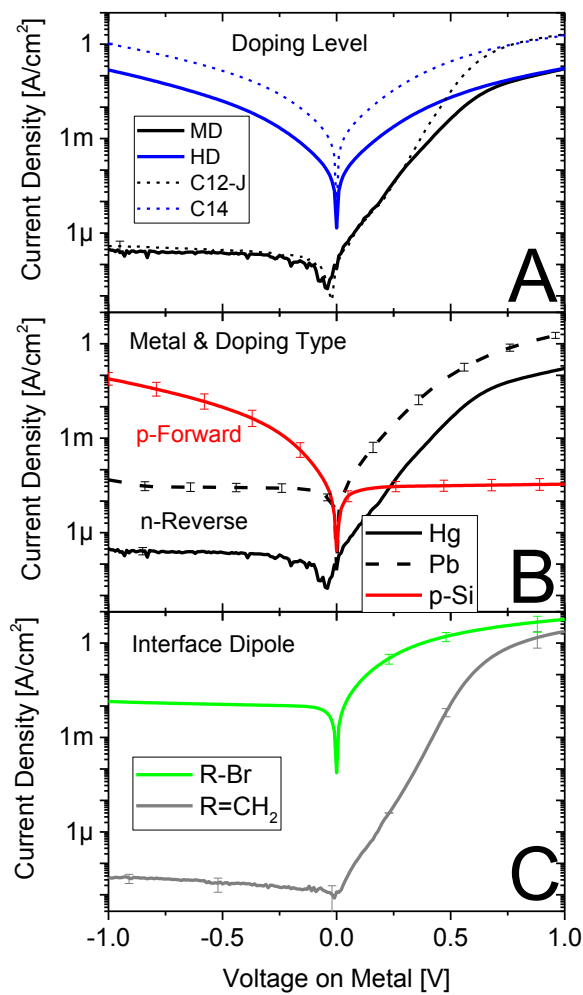

**Figure 13**: Factors controlling the Metal-Insulator-Semiconductor, MIS dual barrier, demonstrated for current-voltage curves across Hg / alkyl-Si(111) junctions of differ-ent types. All panels show current *density* on a semi-log scale (Y-axis) vs. the bias ap-plied to the metal, with the Si grounded (X-axis). The monolayer is composed of 16 carbon long alkyl chains, except for (C) and dotted lines in (A).

  **A.** Effect of **doping level**: black line: moderately-doped, MD n-Si ($N_D \approx 10^{15}$ cm$^{-3}$); blue line: highly doped, HD n-Si ($N_D \approx 10^{19}$ cm$^{-3}$); dashed lines: slightly thinner C12 (MD) or C14 (HD) insulating monolayers.

  **B.** Effect of doping **type** and metal **WF** for **moderately**-doped Si: n-Si (black, same as in a), p-Si (red), both contacted by Hg, and n-Si contacted by Pb (dash).

  **C.** Effect of molecular **dipole**: using an identical junctions of Hg/X-R-Si ($N_D \approx 10^{15}$ cm$^{-3}$), the terminal X group is either −C(H)=CH$_2$ for positive dipole (gray) or −Br for negative dipole (green) where R is a C10 alkyl chain;

In all cases values and error bars are log-averages and log standard deviations, re-spectively. Data sources are refs. [188] (p-Si); [186] (Pb-contact); [93] (panel  c); [180] (all others).

Dashed lines in Fig. 13.A show transport across identical junctions made with slightly shorter molecules (thinner insulators) of 12 or 14, instead of 16 carbon alkyl chains. Decreasing the thickness by $\sim 3\text{Å}$ increases the current by ×10 over the full bias range for highly-doped Si, but for moderately doped ones the current increases only at high forward voltage (V>+0.7V), while it is thickness-independent over most of the bias range. Thickness attenuation is another characteristic of tunneling -dominated transport (Table 1). Actually, the full thickness attenuation occurs where the currents for both doping levels become comparable, i.e., when the current is limited by the insulator rather than by the semiconductor. Thus we argue that for highly-doped junctions, transport is tunneling-dominated at any bias, while for moderately-doped junctions the applied bias shifts the transport from Schottky barrier-dominated at negative to low-positive bias into tunneling - dominated transport at high positive bias.

The saturation current, namely the roughly constant current in the 'off-state' of the diode, is exponentially proportional to (-SBH). As discussed at length in sections 2,4 changing the metal's WF should change the SBH. Fig. 13.B compares the same surface modification (Si-C16) contacted by either Hg (WF $\cong$ 4.5 eV; solid black curve, identical to the one in panel a) or Pb (WF $\cong$ 4.05 eV dashed line). The SBH values, extracted for these curves (see section 7), equal 0.9 eV[148,180] and 0.47 eV,[186] for Hg and Pb, respectively. Thus, a net difference of 0.45 eV in metal's WF yielded a difference of 0.43 eV in the resulting SBH, which is an almost ideal Schottky-Mott dependence (S→1, see Figs. 2 and 5). Thus, a molecular insulator often un-pins the Fermi level and allows using the metal WF as a handle to tune the SBH and its relative contribution with respect to the tunneling barrier.

Another way to verify the stability of the molecular dipole is by comparing the type of doping (cf. level of doping in Fig. 13.A). The red curve of Fig. 13.B shows the log(J)-V curve for a junction identical to the black curve, except that the Si is now p-doped. As expected, altering the doping type inverts the direction of rectification, because the 'on' bias polarity (so-called 'forward' polarity) is obtained for a positive bias on the contact that has the higher work-function (WF). Here, altering the doping type changes the WF of Si from $\sim$ 4.1 eV for n-Si (i.e., near the top of the forbidden energy gap) to $\sim$4.7 eV for p-Si (bottom of energy gap). Compared to Hg (WF $\approx$ 4.5 eV) the high WF side (forward bias direction) is the Hg for n-Si while for p-Si it is the Si side (therefore negative bias on Hg is the forward direction of the diode).

Thus, both the doping type and level, and the metal WF affect the tunneling / Schottky barrier interplay. The fourth handle to tune the interface character is the interface dipole induced by the molecules, as illustrated already in Fig. 5 and, again here in Fig. 13.C. Here the alkyl chain has two different terminal groups (at the contact with the metal), and the chain is only 10 carbons long. The original terminal group was a double bond (-C(H)=CH$_2$); this is the same functionality that is used to form the Si-C bond between the alkyl to the Si. The resulting J-V curve (Fig. 13.C gray line) is very similar to that for the junction with the C16 (black lines in Fig. 13.A,B) because both –CH$_3$ (standard alkyl) and –C(H)=CH$_2$ have a similar dipole. A C=C terminated monolayer can be further reacted *insitu* with Br to decorate it with one or two Br atoms per chain[93] (green line in Fig. 13.C). The high electronegativity of Br withdraws negative charge, which induces an interface dipole that reduces the Schottky barrier in an n-type semiconductor (see more in section 5). As a result the Br-modified MIS junction has a rectification ratio of $\sim$400 while that of the C=C one is more than 10$^7$. This is basically the same effect as that discussed in section 4 above, again demonstrating that the Schottky barrier of MIS interfaces is an extremely sensitive amplifier to minute changes in molecular dipole.

The division of Table 1 between tunneling- and SCR-dominated transport regimes provides useful conceptual guidelines, but reality is obviously more complicated: tunneling and semiconductor contributions are simultaneously effective in almost any MIS regime, as further illustrated in Fig.

14. First, it is often assumed that 'degenerately doped Si' behaves practically like a metallic contact. Formally speaking, 'degenerate' means that the Fermi level resides outside the forbidden energy gap, within one of the allowed bands. Due to material limitations, in Si such a state is impossible without changing the material beyond just doping and Si can only be 'highly-doped' (HD), where the Fermi level is some 50 meV within the forbidden gap (away from the band edge).

Second, even for an ideal degenerated semiconductor, Esaki has predicted that a negative differential resistance (NDR, much like the well-known p-i-n Esaki diode[207,208]) will occur for a p-type semiconductor next to an insulator, which is an electron conductor (i.e., LUMO is the frontier level) or *vice versa*. This is demonstrated in Fig. 14.A, which compares two highly-doped Si substrates, where the only difference is the type of doping: n- or p-doping, shown by black and red curves, respectively. The contact and monolayers are the same Hg drop and C16, such that the black curve represents the same raw data as the blue curve in Fig.13.A. Here, to magnify the differences, the voltage derivative of the current (conductance) on a linear scale is shown instead of the current on a semi-log scale. Although an actual negative derivative (NDR) was not achieved we see that the HD-p-Si – based junction has an almost 1 V gap (plateau) in conductance, extended toward negative bias, while the HD-n conductance is a rather symmetric parabola around 0V.[209] A plateau instead of NDR is actually the best that was achieved with traditional oxide MIS junctions,[208,210] a shortcoming that is attributed to interface states within the Si gap that facilitate tunneling and smear the predicted NDR.[208,210]

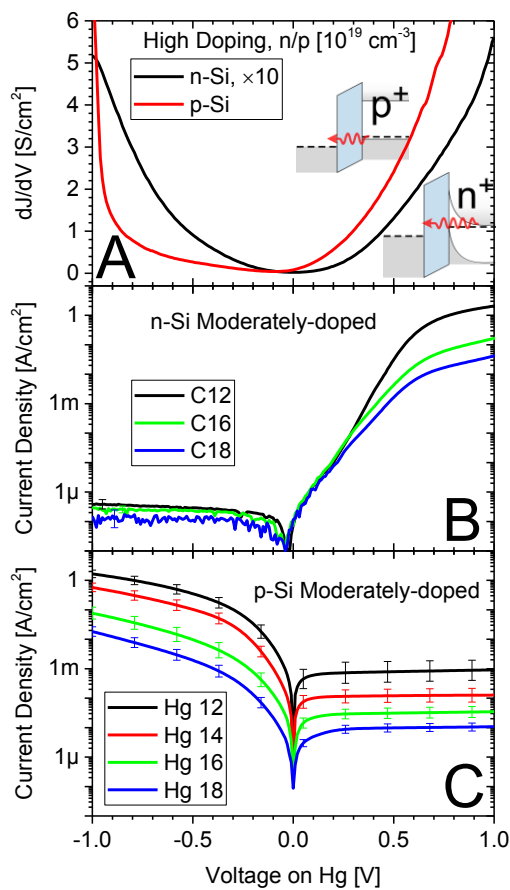

**Figure 14**: Entangled SCR / tunneling effects on MIS transport, demonstrated for Hg / alkyl-Si(111) junctions. In all cases the measurements were done with voltage applied to the Hg contact.

A. Doping **type** effect, illustrated for highly-doped Si, where transport is dominated by tunneling. The differential conductance of HD-Si-C16/Hg for either n-Si (black) or p-Si (red) is shown. The n-Si curve is multiplied by 10. Insets illustrate, schematically, the residual band bending in HD n-Si.

B. Semi-log plots of current density against applied bias voltage for MD **n**-Si-alkyl/Hg for three different lengths of alkyl chains;

C. Same as (b) only for MD **p**-Si.

Data from Refs. [209] (panel **A**); [180] (panel **B**) and [188] (panel **C**).

However, also the symmetric HD-n junction is not purely tunneling-controlled. The conductance (and net current) of the HD-n-Si junction is about 10× *smaller* than that of the HD-p-Si one, for the same molecular length. The reason is a residual barrier within the n-Si space charge region, because of the closer energy alignment between the Hg work-function and the $E_F$ position in p-Si than in n-Si (same argument as for medium doping, Fig. 13.B).[199,209] The high doping considerably narrows the space-charge region, and therefore carriers can cross this extra barrier by tunneling (symmetric, non-rectifying), as demonstrated by the red arrows in the schemes, that appear as insets to Fig. 14.A. Overall, the tunneling distance is longer with HD-n-Si than with HD p-Si, because with the n-Si it includes the SCR, which is absent with the p-Si. There is no simple, analytical description of the current-voltage relations in this regime.[199] A rough relation could be:

$$J_{HD} = e^{-\beta L} \cdot J_{SC}(V_{FB}, N_D, D_{IS}, T) \cdot f(V) \tag{28}$$

where β is the tunneling attenuation factor (1/Å), $J_{SC}$ is the semiconductor-limited current, expected to be a function of the flat-band potential, $V_{FB}$ (V), the doping level, $N_D$ (cm$^{-3}$), the density of interface states, $D_{IS}$ (cm$^{-2}$eV$^{-1}$) and the temperature, $T$ (K). Increasing $T$ enhances the semiconductor current, but for high-doping (HD) this is not expected to be a major effect, and certainly far less than in moderately doped, MD-MIS junctions. The bias function, $f(V)$, is roughly *polynomial*, in contrast to the strong exponential current dependence on bias voltage for MD-MIS (see Eqs. 29-29 below); still the current is generally very responsive to the applied bias, much more than is predicted for pure tunneling.[7,211] Thus any attempts to apply pure tunneling current-voltage (e.g., 'transition voltage spectroscopy') analyses to HD-MIS junctions[212,213] are incorrect (see more in Ref. 7). Moreover, the above mentioned "Esaki effect" as well as sharp conductance onsets at a forward bias exceeding the forbidden energy gap,[186] would further complicate $f(V)$.

The important message is that MIS junctions at highest-possible doping levels are never purely tunneling-dominated and the semiconductor band structure affects both the net current *magnitude* and its *variation with bias*. Interestingly a poor interface, with a large density of gap-states, $D_{IS}$, could practically diminish the role of the forbidden energy gap and with it the unique semiconductor effects (e.g., Esaki effect). Thus, increasing the doping to nearly degenerate levels magnifies the direct molecular barrier to the current but, strictly speaking, never really eliminates the contribution of the semiconductor band structure.

The opposite extreme of pure SCR barrier control is possible, though for rather specific cases, as demonstrated in panels B,C of Fig. 14, for moderately doped (MD) n- and p-Si, respectively. The moderate doping implies an active Schottky barrier as is evident by the asymmetric, diode behavior (the flipped polarity is explained above, see Fig. 13.B with the same C16 data as in Fig.14.B,C). The contribution of the second, tunneling barrier posed by the insulator is tested by varying the length of the alkyl chain (C12 to C18). For MD-n-Si (Fig. 14.B) the transport is independent of insulator thickness over much of the bias range; therefore we argue that this junction shows a pure Schottky-limited current. However, the same monolayers and doping level, with only inverting the doping type (MD-p, Fig. 14.C) yields an exponentially attenuated current over the full bias range. Thus transport across MD-p junction is simultaneously controlled by both Schottky (asymmetric rectification) and tunneling (exponential length attenuation) barriers.

The reason for this drastic difference lies in the nature of the charge carriers. In the case of MD-n-Si, the difference in work-function of the two contacting materials is so large (see Eqs. 25,26) that it induces the formation of a p-n junction within the Si (inversion),[23] as explained in section 4.d. In that case the interface with the metal is crossed by minority carriers and the barrier that controls the net transport is the p-n junction, induced by the strong band bending and buried within the space-charge region of the Si; this is the only case where a pure Schottky behavior, independent of insulator thickness, can be observed. The current across such minority carrier MIS is:[214]

$$J_{MD,min} = J_{Sat} \cdot exp(-E_g/nkT) \cdot [exp(V/nkT) - 1] \tag{29}$$

Eq. 29 applies to transport of minority carriers either by diffusion or by recombination / generation of electron-hole pairs (see green arrows in Fig. 12.A). Each case has a different definition for the saturation current, $J_{sat}$ (A/cm$^2$) and for the ideality factor ($n$=1 for diffusion and 2 for generation-recombination).[2,41]

Finally, the more common case is that of a majority-carrier transport, as shown in Fig. 14.C for MD-p-Si. Here, the current is described by thermionic emission over the Schottky barrier (red arrow in Fig. 12.A), which is attenuated by the tunneling barrier:[51]

$$J_{MD,maj} = A^*T^2 \cdot exp(-\beta L - SBH/kT) \cdot [exp(V/nkT) - 1] \qquad (29)$$

where $A^*$ is the so-called Richardson coefficient (=120 A/cm$^2$/K$^2$, for n-Si) and *SBH* is the Schottky barrier height (eV). The ideality factor, $n$, is in principle unity for pure thermionic emission, but in practice is often higher. The exponential term outside the square brackets dictates the saturation current at reverse voltage. It is attenuated by both tunneling decay and the SBH. The relative scaling of these two contributions can be compared by considering typical values in dimensionless units. Common molecular insulators are in the range of $L$ = 1-3 nm with a tunneling decay coefficient of 5-10 nm$^{-1}$,[215] yielding a net dimensionless $\beta L$ range of 5 to 30. This can be compared to SBH/kT by considering that the SBH of common semiconductors cannot be more than ~1.5 eV (ca. the forbidden band-gap); if we multiply this by 1/kT ≈ 40/eV at room temperature, we find SBH/kT ≤ 60. Thus the $\beta L$ and the SBH/kT contributions are of comparable magnitude. Still, adding a substituent group to a molecule (e.g. Fig. 13.C) hardly increases $\beta L$ (e.g., few Å), while it can easily alter the SBH by 0.2–0.5 eV, which after scaling equals a difference of ~ 8–20. This back of the envelope comparison demonstrates why molecular substitution is often far more significant for dipole tuning than for tunneling attenuation.

To summarize, this section provides a qualitative description of the rich transport behavior across MIS junctions of very thin insulators, like those made with insulating molecular monolayers. The tunneling/Schottky distinction is a conceptually helpful simplification but should not be taken too literally. More specific predictions require numerical simulations,[87,201,202,216] though some analytical tools are provided in section 7.

### c.  Inhomogeneity in monolayer coverage

The molecular / SCR interplay is also critical with respect to inhomogeneity in monolayer coverage. Let us consider a realistic case where the monolayer covers the majority of the interface area, but the molecules are missing from nm-wide patches where the metal contacts the semiconductor directly. Such 'pinholes' affect both the direct molecular barrier to transport and the SCR, via the molecular dipole. However, defects affect the two barriers very differently. The direct molecular barrier is completely absent within the pinhole, and the relative reduction in barrier increases exponentially for longer and more saturated molecules. However, in terms of the SCR, the absence of molecules can either increase or decrease the SBH, depending on the direction of the molecular dipole. Thus, if the molecular dipole acts to decrease the SBH, the pinhole reduces both barriers and attracts current (Kirchoff's law); however, in the opposite case, the reduction in tunneling barrier can be compensated for by the increased SCR barrier (see discussion below Eq. 29).

The width of the barrier is also critical. The SCR extends 100's of nm deep into the semiconductor, far more than the monolayer thickness, which turns the SCR inhomogeneity into a three-dimensional problem. In addition to the usual potential profile from the surface into the bulk (e.g., the one illustrated in Figs. 1, 6 or 12), there is *lateral* 'band-bending' from the pinhole center to its surroundings. Therefore, even if the potential energy at a pinhole is much smaller than around it on the interface plane, along the carrier's path into the semiconductor the potential approaches that of its surrounding. This saddle-like shape is known as 'pinch-off'[217] and it turns the SCR effect on charge transport to be far more robust, in terms of tolerance to defects than the direct molecular barrier (e.g., in metal/monolayer/metal junctions[218]). The pinch-off scales with the ratio between the pinhole diameter and the depletion width of the SCR region ($W_D$, see Eq. 36). A low semiconductor doping level and high SBH increase the depletion width, and therefore pinholes of larger area can be pinched-off. In such a scenario, a molecule-missing pinhole region would withdraw the current because of its lower tunneling barrier, but the SCR barrier would still be almost identical to the dipole-dictated SBH.[170]

The opposite scenario, where the pinhole has a larger SBH than its surroundings, is more complicated, because of the opposite effect on tunneling barrier and on SBH. It was shown that in such cases the edges of the pinhole present the lowest net barrier.[170] The applicability of the pinch-off inhomogeneous model was demonstrated using spatially resolved ballistic electron emission microscopy (BEEM) studies of molecular-modified MIS junctions.[119,170,219,220] Lateral inhomogeneity can account for an ideality factor, n > 1 (see section 7.a) and generally fit well the results obtained for junctions of metal/di-carboxylic acids/GaAs junctions.[119,161,170] However, junctions made by Hg top contact to monolayers directly bound to oxide-free Si, seem to follow almost perfectly the MIS theory described in section 6.a. Specifically, analysis of current-voltage traces, with or without temperature variation (see section 7.c and 7.d below) suggest that the actual area for transport is very close to the nominal contact area.[216,221] This is in marked difference to metal / monolayer / metal junctions where the current passes through a very small fraction, ca. $10^{-4}$, of the nominal contact area.[222] Hg/monolayer-Si junctions also appear to have more reproducible I-V traces than metal / monolayer / metal junctions. It is probable that SCR pinch-off and lateral potential leveling is responsible for the good match in contact area and high reproducibility of semiconductor-based molecular junctions.

## 7. Extraction of the dual barriers

We now present a practical overview of how to experimentally separate between the tunneling and Schottky barriers using standard electrical characterizations. The case of highly-doped Si (Eq. 28, section 6.b) is ignored, because often the semiconductor contribution can be neglected in such junctions (i.e., contacts made of highly-doped semiconductors are approximated as metallic). As explained in the previous section, the separation between the molecular, tunneling barrier and the semiconductor's SBH is not trivial. The first sub-section (7.a) describes how to extract the combined, effective SBH which includes both contributions; complementary impedance measurements (7.b) or temperature-current dependence (7.c) are two methods to isolate the real SBH contribution, and then the molecular barrier accounts for the difference between the effective and real SBH; finally we present two more recent, derivative-based methods (7.d) that attempt to separate the components of the dual barrier using only single I-V traces, without the need for complementary measurements (impedance or temperature-dependence).While these methods appear somewhat specific to certain conditions, they are very appealing when adequate. Other methods that can provide information on interface electronic states are internal photo-emission (IPE)[223] and ballistic electron-emission spectroscopy (BEEM)[170,219,224-227] which is an STM-based technique. As those techniques are not widely available they will not be covered here.

### a. Extraction of net barrier from a single I-V trace

Eqs. 29, 29 describe a rectifying process (as all diodes) where the current under reverse bias (we use V<0 to mark the 'reverse' direction or the 'off' state of the diode) is constant and equals a certain 'saturation current' ( $J_{V<0} \approx J_{Sat}$ ), while at forward bias (V>0) the current increases exponen-tially with the applied bias ($J_{V>0} \approx J_{Sat} \cdot exp(V/nkT)$). Such a distinct behavior is clearly seen in a semi-log plot of the current as function of bias (e.g., Figs. 5.C and 14.C). Therefore, $J_{sat}$ is the key parameter that we seek in analysis of diode I-V traces. Depending on various assumptions, $J_{sat}$ can be further translated into the SBH. Traditionally, $J_{sat}$ is extracted from the intercept (extrapolation to 0 V) of exponentially fitting the forward bias of the curve. In principle, $J_{sat}$ can be also identified with the reverse current, but, practical diodes suffer from various spurious effects[41] which make the reverse current not really constant. Similarly, quite often also the forward current is not purely exponential with the bias. This is exemplified in Fig. 15 for an arbitrary-chosen I-V trace measured across a p-Si(111)-C12/Hg junction.[188] Black symbols are current values and as can be seen they yield a rather curved dependence, and therefore, different fitting ranges could significantly shift the extrapolation to 0 V.

The red-dots of Fig. 15 exemplify a robust, simple alternative. Today's widespread adoption of digital data recording and processing offers a variety of analysis tools, which were impractical half a century ago, when the theory of semiconductor devices was established. Numerical differentiation is one such tool, although, naturally, direct recording of the differential conductance is preferable (better signal to noise ratio). Still, numerical differentiation is acceptable, because we are looking for trends rather than high-resolution extrema. We start from a generic form of Eqs. 29,30, where $V_{sc}$ is the fraction of the applied bias that falls over the space charge region of the semiconductor (in the terminology of Eq. 27, $V_{sc}$ = − dBB):

$$J = J_{Sat} \cdot \left[ exp\left(\frac{V_{sc}}{kT}\right) - 1 \right] \tag{31}$$

Commonly the non-ideality (n > 1) can be interpreted as a partition of the bias between the SCR and the above-mentioned spurious effects. Thus, Eq. 31 uses $V_{sc} = V/n$,[51] to adequately reflect any generic variation of $V_{sc}$ with the applied bias. This approach ignores transport by generation-recombination of minority carriers, where $n \equiv 2$ by theory, a unique case that can be treated specifically.

Differentiation of Eq. 31 eliminates the '-1' term from the brackets and yields a single continuous curve through 0 V:

$$\frac{dJ}{dV} = \frac{1}{kT} \cdot J_{Sat} \cdot exp\left(\frac{V_{sc}}{kT}\right) \cdot \frac{dV_{sc}}{dV} \tag{32}$$

Now, returning to the simplistic case of $V_{sc}$ = $V/n$ → d$V_{sc}$ = d$V/n$, and taking the log of both sides, yields:

$$\ln\left(\frac{dJ}{dV}\right) = \ln\left(\frac{J_{Sat}}{nkT}\right) + \frac{V}{nkT} \tag{33}$$

The red dots in Fig. 15 are the numerically obtained dI/dV values of the I (current) values shown by the black symbols. A semi-log plot of such dI/dV yields an extended linear range, which is continuous through 0 V (i.e., $J_{sat}$ is extracted by interpolation rather than extrapolation). The slope of such a fit is identical to standard fit (1/nkT; the two fitting lines in Fig. 15 are parallel), though the intercept is multiplied by 1/nkT (~20-40 at room temperature).

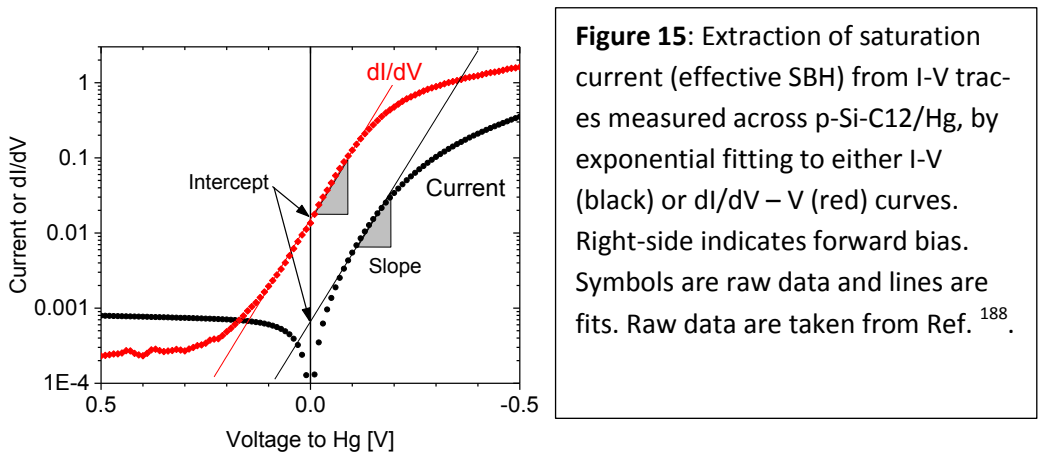

**Figure 15**: Extraction of saturation current (effective SBH) from I-V traces measured across p-Si-C12/Hg, by exponential fitting to either I-V (black) or dI/dV − V (red) curves. Right-side indicates forward bias. Symbols are raw data and lines are fits. Raw data are taken from Ref. [188].

The two top panels of Fig. 16 repeat the analysis of Eq. 33 for two systematic data sets: Fig. 16.A analyzes the I-V traces of a series of alcohols on Si(100) (same raw data as Fig. 5.C), while Fig. 16.B shows a series of alkenes on Si(111) (same raw data as Fig. 14.C). The metal contact for both junctions is Hg. As far as we can tell, the difference in Si orientation between the two sets does not have a major effect on monolayer quality or electronic properties, but in addition to orientation, the type of doping differs between the two sets. For clarity the direction of the X-axis for p-Si (Fig. 16.B,D) is inverted, so that in both panels *rightward* indicates higher *forward* bias. We focus on p-Si because the same system on n-Si (e.g., Fig. 14.B) is under inversion where $J_{Sat}$ is not related directly to the SBH (see section 4.d).

As can be seen in both Fig. 15 and 16.A,B, moving from semi-log I-V presentation to semi-log (dI/dV) − V format reveals sharp transitions between three regimes, which helps identifying the relevant fitting range. Yet, in contrast to the traditional ln(J)-V presentation (black symbols in Fig. 15) where the deviation of the slope near 0 V originates from neglecting the '-1' term of Eq. 31, Eq. 33 is formally accurate, and in principle should extend to any bias. Therefore the distinct regimes of the ln(dJ/dV) presentation (Fig. 16.A,B) reveal genuinely different transport regimes, which are hidden under the mathematical artifact of the traditional presentation. The voltage span of the middle exponential regime depends on the molecular identity as evident by comparing panels A and B of Fig. 16, as further considered in section 7.d below.

In summary, the exponential fitting to dI/dV is technically preferred over fitting the current directly because it extends the linear range, sharpens its onsets and brings the linear range to the vicinity of 0 V to allow interpolation rather than extrapolation. Next, we consider the physical meaning of the extracted parameters.

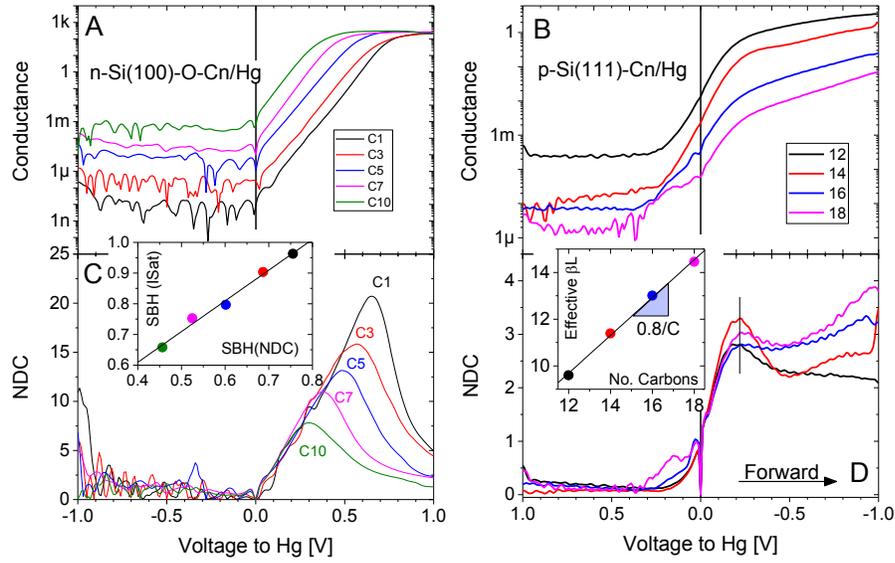

**Figure 16**: Analysis of current-voltage traces of molecularly modified Hg/Si junctions for linear alkyl alcohols of varying lengths, on n-Si (A,C), and alkenes on p-Si (B,D), for extraction of the saturation current from semi-log plots of conductance, G (A,B), or plots of normalized differential conductance, NDC (C,D), against the applied bias to Hg (rightward = forward). Both G (=dI/dV) and NDC are numerically computed. See text for insets. Raw data are taken from Refs. [74,188].

As noted above the ideality factor is basically a measure of the voltage partition (ignoring fundamental n≡2, as in generation-recombination). Using first Eq. 27 (with $dV_{sc} \equiv -dBB$) and then Eq. 24 yields:[51]

$$\frac{1}{n(V)} = \frac{dV_{sc}}{dV} = 1 + \frac{d\Delta}{V} = \frac{L}{\varepsilon_0 \varepsilon_i A}\left(\frac{dQ_{IS}}{dV} + \frac{dQ_{SCR}}{dV}\right) \qquad (34)$$

The amount of charge stored in the space charge region ($Q_{SCR}$) is a direct function of the depletion width, $W_D$:

$$\frac{dQ_{SCR}}{dV} \cdot \frac{1}{A} = \frac{\varepsilon_0 \varepsilon_{sc}}{W_D(V)} \qquad (35)$$

$$W_D{}^2 = BB(V) \cdot 2\varepsilon_0 \varepsilon_{sc}/qN_D \qquad (36)$$

Predicting the change in population of interface states ($Q_{IS}$) with bias voltage cannot be expressed in such a simple way and depends on the energy distribution of these states within the gap ($D_{IS}$) and their coupling strength to each contact (metal or semiconductor). A rigorous analysis employs exact tunneling rates and cross sections;[201,202] a common simplified approach divides the interface states into two populations: $D_{IS-M}$ which equilibrates with the metal side (i.e., vary as a function of

d$\Delta$) and $Q_{IS\text{-}SC}$ that is coupled to the semiconductor side (i.e., $\propto V_{sc}$). Within this rough approximation, Eq. 34 can be re-written as :[51]

$$n(V) = 1 + \frac{\frac{\varepsilon_0 \varepsilon_{SC}}{W_D(V)} + q D_{IS-SC}}{\frac{\varepsilon_0 \varepsilon_i}{L} + q D_{IS-M}} \qquad (37)$$

Eq. 37 can be simplified for the two extreme cases of only one type of interface states: [51]

$$n(V) - 1 = \begin{cases} \frac{L}{W_D(V)} \cdot \frac{\varepsilon_{SC}}{\varepsilon_i + \frac{q}{\varepsilon_0} L D_{IS-M}} & D_{IS-SC} \to 0 \\ \frac{L}{\varepsilon_i} \cdot \left[ \frac{\varepsilon_{SC}}{W_D(V)} + \frac{q}{\varepsilon_0} D_{IS-SC} \right] & D_{IS-M} \to 0 \end{cases} \qquad (38)$$

In the case of monolayers, the thickness of the monolayer (insulator) is commonly much smaller than that of the depletion width ($L\sim$1-3 nm cf. $W_D\sim$100's of nm; $\varepsilon_{sc}/\varepsilon_i \sim 4$) and therefore the first option of Eq. 38 yields a negligible increase in the ideality factor. Moreover, $D_{IS\text{-}SC}>0$ acts to further reduce $n$, and therefore in the context of monolayers, the ideality factor can be used to approximate the density of interface states based on the second option of Eq. 38.[228] Still, Eqs. 37,38 are based on a gross simplification to d$D_{IS}$/dV and, therefore, evaluating $D_{IS}$ using alternative methods such as impedance spectroscopy,[86,228,232,233] are more trusted.

Here our focus is mainly on the ability of the molecular monolayer at the interface to alter the net SBH via its dipole. The SBH affects the junction's rectification or how small is the saturation current at reverse bias: the higher the SBH is, the lower is the saturation current, $J_{Sat}$. While the exact translation of $J_{Sat}$ into the SBH varies for different situations (e.g., minority / majority carriers, Eq. 29 cf. Eq. 30 and section 4.d), it is convenient to compare generic, effective interface barrier, $SBH_{IV}$, values rather than $J_{Sat}$ values (where the subscript 'IV' serves to stress that it is extracted from analysis of I-V traces and not by alternative methods discussed below). Therefore, regardless of details, $J_{Sat}$ is translated into $SBH_{IV}$ by imposing the thermionic emission transport (see Eq. 30):

$$SBH_{IV} = kT \cdot \left[ \ln(A^* T^2) - Itrc_{lnJ-V} \right] \qquad (39.a)$$

$$SBH_{IV} = kT \cdot \left[ \ln(A^* T^2) - Itrc_{lnG-V} + Slp_{lnG-V} \right] \qquad (39.b)$$

where *Itrc* and *Slp* are the intercept and the slope of the linear fit (y = *Slp*·x + *Itrc*), to either ln(J)-V (Eq. 39.a, derived from Eq. 30) or ln(dJ/dV) to V (Eq. 39.b, derived from Eq. 33).

Thus, the saturation current is a single experimental observation that can provide only one piece of information, the net effective barrier, $SBH_{IV}$. As described in section 6.a, in thin MIS junctions there are two significant barriers: the space charge region within the semiconductor ('SBH') and the insulator (molecular layer) tunneling attenuation ('$\beta L$') and generally both add to the net observed $SBH_{IV}$:[228]

$$SBH_{IV} = SBH + kT \cdot \beta L \qquad (40)$$

The *kT* term in Eq. 40 is slightly misleading, because it is the other term (SBH), which is actually temperature-dependent (Eq. 30). This odd format stems from the choice of comparing effective SBH values (Eq. 39). As explained above (section 6.b), adsorbed molecules can affect both *SBH* and $\beta L$, as further illustrated by the systematic length variation in both systems presented in Fig. 16. In both cases (Fig. 16.A,B) there is an exponential change in $J_{Sat}$ ($\to SBH_{IV}$) with length, however of opposite trend. For Hg/Cn-O-Si ("alkoxy", Fig. 16.A) the chain length affects the net induced dipole, while for Hg/Cn-Si ("alkyls", Fig. 16.B) elongating the chain increases the tunneling barrier ($\beta L$). It is not clear though if this difference is due to the presence of the O linker and residual hydroqui-none[74,85] in the Hg/Cn-O-Si junctions (Fig. 16.A) or simply because the alkoxy experiment tests twice shorter alkyls than the alkyl one (1 to 10 carbons cf. 12 to 18 in Fig. 16.A and B respectively). Experimental support for the different molecular role in these systems cannot come from the saturation current alone. Such a differentiation of the dual barrier can be done using complementary measurements and /or other analyses, as discussed in the remainder of this section.

### b. Capacitance – Voltage

Impedance measurements are very common in analysis of metal-oxide-semiconductor structures (MOS, also known as gate dielectrics), where scanning the frequency and bias reveals the density of trap states.[205,230,231] Impedance spectroscopy was used in the context of metal-molecule-semiconductor junctions to estimate the distribution of surface states,[86,228,232,233] as well as to explore polarization of various molecular segments.[234-236] Care should be taken in translating standard MOS models into ultra-thin interfacial layers like molecular monolayers, because the insulator can no longer be approximated as a pure capacitor, because of the significant tunneling current ('leakage current') that adds a non-negligible real component to the net impedance.[86]

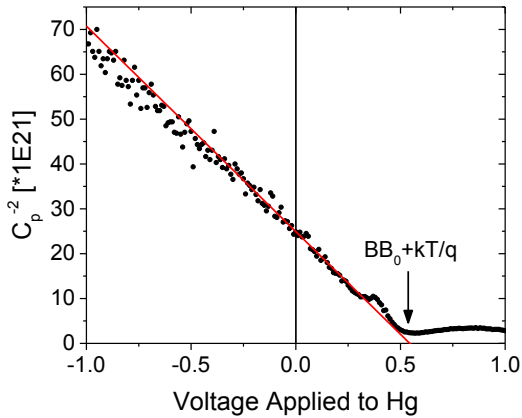

**Figure 17**: Extraction of equilibrium BB from Mott-Schottky plots, demonstrated for n-Si(111)-Styrene/Hg junction (similar to Fig. 8). The net impedance is modeled as a parallel $R_p$-$C_p$ circuit and $C_p^{-2}$ is plotted against the net applied voltage (symbols). Line is linear fit over the range of –0.4 to +0.2V. The voltage of where the fit crosses the X-axis gives the band bending at zero-voltage (BB₀), See Eq. #41. Data are taken from Ref. [245].

However, here we focus on a narrow aspect of impedance measurements, estimation of the *SBH* based on the junction capacitance. The capacitance is measured at relatively high frequencies ($\geq$ 0.5 MHz) to avoid (dis)charging of interface states.[228] As illustrated in Fig. 12.B, the charge accumulated in the space-charge region (green circles in Fig. 12.B) with its counter charge on the metal (yellow circles) makes a capacitor, where its thickness is dictated by the width of the depletion layer ($W_D$). The resulting depletion capacitance, $C_D$, is:[41]

$$C_D = A \frac{\varepsilon_0 \varepsilon_{sc}}{W_D} = A \sqrt{\frac{q \varepsilon_0 \varepsilon_{sc} N_D}{2\left(BB_0 - V_{sc} - \frac{kT}{q}\right)}} \tag{41}$$

Assuming minor losses on the insulator, $C \approx C_D$, and $V \approx V_{sc}$ implies that a plot of $1/C^2$ is expected to decrease linearly with the applied bias, known as Mott-Schottky plot. Such analysis is illustrated in Fig. 17 for a styrene monolayer on n-Si(111), contacted by Hg from top (system presented in Fig. 8).[163] The fit extrapolates to zero (crosses the X-axis, or $C_D \to \infty$) at $V_\infty$ when the denominator of Eq. 41 nulls:

$$BB_0 = \frac{V_\infty}{n(V)} - \frac{kT}{q} \tag{42}$$

where the ideality factor, $n$, is used again to stress the possible difference between the applied voltage, $V$, and the actual potential drop over the space-charge region, $V_{sc}$ (see Eq. 34). For this reason we use the terminology of $BB_0$ instead of the commonly used 'flat-band potential' ($V_{FB}$) since the latter includes also the (small) potential drop over the interfacial insulator ($V_{FB}=BB_0+\Delta_0$, see section 6.a). Notice, though, that the ideality factor, especially for a molecularly-modified MIS, is strongly bias-dependent: the bias range where the Mott-Schottky plot is linear occurs when $n$ is roughly constant with bias. Therefore, it is difficult to get a practical estimate for $n(V)$ for use in Eq. 42 and in practice, Eq. 42 uses $n=1$. However, we often find that C-V traces on repeating junctions produce much better reproducibility in $V_\infty$ than in the slope of the Mott-Schottky plot. Eq. 41 shows that, within the approximations used to arrive at the expression, such a slope depends purely on the semiconductor parameters. It is possible that the varying slope is due to local variations in $Q_{is}$ (Eq. 34) though it is impossible to isolate this effect from some variation in the doping density ($N_D$ in Eq. 41).

Another possible source for the varying slope in Fig. 17 is the monolayer's own capacitance, $C_{SAM}$. The net effective capacitance, $C_{tot}$, is made up of two capacitors in series:

$$C_{tot} = \frac{C_D \cdot C_{SAM}}{C_D + C_{SAM}} \xrightarrow{C_{SAM} \gg C_D} \approx C_D \qquad (43)$$

Because the net capacitance of two capacitors in series is dominated by the smaller one, Eq. 41 can be used only as long as the monolayer is very thin ($C_{SAM} = \frac{\varepsilon_{SAM}}{L}$), to fulfill the condition of $C_{SAM} \gg C_D$. We illustrate this reasoning with some practical value: the capacitance of the space-charge region is in the order of 0.01 to 0.1 $\mu F/cm^2$, for common values of band bending (0.2 to 0.5 eV) and Si doping (1E14 to 1E16 cm$^{-3}$). In comparison, a $1-3$ nm thick monolayer, with a dielectric constant of $2.5-3.5$, yields $C_{SAM}$, in the order of 0.8 to 3 $\mu F/cm^2$. Thus, the use of Eq. 41 is limited to moderate Si doping and relatively thin monolayers. In cases where $C_{SAM}$ is similar or smaller than $C_D$, concepts describing gate dielectrics are applicable, as described in Refs. 41,205,231.

Returning to the main goal of separating the effective SBH$_{IV}$ into its two generic sources (Eq. 40), we can use the C-V derived $BB_0$ (Eq. 42) to deduce SBH$_{CV}$ , based on Eq. 3, and using manufacturer resistivity values to compute ξ. Such SBH$_{CV}$ is expected to be closer to the actual SBH, as it does not include the unknown tunneling attenuation, $\beta L$.

Therefore, trends in *SBH* or $BB_0$ are identical; they only differ by a constant shift of ξ (see left and right Y-axis to Fig. 8.B). For a Si doping level of 1E15 cm$^{-3}$, used in Figs. 8 and 17, ξ = 0.25 eV.

Fig. 8.B compares the *effective SBH*$_{IV}$ (black circles, Eq. 40) with the *SBH*$_{CV}$ (red triangles) extracted from C-V measurements using Eqs. 41, and 44. In the absence of a monolayer (100% transmission probability across the molecules or $\beta L\rightarrow$0), the extracted SBH should be independent of whether it was extracted from C-V or I-V measurements. Looking at Fig. 8.B, such overlap occurs for the methyl-styrene (Me-Sty) junction, but not for simple styrene (H-Sty). The C-V data for the Br styrene junction are missing, because its low capacitance (nearly Ohmic junction) does not allow extraction of meaningful data. The difference in the $\beta L$ contribution to H-Sty and Me-Sty does not originate in the monolayer transmission probability, but reflects a change in the type of dominant carriers (cf. panels B and C of Fig. 14). For H-Sty the space charge is a depletion layer and, thus, transport is by majority carriers, which is tunneling-attenuated (Eq. 30). In contrast, Me-Sty drives the junction into inversion, so that the current is mostly carried by minority-carriers (Eq. 29), and is limited by their supply to the surface (the buried p-n junction), rather than by their probability to cross the interface. For this reason, the tunneling attenuation, $\beta L$, does not add to the effective SBH in this junction (see section 6.c). Fig. 8.B suggests an over-expression of *SBH*$_{CV}$ with respect to the measured surface dipole (X-axis) with a slope larger than the theoretical limit of 1. Such an over-expression may hint at that the interfacial dipole is not identical to the surface one, due to charge-rearrangement with the top contact, though this requires further studies, using larger sets of molecules.

Lateral inhomogeneity of the potential profile (section 6.c) is another reason for differences between SBH values, extracted from C-V and I-V data.[217,237,238] While the capacitance scales linearly with the barrier height, the current depends exponentially on that height. It follows that if the SBH is spatially varying, C-V yields a linear average while the I-V is sensitive to even extremely small patches of low barrier height. Small patches of high barrier would not balance this because the current, at least at low flux would go through the lower barrier patches.[119,217] Thus inhomogeneity generally leads to $SBH_{IV} \leq SBH_{CV}$,[237] which is opposite to what is seen in Fig. 8.B. Thus we infer a minor role layer/ interface inhomogeneity for the styrene-induced SBH, and the higher $SBH_{IV}$ is because of the additional tunneling term ($kT \cdot \beta L$, similar to e.g., the length attenuation in Hg/alkyl-**p**-Si in Fig. 14.C). This term is missing for methyl styrene ($SBH_{IV} \approx SBH_{CV}$), because these junctions are dominated by minority carriers (e.g., length-independence in Hg/alkyl-**n**-Si in Fig. 14.B).

### c. Temperature variation

Measuring the current as a function of temperature is a different way for separating the effective $SBH_{IV}$ (Eqs. 39, 40) into its two components. It is exemplified here for a junction made of a monolayer of a 16 carbon long alkyl chain on oxide-free Si(111) and contacted on the top by thermally evaporated Pb. The area of the junctions was defined by etching circular wells within a high-quality thick $SiO_2$ film and adsorbing the monolayer on freshly exposed H-Si at the bottom of the well. The data of Fig. 18 were recorded for 100 μm diameter wells, though successful junctions were made with diameters down to 3 μm.[186] Such evaporated contact considerably facilitates cryostat-based temperature-dependence experiments. Fig. 18.A shows that the cooling / heating cycle yields reproducible current – voltage traces.

The fundamental Eq. 30 can be re-written with respect to temperature:[186,228,239]

$$ln\left(\frac{I}{T^2}\right) = ln(A \cdot A^*) - \beta L - \frac{q}{k_B T} \cdot \begin{cases} SBH & V < 0 \\ SBH - V/n & +3k_B T < V < SBH \\ 0 & V > SBH \end{cases} \quad (44)$$

A semi-log plot of the current density against the reciprocal temperature (Arrhenius plot) is shown in Fig. 18.B for a few selected voltage values. Eq. 44 predicts that the intercept of the fits includes the tunneling decay coefficient, $\beta L$, while its slope or activation energy includes the SBH. Yet, the bias can also add to the activation energy for the middle of the three regimes within the curly bracket of Eq. 44. These regimes are similar to those that appeared in Fig. 16.A,B for the variation of the conductance with applied bias.

Although Eq. 44 predicts a bias-independent activation energy for V<0, in practice there is a mild decay from 0 V toward high reverse bias, which probably reflects some non-ideality effects like image-force reduction of the SBH.[41] At forward bias there is a very sharp reduction in the activation energy, because it includes a (–V/n) term. Here the slope in activation energy with respect to forward bias (red line in Fig. 18.C) yields an ideality factor of 1.7 in agreement with the ideality factor extracted from the I-V curve directly.[186] In principle extracting the SBH from the activation energy at 0 V would give the least perturbed value. Yet, the current near 0 V is very noisy which explains the data scattering near 0V. Therefore we set the extrapolation of the fitted red line to 0V as the best approximation for the real SBH = 0.47 eV. As the applied forward bias exceeds the SBH, the semiconductor surface turns into accumulation, which saturates the activation energy, to a value around 0.1 eV. This region will be further discussed in section 7.d below.

The numerical result of SBH=0.47 eV can be verified with respect to the Schottky-Mott rule (Eq. 4). With a 4.05 eV work function of Pb and 3.6 eV effective electron affinity of the alkyl-modified Si ($\chi$ + $\Delta$)[180] a SBH of 0.45 eV is calculated, in excellent agreement with the temperature-derived value.

Namely, the Pb/alkyl-Si junction perfectly follows the Schottky-Mott rule, which is not commonly observed for intimate metal/semiconductor junctions, but more often reported for molecularly modified ones.[74]

Eq. 44 suggests that the intercept of the Arrhenius plot contains information on the tunneling decay constant. This parameter is plotted in Fig. 18.D, where the left Y-axis shows the $\beta L$ product ($\beta L = \ln(A \cdot A^*) -$ intercept, using $\ln(A \ A^*) = -4.7$, based on nominal A & $A^*$ values) and the right Y-axis gives its translation to tunneling decay coefficient using $L = 19\text{Å}$, as measured by ellipsometry for these monolayers. The $\beta$ values extracted in this way are around 0.8 $\text{Å}^{-1}$, which are in the center of the range of values commonly reported for tunneling through alkyl chains,[180,215] confirming the validity of this analysis. The $\beta L$ values appear as highly bias sensitive, which probably originates in the complicated balance between the tunneling and space-charge transport barriers (section 6) and does not represent than a genuine bias effect on the tunneling probability.

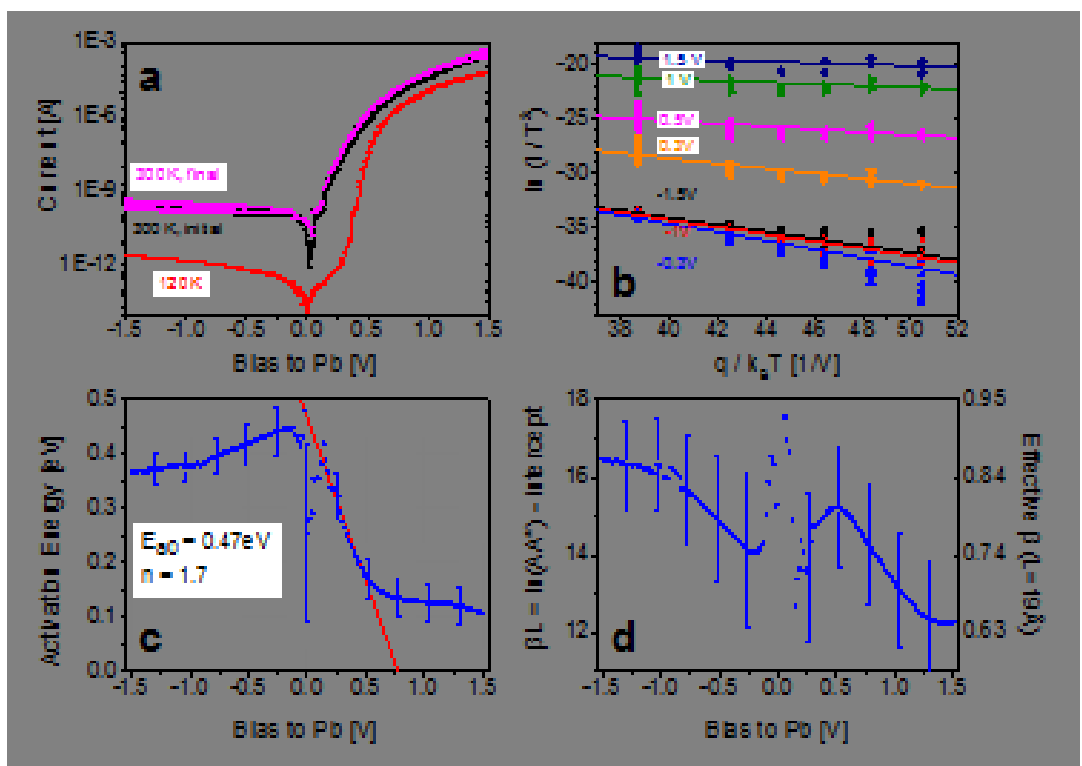

**Figure 18**: Temperature analysis of junctions of Pb/H(CH$_2$)$_{16}$-Si(111), using moderately doped n-Si, showing (**a**) reproducibility in cooling / heating cycle of I-V curves; (**b**) Arrhenius plot of log($I/T^2$) vs. $q/k_BT$ (reciprocal temperature in eV units) for a few bias values; (**c**) thermal activation energy, $E_a$, as a function of bias, and (**d**) tunneling attenuation, $\beta L$, as a function of bias. Curves in (**a**) are averaged over five junctions, and error bars are plotted, but might be too small to note. In (**b**) each symbol is for a different bias, and lines are linear fits to the data between 300 and 230K. Fitting was done on multiple data (different junctions, cooling and heating) without averaging. The slope of such fit equals $-(E_a)$ in [eV], as plotted in (**c**), and the intercept of the fit relates to $-\beta L$, as shown in (**d**). Error bars in (**c, d**) are the regression uncertainty in the extracted fitting values. The red line in (**c**) is a fit to Eq. #44. The right Y-axis of (**d**) gives $\beta$ values, calculated from the intercepts for a monolayer width, $L = 19$ Å. Figure was reproduced with permission from Ref. [186].

### d. Advanced derivative-based methods

So far we relied on electrical characterization, in addition to room temperature I-V characteristics, such as impedance measurements (section 7.b) or temperature variation (section 7.c) to distinguish between the semiconductor and the tunneling contributions to the net barrier. Alternatively, additional information can be gained by various manipulations on the current-derivative with respect to bias, as described in this section. These approaches are by definition more noisy and rough, but are appealing as they save the extra characterization. As mentioned above, all analyses presented here are based on numerical differentiation; the results are expected to be identical to those based on direct recording of differential conductance, though the latter might well be less noisy.

We consider two methods: one uses normalized differential conductance (NDC) and the other exploits the correlation between the current derivative and the current at high signals. NDC is often used for analysis of tunneling spectroscopy,[240] and Godet et al. have adopted it to analyze molecularly modified MIS junctions.[241] Mathematically, NDC, is defined as the derivative divided by the ratio (NDC = (dI/dV)/(I/V) or NDC = d(Log I)/d(Log V)); for a standard diode equation (regardless of the exact semiconductor transport type), using Eq. 31 for the current and Eq. 32 for the current derivative, we get:

$$NDC_{sc} = \frac{dJ_{sc}}{dV} \cdot \frac{V}{J_{sc}} = \frac{V}{kT} \cdot \frac{dV_{sc}}{dV} \left[ 1 - exp\left(-\frac{V_{sc}}{kT}\right) \right]^{-1} \qquad (45)$$

where the subscript $sc$ stresses the assumption of semiconductor-limited current and the concept of $V_{sc}$ was introduced in Eq. 34 above. For the commonly used assumption of $V_{sc} = V/n$ (or any other polynomial relation), $V \cdot dV_{sc}/dV = V_{sc}$, and therefore, Eq. 45 simplifies to:[241]

$$NDC_{sc}\left(V_{sc} = \frac{V}{n}\right) = \begin{cases} 0 & reverse \\ |V_{sc}|/kT & forward \end{cases} \qquad (46)$$

Such NDC analysis is shown in Fig. 16.C,D for the same raw data of the corresponding top-panels. For both systems the NDC approaches zero under reverse bias (left-ward) and starts increasing linearly with the bias as the bias polarity switches, in agreement with Eq. 46. However, at a certain point this trend is broken and the NDC starts decreasing. Because Eq. 46 ignores the interfacial insulator (i.e., it is valid for an abrupt diode), this peak is easily understood as the cross-over from bias drop over the space-charge region (dV≅dV$_{sc}$, diode-dominated) to bias drop over the monolayer (dV≅dΔ, tunneling-dominated; see Fig. 12, section 6.a and Eq. 34). Such a transition occurs roughly at the flat band potential ($V_{FB}$),[216,241] though it is difficult to derive an exact relation between the NDC peak position (NDC$_P$) an $V_{FB}$. Numerical simulations (following Ref. 202) suggest that the ratio NDC$_{MX}/V_{FB}$ = 1 − 1.5, where the ratio increases for low doping levels and decreases with temperature. A reasonable approximation is $V_{FB} \cong NDC_{MX}/1.15$ for room temperature and moderate doping level of 1E15 cm$^{-3}$. While this is admittedly a crude assumption it can be employed for separating the effective SBH (Eq. 39) into its two components (Eq. 40) for each individual I-V trace without any additional characterization.

The use of NDC peak for separating the effective SBH$_{IV}$ into its two sources (Eq. 40) is demonstrated for the two molecular systems in Fig. 16. For the alkoxy system (Fig. 16.C), the NDC peak shifts to higher bias for shorter length, while for alkyl monolayers (Fig. 16.D) the peak is at a constant position. This agrees with our understanding that varying the length of the alkoxy alters their surface- induced dipole by ca. 0.4 eV and therefore the net SBH$_{IV}$ (Fig. 5.D). The inset in Fig. 72C,D shows the correlation between the effective SBH$_{IV}$ (extracted from the saturation current and Eq. 39.b) and the real SBH, extracted from the position of NDC maximum (SBH$_{NDC}$ = NDC$_{MX}$/1.17 + ξ; ξ=0.2[74]). The line is a linear fit with slope of unity achieved by slightly adjusting the ratio to 1.17 (from 1.15). The intercept of 0.2 eV, is possibly due to some residual $kT \cdot \beta L$ (≈ 8, see Eq. 40).

In contrast, elongating the length of alkyl monolayers from 12 to 18 carbons has no effect on $NDC_{MX}$ (Fig. 16.D). This is because the surface dipole induced by alkyls is not sensitive to their length[148,180] and therefore their actual SBH is constant. Yet, their reverse current is length-dependent because of the tunneling attenuation. The $\beta L$ contribution can be now evaluated from the difference between the effective $SBH_{IV}$ ($I_{sat}$ + Eq. 39.b) and the NDC-derived SBH ($NDC_{MX}$ /1.15 +0.13)[242] The $\beta L$ values extracted in this way are shown in the inset to Fig. 16.D as a function of number of carbons. The black line is a linear fit, with a slope of $\beta$ =0.81/C (0.66/Å) which is still within, though at the low end of the range of values found for saturated alkyl chain-based mono-layers.[180,215]

The height of the NDC peak is simply a function of how large is the SBH: the larger is the SBH, the higher the forward bias at which the peak occurs and the NDC is able to reach huge values (see C1 cf. C10 in Fig. 16.C). In comparison, tunneling I-V relations yield maximal NDC in the order of $\sim 3$.[7] Godet et al. derived an approximate expression for NDC(V) in the tunneling regime based on the Simmons model ($NDC_{Simmons} = LqV/\sqrt{BH - qV}$) and used it to characterize the tunneling barrier (BH) within similar Hg/SAM-Si junctions.[241] As discussed next, even at high forward bias the transport characteristics are not purely those of tunneling, and therefore we hesitate from fully following this approach. Still, Fig. 16.D shows that the saturation NDC increases for longer alkyl chains, in qualitative agreement with Godet's prediction.[241]

A different derivative approach focuses on the high forward current and plots the inverse derivative against the current (rather than vs. bias as commonly used). Since the current increases exponentially until roughly the flat-band potential (ca. $NDC_{MX}$), such a presentation naturally obscures the low signal range (V<$NDC_{MX}$). This analysis was originally derived[243,244] to account for a resistance in series to the diode, $IR_S$. It is straightforward to extend this analysis to insulator-containing diodes by adding the potential drop on the insulator, $R_{ins}$[V/A$^{1/2}$] , which yields:[216]

$$\frac{dV}{d\ln I} = nkT/q + R_{ins}\sqrt{I} + IR_S \qquad (47)$$

In difference to former views,[203,204,241] where the potential drop on the insulator was attributed to tunneling resistance, Eq. 47 assumes that charge balance dictates the potential drop on the insulator (Δ see Fig. 12 and Eq. 24). As the applied bias crosses the flat-band potential (ca. $NDC_{MX}$) the semiconductor's space-charge region turns into accumulation and its net charge, $Q_{SCR}$, increases exponentially with $\sqrt{V_{sc}}$; the accumulating $Q_{SCR}$ increases linearly the potential drop on the insulator, Δ (Eq. 24). Assuming that this effect is much larger than the tunneling resistance, $R_{ins}$ is derived from the classical expression for charge stored in a semiconductor space charge region under accumulation:[216]

$$R_{ins} = \frac{L}{\varepsilon_i D} \sqrt{\frac{2\varepsilon_{sc}kN_C}{\pi A^* \varepsilon_0 T}} \cdot e^{\beta L/2} \qquad (48)$$

where the square root dependence on current as well as the scaling with diameter rather than area, qualitatively originate in the dependence: $\Delta \propto Q_{SCR} \propto \exp(V_{sc}/2) \propto \sqrt{I}$ (Eq. 31). Therefore, accounting for the insulator adds a $\sqrt{I}$ term to Eq. 47 and scaling by the diameter rather than the area, both are not expected for pure series resistance.[243,244] This was tested on two types of Pb/monolayer-Si(111) junctions, for monolayers made of either methyl-styrene (Me-Sty, Fig. 19.A,C) or 16-carbons long alkyl chains (C16, Fig. 19.B,D). Junctions were fabricated by etching wells within $SiO_2$ (same as in Fig. 18), such that each curve in Fig. 19.A,B refers to a junction of different lateral size (contact area). For the C16 junction (Fig. 19.B) is linear with $\sqrt{I}$, indicating a negligible $IR_S$ contribution while the Me-Sty monolayer (Fig. 19.A) includes a significant linear term ($\sqrt{I}$), though some $IR_S$ contribution turns them less linear than the C16 plots. In addition, all curves intersect the Y-axis (*I*=0) roughly at the same value, in accordance with the diameter-independent constant term of Eq. 47. The mean intercept yields ideality-factor values of *n* = 1.3 and 2 for Me-Sty and C16, respectively (the high *n* value for C16 agrees with the high *n*-values deduced from the temperature-dependence (Fig. 18.C).

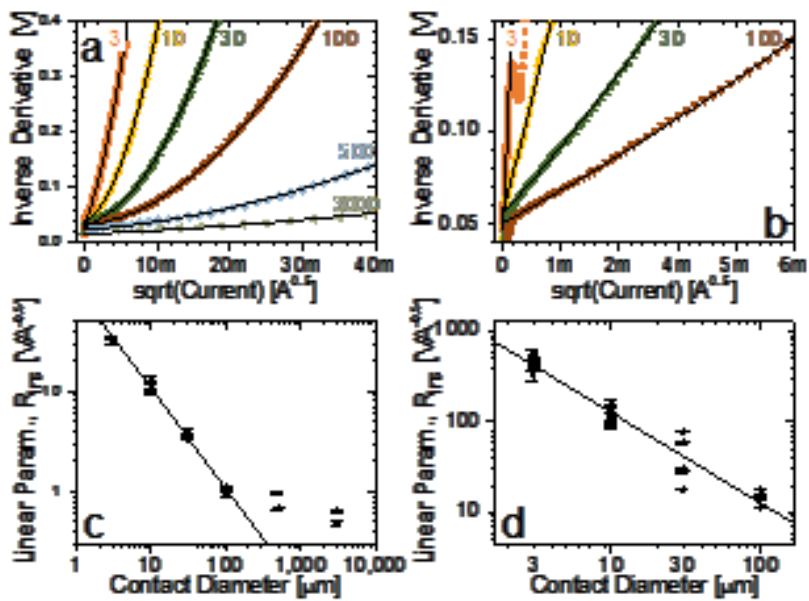

**Figure 19.** Inverse derivative analysis (Equation 77) of transport across molecular junctions made of Pb/Me-Sty-Si **(a,c)** and Pb/C16-Si **(b,d)**. The top panels **(a,b)** show the inverse derivative, d*V*/dln(*I*), against the square root of the current where each curve is for a different contact diameter (see legend) with symbols for data; lines are fits to Equation #47. The bottom panels **(c,d)** show the linear coefficient of the fit (*i.e.*, *R*ins) against the contact diameter on a log-log plot; multiple points refer to coefficients, extracted from *I* − *V* sets measured on different contacts. Lines in **(c,d)** are power fits to *R*ins ∝ *D*⁻¹ (Eq. #48). Figure was reproduced with permission from Ref. [216].

However the physically interesting quantity is $R_{ins}$ which is the coefficient of the linear term in the fitting of Fig. 19.A,B. To support the validity of the charging interpretation of $R_{ins}$ (i.e., Eq. 48), the bottom two panels of Fig. 19 show a power plot the experimentally extracted $R_{ins}$ as a function of the junction diameter, *D*. The solid lines show that almost all data points (extracted from few junctions for each size) fall on the expected slope of -1. The only exceptions are the largest contacts, possibly due to too high net currents. Beyond the physical constants and materials properties, $R_{ins}$ is dominated by the exponential dependence on $\beta L$. Thus the point where the fitted lines of Fig. 19.C or 19.D crosses D=1 can be used to extract $\beta$ by using L= 8Å for Me-Sty and 19Å for C16 and an insulator dielectric constant, $\varepsilon_i$ = 2.7, gives tunneling decay constants, $\beta$, of 1.45 Å⁻¹ for Me-Sty and 0.78 Å⁻¹ for C16.[216]

In short, section 7 described the key parameters that characterize the performance of Metal-Insulator-Semiconductor junctions with thin (< 3 nm) insulator in general, and specifically, when the insulator is made of a molecular monolayer. These parameters are the SBH, ideality factor and tunneling barrier. We reviewed several possible approaches, which overall yield similar values for the key parameters. Therefore, for well-behaved, close to ideal junctions, the different methods are fairly comparable, but for less ideal junctions (e.g., inhomogeneous ones, or junctions with considerable density of interface traps) the different models can show a significant spread in parameter values. The derivative techniques presented in 7.a and 7.d are rather novel, and hold potential to become handy, insightful tools.

## 8. Summary

We discussed the power of adsorbed molecular monolayers to control interface electrostatics. Harnessing such control is fundamental to numerous technical fields, mostly electronics, but in a wider sense, it also relates to catalysis and friction. Interface electrostatics is fascinating, because it emerges by, and reflects on, the illusive border between localized and continuum systems. Localized, inter-molecular polarization translates into very strong electric fields when many molecules form a 2-dimensional array as in a monolayer. Although just 1-2 nm wide, this abrupt potential step can easily control the surface charging of the semiconductors, often extending 100's of nm into the semiconductor. At the same time, the localized, molecular polarization is affected by long-range interactions with the substrate, beyond the formal chemical bond, which induce hybrid energy-states. The relation between the polarizability of the organic monolayers and the establishment of electronic equilibrium between a closed-shell molecular species and solid, bulk material, is a key toward predictable use of molecular modification, specifically, and interfacial energy alignment in general. Lastly we demonstrate the effect of surface modifications on transport across metal / molecular-monolayer / semiconductor junctions. The combination of molecular, tunneling barrier in series with a semiconductor space-charge barrier, leads to rich transport behavior. We described the various chemical handles available for tuning this seemingly chaotic interplay and provided various analysis tools for distinguishing between the direct molecular barrier and the induced, semiconductor barrier. The study of molecularly modified interface electrostatics extends back a few decades and has now reached a level of understanding that allows chemists to control the effects to a level where the field is ripe for application.

## AUTHOR INFORMATION

### Corresponding Author: Ayelet Vilan, Ph.D.   E-mail: ayelet.vilan@weizmann.ac.il

### Present Addresses

### Notes

The authors declare no competing financial interest.

### Acknowledgments

For this review we used, in addition to data published in journals, data from the M.Sc. thesis of Nir Stein. We thank Antoine Kahn (Princeton Univ.) and Leeor Kronik (Weizmann Inst.) for fruitful discussions, and the Israel Science Foundation via its centers of Excellence program and the Kimmel Centre for Nanoscale Science for partial support. This research was made possible in part via the historic generosity of the Harold Perlman family. DC holds the Rowland and Sylvia Schaefer chair in Energy Research.

### Biographies

**Ayelet Vilan** received her Ph.D. in the Materials and Interfaces department of the Weizmann Institute of Science, studying dipolar monolayers at the Au/GaAs surface. She did post-doctoral studies at Philips Research and Ben-Gurion University of the Negev, after which she returned to the Weizmann Institute in 2005, adding a specialization in surface science. In 2013 she spent a year at Texas A&M. Her research interests focus on molecular electronics, until recently primarily on large-area molecular junctions, especially those on semiconductor electrodes and more recently also include single-molecule break junctions.

**David Cahen** studied chemistry & physics at the Hebrew Univ. of Jerusalem (HUJ), Materials Research and Phys. Chem. at Northwestern Univ, and Biophysics of Photosynthesis (postdoc) at HUJ and the Weizmann Institute of Science, WIS. After joining the WIS faculty he focused on alternative sustainable energy resources, in particular on various types of solar cells. Today his work in this area focuses on the materials and device chemistry and physics of high-voltage (mainly halide perovskite-based) cells. In parallel he researches hybrid molecular/non-molecular systems, nowadays focusing on peptide & protein bio-optoelectronics and implications for electron transport across biomolecules. A fellow of the AVS and MRS, he heads WIS' Alternative, sustainable energy research initiative.

## 9. References

(1)     Aswal, D. K.; Lenfant, S.; Guerin, D.; Yakhmi, J. V.; Vuillaume, D. Self assembled monolayers on silicon for molecular electronics. *Anal. Chim. Acta* **2006,** *568* (1-2), 84.

(2)     Vilan, A.; Yaffe, O.; Biller, A.; Salomon, A.; Kahn, A.; Cahen, D. Molecules on Si: Electronics with Chemistry. *Adv. Mater.* **2010,** *22* (2), 140.

(3)     Metzger, R. M. Unimolecular electronics. *Chem. Rev.* **2015,** *115* (11), 5056.

(4)     Nichols, R. J.; Higgins, S. J. Single-Molecule Electronics: Chemical and Analytical Perspectives. *Annu. Rev. Anal. Chem.* **2015,** *8*, 389.

(5)     Xiang, D.; Wang, X.; Jia, C.; Lee, T.; Guo, X. Molecular-Scale Electronics: From Concept to Function. *Chem. Rev.* **2016,** *116* (7), 4318.

(6)     Su, T. A.; Neupane, M.; Steigerwald, M. L.; Venkataraman, L.; Nuckolls, C. Chemical principles of single-molecule electronics. *Nat. Rev. Mater.* **2016,** *1*, 16002.

(7)     Vilan, A.; Aswal, D.; Cahen, D. Large-area, ensemble molecular electronics: Motivation and challenges *submitted* **2016**.

(8)     ε0=1 in the Gaussian unit system.

(9)     Kroemer, H. Nobel Lecture: Quasielectric fields and band offsets: teaching electrons new tricks. *Rev. Mod. Phys.* **2001,** *73* (3), 783.

(10)    Cahen, D.; Kahn, A.; Umbach, E. Energetics of molecular interfaces. *Mater. Today* **2005,** *8* (7), 32.

(11)    Kahn, A.; Koch, N. In *The Molecule–Metal Interface*; Wiley-VCH Verlag GmbH & Co. KGaA, 2013.

(12)    Celle, C.; Suspène, C.; Ternisien, M.; Lenfant, S.; Guérin, D.; Smaali, K.; Lmimouni, K.; Simonato, J. P.; Vuillaume, D. Interface dipole: Effects on threshold voltage and mobility for both amorphous and poly-crystalline organic field effect transistors. *Org. Electron.* **2014,** *15* (3), 729.

(13)    Onclin, S.; Ravoo, B. J.; Reinhoudt, D. N. Engineering Silicon Oxide Surfaces Using Self-Assembled Monolayers. *Angew. Chem. Int. Ed.* **2005,** *44* (39), 6282.

(14)    Pujari, S. P.; Scheres, L.; Marcelis, A.; Zuilhof, H. Covalent surface modification of oxide surfaces. *Angew. Chem. Int. Ed.* **2014,** *53* (25), 6322.

(15)    Paniagua, S. A.; Giordano, A. J.; Smith, O. N. L.; Barlow, S.; Li, H.; Armstrong, N. R.; Pemberton, J. E.; Brédas, J.-L.; Ginger, D.; Marder, S. R. Phosphonic Acids for Interfacial Engineering of Transparent Conductive Oxides. *Chem. Rev.* **2016,** *116* (12), 7117.

(16)    Li, Y.; Calder, S.; Yaffe, O.; Cahen, D.; Haick, H.; Kronik, L.; Zuilhof, H. Hybrids of organic molecules and flat, oxide-free silicon: high-density monolayers, electronic properties, and functionalization. *Langmuir* **2012,** *28* (26), 9920.

(17)    Fabre, B. Functionalization of Oxide-Free Silicon Surfaces with Redox-Active Assemblies. *Chem. Rev.* **2016,** *116* (8), 4808.

(18)    Hu, M.; Liu, F.; Buriak, J. M. Expanding the Repertoire of Molecular Linkages to Silicon: Si–S, Si–Se, and Si–Te Bonds. *ACS Appl. Mater. Interfaces* **2016,** *8* (17), 11091.

(19)    Loscutoff, P. W.; Bent, S. F. Reactivity of the Germanium Surface: Chemical Passivation and Functionalization. *Annu. Rev. Phys. Chem.* **2006,** *57* (1), 467.

(20)    Teplyakov, A. V.; Bent, S. F. Semiconductor surface functionalization for advances in electronics, energy conversion, and dynamic systems. *J. Vac. Sci. Technol., A* **2013,** *31* (5), 050810.

(21)    Haick, H.; Cahen, D. Making contact: Connecting molecules electrically to the macroscopic world. *Prog. Surf. Sci.* **2008,** *83* (4), 217.

(22)    Walker, A. V. Toward a new world of molecular devices: Making metallic contacts tomolecules. *J. Vac. Sci. Technol., A* **2013,** *31* (5), 050816.

(23)     Har-Lavan, R.; Cahen, D. 40 Years of Inversion Layer Solar Cells: From MOS to Conducting Polymer/Inorganic Hybrids. *IEEE J. Photovolt.* **2013,** *3* (4), 1443.

(24)     Tung, R. T. The physics and chemistry of the Schottky barrier height. *Appl. Phys. Rev.* **2014,** *1* (1), 011304.

(25)     Schottky, W. Zur Halbleitertheorie der Sperrschicht- und Spitzengleichrichter. *Z. Physik* **1939,** *113* (5-6), 367.

(26)     Mott, N. F. The Theory of Crystal Rectifiers. *Proc. R. Soc. London, Ser. A* **1939,** *171* (944), 27.

(27)     Mönch, W. *Semiconductor surfaces and interfaces*; Springer, 2001.

(28)     Zhang, Z.; Yates, J. T., Jr. Band bending in semiconductors: chemical and physical consequences at surfaces and interfaces. *Chem. Rev.* **2012,** *112* (10), 5520.

(29)     Trasatti, S. The "absolute" electrode potential—the end of the story. *Electrochim. Acta* **1990,** *35* (1), 269.

(30)     Often *e* is used as sub- or superscript to distinguish it from chemical potentials of other components of the system, but throughout this review we will omit this sub-or superscript.

(31)     In electrical potential we refer to that of an electron, which has a sign opposite to the conventional electrostatic potential, as the latter is defined for a positive test charge.

(32)     Gerischer, H.; Ekardt, W. Fermi levels in electrolytes and the absolute scale of redox potentials. *Appl. Phys. Lett.* **1983,** *43* (4), 393.

(33)     Bockris, J. O. M.; Khan, S. U. M. Comment on ''Fermi levels in electrolytes and the absolute scale of redox potentials''. *Appl. Phys. Lett.* **1984,** *45* (8), 913.

(34)     Reiss, H. The Fermi level and the redox potential. *J. Phys. Chem.* **1985,** *89* (18), 3783.

(35)     Riess, I. What does a voltmeter measure? *Solid State Ionics* **1997,** *95* (3), 327.

(36)     Cahen, D.; Kahn, A. Electron Energetics at Surfaces and Interfaces: Concepts and Experiments. *Adv. Mater.* **2003,** *15* (4), 271.

(37)     Marshak, A. H. Modeling semiconductor devices with position-dependent material parameters. *IEEE Trans. Elect. Dev.* **1989,** *36* (9), 1764.

(38)     Kronik, L.; Shapira, Y. Surface photovoltage phenomena: theory, experiment, and applications. *Surf. Sci. Rep.* **1999,** *37* (1), 1.

(39)     Brabec, C. J.; Cravino, A.; Meissner, D.; Sariciftci, N. S.; Fromherz, T.; Rispens, M. T.; Sanchez, L.; Hummelen, J. C. Origin of the Open Circuit Voltage of Plastic Solar Cells. *Adv. Funct. Mater.* **2001,** *11* (5), 374.

(40)     Braun, S.; Salaneck, W. R.; Fahlman, M. Energy-Level Alignment at Organic/Metal and Organic/Organic Interfaces. *Adv. Mater.* **2009,** *21* (14-15), 1450.

(41)     Sze, S. M.; Ng, K. K. *Physics of Semiconductor Devices*; third ed.; A JOHN WILEY & SONS, 2007.

(42)     Rhoderick, E. H. *Monographs in Electrical and Electronic Engineering. Metal-Semiconductor Contacts*; Clarendon press: Oxford, 1988.

(43)     Campbell, I. H.; Rubin, S.; Zawodzinski, T. A.; Kress, J. D.; Martin, R. L.; Smith, D. L.; Barashkov, N. N.; Ferraris, J. P. Controlling Schottky energy barriers in organic electronic devices using self-assembled monolayers. *Phys. Rev. B* **1996,** *54* (20), R14321.

(44)     Campbell, I. H.; Kress, J. D.; Martin, R. L.; Smith, D. L.; Barashkov, N. N.; Ferraris, J. P. Controlling charge injection in organic electronic devices using self-assembled monolayers. *Appl. Phys. Lett.* **1997,** *71* (24), 3528.

(45)     Ishii, H.; Sugiyama, K.; Ito, E.; Seki, K. Energy level alignment and interfacial electronic structures at organic/metal and organic/organic interfaces. *Adv. Mater.* **1999,** *11* (8), 605.

(46)     Vilan, A.; Shanzer, A.; Cahen, D. Molecular control over Au/GaAs diodes. *Nature* **2000,** *404* (6774), 166.

(47)     Tung, R. T. Chemical Bonding and Fermi Level Pinning at Metal-Semiconductor Interfaces. *Phys. Rev. Lett.* **2000,** *84* (26), 6078.

(48)     Tung, R. T. Recent advances in Schottky barrier concepts. *Mater. Sci. Eng., R* **2001,** *35* (1), 1.

(49)     Bardeen, J. Surface States and Rectification at a Metal Semi-Conductor Contact. *Phys. Rev.* **1947,** *71* (10), 717.

(50)     Cowley, A. M.; Sze, S. M. Surface States and Barrier Height of Metal-Semiconductor Systems. *J. Appl. Phys.* **1965,** *36* (10), 3212.

(51)     Card, H. C.; Rhoderick, E. H. Studies of tunnel MOS diodes I. Interface effects in silicon Schottky diodes. *J. Phys. D: Appl. Phys.* **1971,** *4* (10), 1589.

(52)     Brillson, L. The structure and properties of metal-semiconductor interfaces. *Surf. Sci. Rep.* **1982,** *2* (2), 123.

(53)     Kurtin, S.; McGill, T.; Mead, C. Fundamental transition in the electronic nature of solids. *Phys. Rev. Lett.* **1969,** *22* (26), 1433.

(54)     Vázquez, H.; Flores, F.; Kahn, A. Induced Density of States model for weakly-interacting organic semiconductor interfaces. *Org. Electron.* **2007,** *8* (2-3), 241.

(55)     Hwang, J.; Wan, A.; Kahn, A. Energetics of metal–organic interfaces: New experiments and assessment of the field. *Mater. Sci. Eng., R* **2009,** *64* (1-2), 1.

(56)     Vázquez, H.; Oszwaldowski, R.; Pou, P.; Ortega, J.; Pérez, R.; Flores, F.; Kahn, A. Dipole formation at metal/PTCDA interfaces: Role of the Charge Neutrality Level. *Europhys. Lett.* **2004,** *65* (6), 802.

(57)     Vázquez, H.; Gao, W.; Flores, F.; Kahn, A. Energy level alignment at organic heterojunctions: Role of the charge neutrality level. *Phys. Rev. B* **2005,** *71* (4).

(58)     Vazquez, H.; Dappe, Y. J.; Ortega, J.; Flores, F. Energy level alignment at metal/organic semiconductor interfaces: "pillow" effect, induced density of interface states, and charge neutrality level. *J. Chem. Phys.* **2007,** *126* (14), 144703.

(59)     Segev, L.; Salomon, A.; Natan, A.; Cahen, D.; Kronik, L.; Amy, F.; Chan, C. K.; Kahn, A. Electronic structure of Si(111)-bound alkyl monolayers: Theory and experiment. *Phys. Rev. B* **2006,** *74* (16), 165323.

(60)     Toledano, T.; Garrick, R.; Sinai, O.; Bendikov, T.; Haj-Yahia, A.-E.; Lerman, K.; Alon, H.; Sukenik, C. N.; Vilan, A.; Kronik, L. Effect of binding group on hybridization across the silicon/aromatic-monolayer interface. *J. Electron Spectrosc. Relat. Phenom.* **2015,** *204,* 149.

(61)     Hofmann, O. T.; Rinke, P.; Scheffler, M.; Heimel, G. Integer versus Fractional Charge Transfer at Metal(/Insulator)/Organic Interfaces: Cu(/NaCl)/TCNE. *ACS Nano* **2015,** *9* (5), 5391.

(62)     Ashkenasy, G.; Cahen, D.; Cohen, R.; Shanzer, A.; Vilan, A. Molecular engineering of semiconductor surfaces and devices. *Acc. Chem. Res.* **2002,** *35* (2), 121.

(63)     Vilan, A.; Cahen, D. How organic molecules can control electronic devices. *Trends Biotechnol.* **2002,** *20* (1), 22.

(64)     Long, W.; Li, Y.; Tung, R. T. Schottky barrier height systematics studied by partisan interlayer. *Thin Solid Films* **2014,** *557,* 254.

(65)     Waldrop, J. R.; Grant, R. W. Metal contacts to GaAs with 1 eV Schottky barrier height. *Appl. Phys. Lett.* **1988,** *52* (21), 1794.

(66)     Nishimura, T.; Kita, K.; Toriumi, A. A Significant Shift of Schottky Barrier Heights at Strongly Pinned Metal/Germanium Interface by Inserting an Ultra-Thin Insulating Film. *Appl. Phys. Express* **2008,** *1,* 051406.

(67)     Tersoff, J. Schottky barrier heights and the continuum of gap states. *Phys. Rev. Lett.* **1984,** *52* (6), 465.

(68)    Murphy, C. J.; Lisensky, G. C.; Leung, L. K.; Kowach, G. R.; Ellis, A. B. Photoluminescence-based correlation of semiconductor electric field thickness with adsorbate Hammett substituent constants. Adsorption of aniline derivatives onto cadmium selenide. *J. Am. Chem. Soc.* **1990,** *112* (23), 8344.

(69)    Lunt, S. R.; Ryba, G. N.; Santangelo, P. G.; Lewis, N. S. Chemical studies of the passivation of GaAs surface recombination using sulfides and thiols. *J. Appl. Phys.* **1991,** *70* (12), 7449.

(70)    Cohen, R.; Bastide, S.; Cahen, D.; Libman, J.; Shanzer, A.; Rosenwaks, Y. Controlling electronic properties of CdTe by adsorption of dicarboxylic acid derivatives: Relating molecular parameters to band bending and electron affinity changes. *Adv. Mater.* **1997,** *9* (9), 746.

(71)    Royea, W. J.; Juang, A.; Lewis, N. S. Preparation of air-stable, low recombination velocity Si(111) surfaces through alkyl termination. *Appl. Phys. Lett.* **2000,** *77* (13), 1988.

(72)    Sieval, A. B.; Huisman, C. L.; Schönecker, A.; Schuurmans, F. M.; van der Heide, A. S.; Goossens, A.; Sinke, W. C.; Zuilhof, H.; Sudhölter, E. J. Silicon surface passivation by organic monolayers: minority charge carrier lifetime measurements and Kelvin probe investigations. *J. Phys. Chem. B* **2003,** *107* (28), 6846.

(73)    Yaffe, O.; Ely, T.; Har-Lavan, R.; Egger, D. A.; Johnston, S.; Cohen, H.; Kronik, L.; Vilan, A.; Cahen, D. Effect of Molecule-Surface Reaction Mechanism on the Electronic Characteristics and Photovoltaic Performance of Molecularly Modified Si. *J. Phys. Chem. C* **2013,** *117* (43), 22351.

(74)    Har-Lavan, R.; Yaffe, O.; Joshi, P.; Kazaz, R.; Cohen, H.; Cahen, D. Ambient organic molecular passivation of Si yields near-ideal, Schottky-Mott limited, junctions. *AIP Adv.* **2012,** *2* (1), 012164.

(75)    Seitz, O.; Bocking, T.; Salomon, A.; Gooding, J. J.; Cahen, D. Importance of Monolayer Quality for Interpreting Current Transport through Organic Molecules: Alkyls on Oxide-Free Si. *Langmuir* **2006,** *22* (16), 6915.

(76)    Wong, K. T.; Lewis, N. S. What a difference a bond makes: the structural, chemical, and physical properties of methyl-terminated Si(111) surfaces. *Acc. Chem. Res.* **2014,** *47* (10), 3037.

(77)    Bhairamadgi, N. S.; Pujari, S. P.; Trovela, F. G.; Debrassi, A.; Khamis, A. A.; Alonso, J. M.; Al Zahrani, A. A.; Wennekes, T.; Al-Turaif, H. A.; van Rijn, C.et al. Hydrolytic and Thermal Stability of Organic Monolayers on Various Inorganic Substrates. *Langmuir* **2014,** *30* (20), 5829.

(78)    Nguyen Minh, Q.; Pujari, S. P.; Wang, B.; Wang, Z.; Haick, H.; Zuilhof, H.; van Rijn, C. J. M. Fluorinated Alkyne-derived Monolayers on Oxide-Free Silicon Nanowires via One-step Hydrosilylation. *Appl. Surf. Sci.* **2016,** *387*, 1202.

(79)    Cohen, R.; Kronik, L.; Shanzer, A.; Cahen, D.; Liu, A.; Rosenwaks, Y.; Lorenz, J. K.; Ellis, A. B. Molecular Control over Semiconductor Surface Electronic Properties:  Dicarboxylic Acids on CdTe, CdSe, GaAs, and InP. *J. Am. Chem. Soc.* **1999,** *121* (45), 10545.

(80)    Cohen, R.; Kronik, L.; Vilan, A.; Shanzer, A.; Cahen, D. Frontier orbital model of semiconductor surface passivation: Dicarboxylic acids on n- and p-GaAs. *Adv. Mater.* **2000,** *12* (1), 33.

(81)    Hunger, R.; Fritsche, R.; Jaeckel, B.; Jaegermann, W.; Webb, L. J.; Lewis, N. S. Chemical and electronic characterization of methyl-terminated Si(111) surfaces by high-resolution synchrotron photoelectron spectroscopy. *Phys. Rev. B* **2005,** *72* (4), 045317.

(82)    Hunger, R.; Jaegermann, W.; Merson, A.; Shapira, Y.; Pettenkofer, C.; Rappich, J. Electronic Structure of Methoxy-, Bromo-, and Nitrobenzene Grafted onto Si(111). *J. Phys. Chem. B* **2006,** *110* (31), 15432.

(83)     Yaffe, O.; Pujari, S.; Sinai, O.; Vilan, A.; Zuilhof, H.; Kahn, A.; Kronik, L.; Cohen, H.; Cahen, D. Effect of Doping Density on the Charge Rearrangement and Interface Dipole at the Molecule-Silicon Interface. *J. Phys. Chem. C* **2013,** *117* (43), 22422.

(84)     Sinton, R. A.; Cuevas, A. Contactless determination of current–voltage characteristics and minority-carrier lifetimes in semiconductors from quasi-steady-state photoconductance data. *Appl. Phys. Lett.* **1996,** *69* (17), 2510.

(85)     Har-Lavan, R.; Schreiber, R.; Yaffe, O.; Cahen, D. Molecular field effect passivation: Quinhydrone/methanol treatment of n-Si(100). *J. Appl. Phys.* **2013,** *113* (8), 084909.

(86)     Godet, C.; Fadjie-Djomkam, A.-B.; Ababou-Girard, S. Effect of high current density on the admittance response of interface states in ultrathin MIS tunnel junctions. *Solid-State Electron.* **2013,** *80* (0), 142.

(87)     Bora, A.; Pathak, A.; Liao, K.-C.; Vexler, M. I.; Kuligk, A.; Cattani-Scholz, A.; Meinerzhagen, B.; Abstreiter, G.; Schwartz, J.; Tornow, M. Organophosphonates as model system for studying electronic transport through monolayers on SiO2/Si surfaces. *Appl. Phys. Lett.* **2013,** *102* (24), 241602.

(88)     Barnes, P. R. F.; Miettunen, K.; Li, X.; Anderson, A. Y.; Bessho, T.; Gratzel, M.; O'Regan, B. C. Interpretation of Optoelectronic Transient and Charge Extraction Measurements in Dye-Sensitized Solar Cells. *Adv. Mater.* **2013,** *25* (13), 1881.

(89)     Moons, E.; Bruening, M.; Shanzer, A.; Beier, J.; Cahen, D. Electron transfer in hybrid molecular solid-state devices. *Synth. Met.* **1996,** *76* (1), 245.

(90)     Vuillaume, D.; Chen, B.; Metzger, R. M. Electron transfer through a monolayer of hexadecylquinolinium tricyanoquinodimethanide. *Langmuir* **1999,** *15* (11), 4011.

(91)     Puniredd, S. R.; Jayaraman, S.; Yeong, S. H.; Troadec, C.; Srinivasan, M. P. Stable Organic Monolayers on Oxide-Free Silicon/Germanium in a Supercritical Medium: A New Route to Molecular Electronics. *J. Phys. Chem. Lett.* **2013,** *4* (9), 1397.

(92)     Aqua, T.; Cohen, H.; Vilan, A.; Naaman, R. Long-range substrate effects on the stability and reactivity of thiolated self-assembled monolayers. *J. Phys. Chem. C* **2007,** *111* (44), 16313.

(93)     Toledano, T.; Biller, A.; Bendikov, T.; Cohen, H.; Vilan, A.; Cahen, D. Controlling Space Charge of Oxide-Free Si by in Situ Modification of Dipolar Alkyl Monolayers. *J. Phys. Chem. C* **2012,** *116* (21), 11434.

(94)     Brown, E. S.; Hlynchuk, S.; Maldonado, S. Chemically modified Si(111) surfaces simultaneously demonstrating hydrophilicity, resistance against oxidation, and low trap state densities. *Surf. Sci.* **2016,** *645*, 49.

(95)     Peng, W.; Rupich, S. M.; Shafiq, N.; Gartstein, Y. N.; Malko, A. V.; Chabal, Y. J. Silicon Surface Modification and Characterization for Emergent Photovoltaic Applications Based on Energy Transfer. *Chem. Rev.* **2015,** *115* (23), 12764.

(96)     Hidetaka, T.; Isao, S.; Ryuichi, S. Quinhydrone/Methanol Treatment for the Measurement of Carrier Lifetime in Silicon Substrates. *Jpn. J. Appl. Phys.* **2002,** *41* (8A), L870.

(97)     Gartsman, K.; Cahen, D.; Kadyshevitch, A.; Libman, J.; Moav, T.; Naaman, R.; Shanzer, A.; Umansky, V.; Vilan, A. Molecular control of a GaAs transistor. *Chem. Phys. Lett.* **1998,** *283* (5-6), 301.

(98)     Ristein, J. Surface Transfer Doping of Semiconductors. *Science* **2006,** *313* (5790), 1057.

(99)     Chen, W.; Qi, D.; Gao, X.; Wee, A. T. S. Surface transfer doping of semiconductors. *Prog. Surf. Sci.* **2009,** *84* (9–10), 279.

(100)   Vilan, A.; Ussyshkin, R.; Gartsman, K.; Cahen, D.; Naaman, R.; Shanzer, A. Real-time electronic monitoring of adsorption kinetics: Evidence for two-site adsorption mechanism of dicarboxylic acids on GaAs(100). *J. Phys. Chem. B* **1998,** *102* (18), 3307.

(101)   Wu, D. G.; Cahen, D.; Graf, P.; Naaman, R.; Nitzan, A.; Shvarts, D. Direct Detection of Low-Concentration NO in Physiological Solutions by a New GaAs-Based Sensor. *Chem. Eur. J.* **2001,** *7* (8), 1743.

(102)  Shvarts, D.; Haran, A.; Benshafrut, R.; Cahen, D.; Naaman, R. Monitoring electron redistribution in molecules during adsorption. *Chem. Phys. Lett.* **2002,** *354* (3–4), 349.

(103)  Stern, E.; Klemic, J. F.; Routenberg, D. A.; Wyrembak, P. N.; Turner-Evans, D. B.; Hamilton, A. D.; LaVan, D. A.; Fahmy, T. M.; Reed, M. A. Label-free immunodetection with CMOS-compatible semiconducting nanowires. *Nature* **2007,** *445* (7127), 519.

(104)  Zheng, G.; Patolsky, F.; Cui, Y.; Wang, W. U.; Lieber, C. M. Multiplexed electrical detection of cancer markers with nanowire sensor arrays. *Nat. Biotechnol.* **2005,** *23* (10), 1294.

(105)  Paska, Y.; Stelzner, T.; Assad, O.; Tisch, U.; Christiansen, S.; Haick, H. Molecular Gating of Silicon Nanowire Field-Effect Transistors with Nonpolar Analytes. *Acs Nano* **2011,** *6* (1), 335.

(106)  Lichtenstein, A.; Havivi, E.; Shacham, R.; Hahamy, E.; Leibovich, R.; Pevzner, A.; Krivitsky, V.; Davivi, G.; Presman, I.; Elnathan, R.et al. Supersensitive fingerprinting of explosives by chemically modified nanosensors arrays. *Nat. Commun.* **2014,** *5*.

(107)  Cheng, X.; Lowe, S. B.; Reece, P. J.; Gooding, J. J. Colloidal silicon quantum dots: from preparation to the modification of self-assembled monolayers (SAMs) for bio-applications. *Chem. Sci. Rev.* **2014,** *43* (8), 2680.

(108)  Buhbut, S.; Itzhakov, S.; Tauber, E.; Shalom, M.; Hod, I.; Geiger, T.; Garini, Y.; Oron, D.; Zaban, A. Built-in Quantum Dot Antennas in Dye-Sensitized Solar Cells. *ACS Nano* **2010,** *4* (3), 1293.

(109)  Weiss, E. A. Organic Molecules as Tools To Control the Growth, Surface Structure, and Redox Activity of Colloidal Quantum Dots. *Acc. Chem. Res.* **2013,** *46* (11), 2607.

(110)  Hines, D. A.; Kamat, P. V. Recent Advances in Quantum Dot Surface Chemistry. *ACS Appl. Mater. Interfaces* **2014,** *6* (5), 3041.

(111)  Roelofs, K. E.; Brennan, T. P.; Bent, S. F. Interface Engineering in Inorganic-Absorber Nanostructured Solar Cells. *J. Phys. Chem. Lett.* **2014,** *5* (2), 348.

(112)  Cahen, D.; Noufi, R. Defect chemical explanation for the effect of air anneal on CdS/CuInSe2 solar cell performance. *Appl. Phys. Lett.* **1989,** *54* (6), 558.

(113)  Cahen, D.; Noufi, R. Surface passivation of polycrystalline, chalcogenide based photovoltaic cells. *Solar Cells* **1991,** *30* (1), 53.

(114)  Nayak, P. K.; Rosenberg, R.; Barnea-Nehoshtan, L.; Cahen, D. O2 and organic semiconductors: Electronic effects. *Org. Electron.* **2013,** *14* (3), 966.

(115)  Zhang, Y.; Zherebetskyy, D.; Bronstein, N. D.; Barja, S.; Lichtenstein, L.; Alivisatos, A. P.; Wang, L.-W.; Salmeron, M. Molecular Oxygen Induced in-Gap States in PbS Quantum Dots. *ACS Nano* **2015,** *9* (10), 10445.

(116)  Zohar, A. M.Sc. thesis, Weizmann Institute of Science, 2014.

(117)  Kedem, N.; Blumstengel, S.; Henneberger, F.; Cohen, H.; Hodes, G.; Cahen, D. Morphology-, synthesis- and doping-independent tuning of ZnO work function using phenylphosphonates. *Phys. Chem. Chem. Phys.* **2014,** *16* (18), 8310.

(118)  Lee, Y.-J.; Wang, J.; Cheng, S. R.; Hsu, J. W. P. Solution Processed ZnO Hybrid Nanocomposite with Tailored Work Function for Improved Electron Transport Layer in Organic Photovoltaic Devices. *ACS Appl. Mater. Interfaces* **2013,** *5* (18), 9128.

(119)  Haick, H.; Ambrico, M.; Ligonzo, T.; Tung, R. T.; Cahen, D. Controlling Semiconductor/Metal Junction Barriers by Incomplete, Nonideal Molecular Monolayers. *J. Am. Chem. Soc.* **2006,** *128* (21), 6854.

(120)  Nouchi, R.; Shigeno, M.; Yamada, N.; Nishino, T.; Tanigaki, K.; Yamaguchi, M. Reversible switching of charge injection barriers at metal/organic-semiconductor contacts modified with structurally disordered molecular monolayers. *Appl. Phys. Lett.* **2014,** *104* (1), 013308.

(121)  Nouchi, R.; Tanimoto, T. Substituent-Controlled Reversible Switching of Charge Injection Barrier Heights at Metal/Organic Semiconductor Contacts Modified with Disordered Molecular Monolayers. *ACS Nano* **2015,** *9* (7), 7429.

(122)  Van Dyck, C.; Geskin, V.; Kronemeijer, A. J.; de Leeuw, D. M.; Cornil, J. Impact of derivatization on electron transmission through dithienylethene-based photoswitches in molecular junctions. *Phys. Chem. Chem. Phys.* **2013,** *15* (12), 4392.

(123)  Natan, A.; Kronik, L.; Haick, H.; Tung, R. T. Electrostatic Properties of Ideal and Non-ideal Polar Organic Monolayers: Implications for Electronic Devices. *Adv. Mater.* **2007,** *19* (23), 4103.

(124)  Heimel, G.; Rissner, F.; Zojer, E. Modeling the Electronic Properties of π-Conjugated Self-Assembled Monolayers. *Adv. Mater.* **2010,** *22* (23), 2494.

(125)  Levine, I.; Weber, S. M.; Feldman, Y.; Bendikov, T.; Cohen, H.; Cahen, D.; Vilan, A. Molecular Length, Monolayer Density, and Charge Transport: Lessons from Al-AlOx/Alkyl-Phosphonate/Hg Junctions. *Langmuir* **2012,** *28* (1), 404.

(126)  Shpaisman, H.; Seitz, O.; Yaffe, O.; Roodenko, K.; Scheres, L.; Zuilhof, H.; Chabal, Y. J.; Sueyoshi, T.; Kera, S.; Ueno, N.et al. Structure Matters: Correlating temperature dependent electrical transport through alkyl monolayers with vibrational and photoelectron spectroscopies. *Chem. Sci.* **2012,** *3* (3), 851.

(127)  Ballav, N.; Schüpbach, B.; Dethloff, O.; Feulner, P.; Terfort, A.; Zharnikov, M. Direct Probing Molecular Twist and Tilt in Aromatic Self-Assembled Monolayers. *J. Am. Chem. Soc.* **2007,** *129* (50), 15416.

(128)  Acton, O.; Dubey, M.; Weidner, T.; O'Malley, K. M.; Kim, T.-W.; Ting, G. G.; Hutchins, D.; Baio, J. E.; Lovejoy, T. C.; Gage, A. H.et al. Simultaneous Modification of Bottom-Contact Electrode and Dielectric Surfaces for Organic Thin-Film Transistors Through Single-Component Spin-Cast Monolayers. *Adv. Funct. Mater.* **2011,** *21* (8), 1476.

(129)  She, Z.; Lahaye, D.; Champness, N. R.; Bühl, M.; Hamoudi, H.; Zharnikov, M.; Buck, M. Accommodation of Lattice Mismatch in a Thiol Self-Assembled Monolayer. *J. Phys. Chem. C* **2013,** *117* (9), 4647.

(130)  Gliboff, M.; Sang, L.; Knesting, K. M.; Schalnat, M. C.; Mudalige, A.; Ratcliff, E. L.; Li, H.; Sigdel, A. K.; Giordano, A. J.; Berry, J. J.et al. Orientation of Phenylphosphonic Acid Self-Assembled Monolayers on a Transparent Conductive Oxide: A Combined NEXAFS, PM-IRRAS, and DFT Study. *Langmuir* **2013,** *29* (7), 2166.

(131)  Lee, H. J.; Jamison, A. C.; Lee, T. R. Surface Dipoles: A Growing Body of Evidence Supports Their Impact and Importance. *Acc. Chem. Res.* **2015,** *48* (12), 3007.

(132)  Natan, A.; Zidon, Y.; Shapira, Y.; Kronik, L. Cooperative effects and dipole formation at semiconductor and self-assembled-monolayer interfaces. *Phys. Rev. B* **2006,** *73* (19), 193310.

(133)  Gershevitz, O.; Sukenik, C. N.; Ghabboun, J.; Cahen, D. Molecular Monolayer-Mediated Control over Semiconductor Surfaces:  Evidence for Molecular Depolarization of Silane Monolayers on Si/SiOx. *J. Am. Chem. Soc.* **2003,** *125* (16), 4730.

(134)  Cornil, D.; Olivier, Y.; Geskin, V.; Cornil, J. Depolarization Effects in Self-Assembled Monolayers: A Quantum-Chemical Insight. *Adv. Funct. Mater.* **2007,** *17* (7), 1143.

(135)  Monti, O. L. A. Understanding Interfacial Electronic Structure and Charge Transfer: An Electrostatic Perspective. *J. Phys. Chem. Lett.* **2012,** *3* (17), 2342.

(136)  Lenfant, S.; Guerin, D.; Tran Van, F.; Chevrot, C.; Palacin, S.; Bourgoin, J. P.; Bouloussa, O.; Rondelez, F.; Vuillaume, D. Electron Transport through Rectifying Self-Assembled Monolayer Diodes on Silicon:  Fermi-Level Pinning at the Molecule–Metal Interface. *J. Phys. Chem. B* **2006,** *110* (28), 13947.

(137) Gozlan, N.; Tisch, U.; Haick, H. Tailoring the Work Function of Gold Surface by Controlling Coverage and Disorder of Polar Molecular Monolayers. *J. Phys. Chem. C* **2008,** *112* (33), 12988.

(138) Gozlan, N.; Haick, H. Coverage Effect of Self-Assembled Polar Molecules on the Surface Energetics of Silicon. *J. Phys. Chem. C* **2008,** *112* (33), 12599.

(139) Topham, B. J.; Kumar, M.; Soos, Z. G. Profiles of Work Function Shifts and Collective Charge Transfer in Submonolayer Metal-Organic Films. *Adv. Funct. Mater.* **2011,** *21* (10), 1931.

(140) Lacher, S.; Matsuo, Y.; Nakamura, E. Molecular and Supramolecular Control of the Work Function of an Inorganic Electrode with Self-Assembled Monolayer of Umbrella-Shaped Fullerene Derivatives. *J. Am. Chem. Soc.* **2011,** *133* (42), 16997.

(141) Verwüster, E.; Hofmann, O. T.; Egger, D. A.; Zojer, E. Electronic Properties of Biphenylthiolates on Au(111): The Impact of Coverage Revisited. *J. Phys. Chem. C* **2015,** *119* (14), 7817.

(142) Duhm, S.; Heimel, G.; Salzmann, I.; Glowatzki, H.; Johnson, R. L.; Vollmer, A.; Rabe, J. P.; Koch, N. Orientation-dependent ionization energies and interface dipoles in ordered molecular assemblies. *Nat. Mater.* **2008,** *7* (4), 326.

(143) Toledano, T.; Sazan, H.; Mukhopadhyay, S.; Alon, H.; Lerman, K.; Bendikov, T.; Major, D. T.; Sukenik, C. N.; Vilan, A.; Cahen, D. Odd-Even Effect in Molecular Electronic Transport via an Aromatic Ring. *Langmuir* **2014,** *30* (45), 13596.

(144) Eckshtain-Levi, M.; Capua, E.; Refaely-Abramson, S.; Sarkar, S.; Gavrilov, Y.; Mathew, S. P.; Paltiel, Y.; Levy, Y.; Kronik, L.; Naaman, R. Cold denaturation induces inversion of dipole and spin transfer in chiral peptide monolayers. *Nat. Commun.* **2016,** *7*, 10744.

(145) Alloway, D. M.; Hofmann, M.; Smith, D. L.; Gruhn, N. E.; Graham, A. L.; Colorado, R.; Wysocki, V. H.; Lee, T. R.; Lee, P. A.; Armstrong, N. R. Interface Dipoles Arising from Self-Assembled Monolayers on Gold: UV-Photoemission Studies of Alkanethiols and Partially Fluorinated Alkanethiols. *J. Phys. Chem. B* **2003,** *107* (42), 11690.

(146) Faber, E. J.; de Smet, L. C. P. M.; Olthuis, W.; Zuilhof, H.; Sudhölter, E. J. R.; Bergveld, P.; van den Berg, A. Si-C Linked Organic Monolayers on Crystalline Silicon Surfaces as Alternative Gate Insulators. *ChemPhysChem* **2005,** *6* (10), 2153.

(147) Maldonado, S.; Plass, K. E.; Knapp, D.; Lewis, N. S. Electrical Properties of Junctions between Hg and Si(111) Surfaces Functionalized with Short-Chain Alkyls. *J. Phys. Chem. C* **2007,** *111* (48), 17690.

(148) Yaffe, O.; Scheres, L.; Puniredd, S. R.; Stein, N.; Biller, A.; Har-Lavan, R.; Shpaisman, H.; Zuilhof, H.; Haick, H.; Cahen, D.et al. Molecular Electronics at Metal/Semiconductor Junctions. Si Inversion by Sub-Nanometer Molecular Films. *Nano Lett.* **2009,** *9* (6), 2390.

(149) Ichii, T.; Fukuma, T.; Kobayashi, K.; Yamada, H.; Matsushige, K. Surface potential measurements of phase-separated alkanethiol self-assembled monolayers by non-contact atomic force microscopy. *Nanotechnology* **2004,** *15* (2), S30.

(150) Magid, I.; Burstein, L.; Seitz, O.; Segev, L.; Kronik, L.; Rosenwaks, Y. Electronic Characterization of Si(100)-Bound Alkyl Monolayers Using Kelvin Probe Force Microscopy. *J. Phys. Chem. C* **2008,** *112* (18), 7145.

(151) Alloway, D. M.; Graham, A. L.; Yang, X.; Mudalige, A.; Colorado, R.; Wysocki, V. H.; Pemberton, J. E.; Randall Lee, T.; Wysocki, R. J.; Armstrong, N. R. Tuning the Effective Work Function of Gold and Silver Using ω-Functionalized Alkanethiols: Varying Surface Composition through Dilution and Choice of Terminal Groups. *J. Phys. Chem. C* **2009,** *113* (47), 20328.

(152) Heimel, G.; Romaner, L.; Zojer, E.; Bredas, J.-L. The Interface Energetics of Self-Assembled Monolayers on Metals. *Acc. Chem. Res.* **2008,** *41* (6), 721.

(153)  Moons, E.; Bruening, M.; Burstein, L.; Libman, J.; Shanzer, A.; Cahen, D. Molecular Approach to Surface Control of Chalcogenide Semiconductors. *Jpn. J. Appl. Phys.* **1993,** *32* (S3), 730.

(154)  Bruening, M.; Moons, E.; Yaron-Marcovich, D.; Cahen, D.; Libman, J.; Shanzer, A. Polar Ligand Adsorption Controls Semiconductor Surface Potentials. *J. Am. Chem. Soc.* **1994,** *116* (7), 2972.

(155)  Hansch, C.; Leo, A.; Taft, R. W. A survey of Hammett substituent constants and resonance and field parameters. *Chem. Rev.* **1991,** *91* (2), 165.

(156)  Bastide, S.; Butruille, R.; Cahen, D.; Dutta, A.; Libman, J.; Shanzer, A.; Sun, L. M.; Vilan, A. Controlling the work function of GaAs by chemisorption of benzoic acid derivatives. *J. Phys. Chem. B* **1997,** *101* (14), 2678.

(157)  Nüesch, F.; Rotzinger, F.; Si-Ahmed, L.; Zuppiroli, L. Chemical potential shifts at organic device electrodes induced by grafted monolayers. *Chem. Phys. Lett.* **1998,** *288* (5–6), 861.

(158)  Krüger, J.; Bach, U.; Grätzel, M. Controlling electronic properties of $TiO_2$ by adsorption of carboxylic acid derivatives. *Adv. Mater.* **2000,** *12*, 447.

(159)  Wu, D. G.; Ghabboun, J.; Martin, J. M. L.; Cahen, D. Tuning of Au/n-GaAs Diodes with Highly Conjugated Molecules. *J. Phys. Chem. B* **2001,** *105* (48), 12011.

(160)  Yip, H.-L.; Hau, S. K.; Baek, N. S.; Ma, H.; Jen, A. K. Y. Polymer Solar Cells That Use Self-Assembled-Monolayer- Modified ZnO/Metals as Cathodes. *Adv. Mater.* **2008,** *20* (12), 2376.

(161)  Vilan, A.; Ghabboun, J.; Cahen, D. Molecule-metal polarization at rectifying GaAs interfaces. *J. Phys. Chem. B* **2003,** *107* (26), 6360.

(162)  Selzer, Y.; Cahen, D. Fine tuning of Au/SiO₂/Si diodes by varying interfacial dipoles using molecular monolayers. *Adv. Mater.* **2001,** *13*, 508.

(163)  Haj-Yahia, A.; Yaffe, O.; Bendikov, T.; Cohen, H.; Feldman, Y.; Vilan, A.; Cahen, D. Substituent Variation Drives Metal/Monolayer/Semiconductor Junctions from Strongly Rectifying to Ohmic Behavior. *Adv. Mater.* **2013,** *25* (5), 702.

(164)  Kretz, B.; Egger, D. A.; Zojer, E. A Toolbox for Controlling the Energetics and Localization of Electronic States in Self-Assembled Organic Monolayers. *Adv. Sci.* **2015,** *2* (3), n/a.

(165)  Abu-Husein, T.; Schuster, S.; Egger, D. A.; Kind, M.; Santowski, T.; Wiesner, A.; Chiechi, R.; Zojer, E.; Terfort, A.; Zharnikov, M. The Effects of Embedded Dipoles in Aromatic Self-Assembled Monolayers. *Adv. Funct. Mater.* **2015,** *25* (25), 3943.

(166)  Barnea-Nehoshtan, L.; Nayak, P. K.; Shu, A.; Bendikov, T.; Kahn, A.; Cahen, D. Enhancing the tunability of the open-circuit voltage of hybrid photovoltaics with mixed molecular monolayers. *ACS Appl. Mater. Interfaces* **2014,** *6* (4), 2317.

(167)  Pujari, S. P.; van Andel, E.; Yaffe, O.; Cahen, D.; Weidner, T.; van Rijn, C. J. M.; Zuilhof, H. Mono-Fluorinated Alkyne-Derived SAMs on Oxide-Free Si(111) Surfaces: Preparation, Characterization and Tuning of the Si Workfunction. *Langmuir* **2013,** *29* (2), 570.

(168)  Cohen, Y. S.; Vilan, A.; Ron, I.; Cahen, D. Hydrolysis Improves Packing Density of Bromine-Terminated Alkyl-Chain, Silicon-Carbon Monolayers Linked to Silicon. *J. Phys. Chem. C* **2009,** *113* (15), 6174.

(169)  Salomon, A.; Berkovich, D.; Cahen, D. Molecular modification of an ionic semiconductor–metal interface: ZnO/molecule/Au diodes. *Appl. Phys. Lett.* **2003,** *82* (7), 1051.

(170)  Haick, H.; Pelz, J. P.; Ligonzo, T.; Ambrico, M.; Cahen, D.; Cai, W.; Marginean, C.; Tivarus, C.; Tung, R. T. Controlling Au/n-GaAs junctions by partial molecular monolayers. *Phys. Status Solidi A* **2006,** *203* (14), 3438.

(171)  Crispin, X.; Geskin, V.; Crispin, A.; Cornil, J.; Lazzaroni, R.; Salaneck, W. R.; Brédas, J.-L. Characterization of the Interface Dipole at Organic/ Metal Interfaces. *J. Am. Chem. Soc.* **2002,** *124* (27), 8131.

(172) Witte, G.; Lukas, S.; Bagus, P. S.; Wöll, C. Vacuum level alignment at organic/metal junctions: "Cushion" effect and the interface dipole. *Appl. Phys. Lett.* **2005**, *87* (26), 263502.

(173) Parr, R. G.; Pearson, R. G. Absolute hardness: companion parameter to absolute electronegativity. *J. Am. Chem. Soc.* **1983**, *105* (26), 7512.

(174) Pearson, R. G. Absolute electronegativity and hardness correlated with molecular orbital theory. *Proc. Natl. Acad. Sci. U.S.A.* **1986**, *83* (22), 8440.

(175) Ley, L.; Smets, Y.; Pakes, C. I.; Ristein, J. Calculating the Universal Energy-Level Alignment of Organic Molecules on Metal Oxides. *Adv. Funct. Mater.* **2013,** *23* (7), 794.

(176) Vázquez, H.; Flores, F.; Oszwaldowski, R.; Ortega, J.; Pérez, R.; Kahn, A. Barrier formation at metal–organic interfaces: dipole formation and the charge neutrality level. *Appl. Surf. Sci.* **2004**, *234* (1–4), 107.

(177) Van Dyck, C.; Ratner, M. A. Molecular Rectifiers: A New Design Based on Asymmetric Anchoring Moieties. *Nano Lett.* **2015,** *15* (3), 1577.

(178) Hill, I. G.; Schwartz, J.; Kahn, A. Metal-dependent charge transfer and chemical interaction at interfaces between 3,4,9,10-perylenetetracarboxylic bisimidazole and gold, silver and magnesium. *Org. Electron.* **2000,** *1* (1), 5.

(179) Greiner, M. T.; Helander, M. G.; Tang, W.-M.; Wang, Z.-B.; Qiu, J.; Lu, Z.-H. Universal energy-level alignment of molecules on metal oxides. *Nat. Mater.* **2012,** *11* (1), 76.

(180) Yaffe, O.; Scheres, L.; Segev, L.; Biller, A.; Ron, I.; Salomon, E.; Giesbers, M.; Kahn, A.; Kronik, L.; Zuilhof, H.et al. Hg/Molecular Monolayer-Si Junctions: Electrical Interplay between Monolayer Properties and Semiconductor Doping Density. *J. Phys. Chem. C* **2010,** *114* (22), 10270.

(181) Zhou, Y.; Fuentes-Hernandez, C.; Shim, J.; Meyer, J.; Giordano, A. J.; Li, H.; Winget, P.; Papadopoulos, T.; Cheun, H.; Kim, J.et al. A Universal Method to Produce Low–Work Function Electrodes for Organic Electronics. *Science* **2012,** *336* (6079), 327.

(182) Hinckley, A. C.; Wang, C.; Pfattner, R.; Kong, J.; Zhou, Y.; Ecker, B.; Gao, Y.; Bao, Z. Investigation of a Solution-Processable, Nonspecific Surface Modifier for Low Cost, High Work Function Electrodes. *ACS Appl. Mater. Interfaces* **2016**, *8* (30), 19658.

(183) Haick, H.; Ghabboun, J.; Niitsoo, O.; Cohen, H.; Cahen, D.; Vilan, A.; Hwang, J. Y.; Wan, A.; Amy, F.; Kahn, A. Effect of molecular binding to a semiconductor on metal/molecule/semiconductor junction behavior. *J. Phys. Chem. B* **2005**, *109* (19), 9622.

(184) Boudinet, D.; Benwadih, M.; Qi, Y.; Altazin, S.; Verilhac, J.-M.; Kroger, M.; Serbutoviez, C.; Gwoziecki, R.; Coppard, R.; Le Blevennec, G.et al. Modification of gold source and drain electrodes by self-assembled monolayer in staggered n- and p-channel organic thin film transistors. *Org. Electron.* **2010,** *11* (2), 227.

(185) Wang, J.; Friedman, C. R.; Cabrera, W.; Tan, K.; Lee, Y.-J.; Chabal, Y. J.; Hsu, J. W. P. Effect of metal/bulk-heterojunction interfacial properties on organic photovoltaic device performance. *J. Mater. Chem. A* **2014,** *2* (37), 15288.

(186) Yu, X.; Lovrincic, R.; Kraynis, O.; Man, G.; Ely, T.; Zohar, A.; Toledano, T.; Cahen, D.; Vilan, A. Fabrication of Reproducible, Integration-Compatible Hybrid Molecular/Si Electronics. *Small* **2014,** *10* (24), 5151.

(187) Vilan, A.; Cahen, D. Soft contact deposition onto molecularly modified GaAs. Thin metal film flotation: Principles and electrical effects. *Adv. Funct. Mater.* **2002,** *12* (11-12), 795.

(188) Stein, N. M.Sc. thesis, Weizmann Institute of Science, 2010.

(189) Stein, N.; Korobko, R.; Yaffe, O.; Lavan, R. H.; Shpaisman, H.; Tirosh, E.; Vilan, A.; Cahen, D. Nondestructive Contact Deposition for Molecular Electronics: Si-Alkyl//Au Junctions. *J. Phys. Chem. C* **2010,** *114* (29), 12769.

(190) Yelin, T.; Korytar, R.; Sukenik, N.; Vardimon, R.; Kumar, B.; Nuckolls, C.; Evers, F.; Tal, O. Conductance saturation in a series of highly transmitting molecular junctions. *Nat. Mater.* **2016,** *15* (4), 444.

(191) Díez-Pérez, I.; Hihath, J.; Lee, Y.; Yu, L.; Adamska, L.; Kozhushner, M. A.; Oleynik, I. I.; Tao, N. Rectification and stability of a single molecular diode with controlled orientation. *Nat. Chem.* **2009,** *1* (8), 635.

(192) de Boer, B.; Hadipour, A.; Mandoc, M. M.; van Woudenbergh, T.; Blom, P. W. M. Tuning of Metal Work Functions with Self-Assembled Monolayers. *Adv. Mater.* **2005,** *17* (5), 621.

(193) Wang, L.; Rangger, G. M.; Ma, Z.; Li, Q.; Shuai, Z.; Zojer, E.; Heimel, G. Is there a Au-S bond dipole in self-assembled monolayers on gold? *Phys. Chem. Chem. Phys.* **2010,** *12* (17), 4287.

(194) Gleason-Rohrer, D. C.; Brunschwig, B. S.; Lewis, N. S. Measurement of the band bending and surface dipole at chemically functionalized Si (111)/vacuum interfaces. *J. Phys. Chem. C* **2013,** *117* (35), 18031.

(195) For the specific case of surface modification with the shortest possible organic moiety, a single $CH_3$, the net negative dipole was attributed to polarization of the terminal C-H groups,(Refs. 81,194) however, a detailed DFT study of thiol-Au based monolayers of a variety of long (8-12 carbons) saturated and conjugated molecules demonstrates that the negative potential step occurs at the S-Au bond,(Ref. 193) in opposite to the core-level charging of the S atom.

(196) Kim, B.; Choi, S. H.; Zhu, X. Y.; Frisbie, C. D. Molecular Tunnel Junctions Based on π-Conjugated Oligoacene Thiols and Dithiols between Ag, Au, and Pt Contacts: Effect of Surface Linking Group and Metal Work Function. *J. Am. Chem. Soc.* **2011,** *133* (49), 19864.

(197) Xie, Z.; Bâldea, I.; Smith, C. E.; Wu, Y.; Frisbie, C. D. Experimental and Theoretical Analysis of Nanotransport in Oligophenylene Dithiol Junctions as a Function of Molecular Length and Contact Work Function. *ACS nano* **2015,** *9* (8), 8022.

(198) Mäkinen, A. J.; Kim, C.-S.; Kushto, G. P. Monolayer-induced band shifts at Si(100) and Si(111) surfaces. *Appl. Phys. Lett.* **2014,** *104* (4), 041601.

(199) Pluchery, O.; Zhang, Y.; Benbalagh, R.; Caillard, L.; Gallet, J. J.; Bournel, F.; Lamic-Humblot, A. F.; Salmeron, M.; Chabal, Y. J.; Rochet, F. Static and dynamic electronic characterization of organic monolayers grafted on a silicon surface. *Phys. Chem. Chem. Phys.* **2016,** *18* (5), 3675.

(200) Capua, E.; Kumar, T. A.; Tkachev, M.; Naaman, R. The Molecular Controlled Semiconductor Resistor: A Universal Sensory Technology. *Isr. J. Chem.* **2014,** *54* (5-6), 586.

(201) Green, M. A.; King, F. D.; Shewchun, J. Minority carrier MIS tunnel diodes and their application to electron and Photo-voltaic energy conversion - I. theory. *Solid-State Electron.* **1974,** *17*, 551.

(202) Tarr, N. G.; Pulfrey, D. L.; Camporese, D. S. An analytic model for the MIS tunnel junction. *IEEE Trans. Elect. Dev.* **1983,** *30* (12), 1760.

(203) Nesher, G.; Vilan, A.; Cohen, H.; Cahen, D.; Amy, F.; Chan, C.; Hwang, J. H.; Kahn, A. Energy level and band alignment for GaAs-alkylthiol monolayer-Hg junctions from electrical transport and photoemission experiments. *J. Phys. Chem. B* **2006,** *110* (29), 14363.

(204) Salomon, A.; Boecking, T.; Seitz, O.; Markus, T.; Amy, F.; Chan, C.; Zhao, W.; Cahen, D.; Kahn, A. What is the Barrier for Tunneling Through Alkyl Monolayers? Results from n- and p-Si–Alkyl/Hg Junctions. *Adv. Mater.* **2007,** *19*, 445.

(205) Nicollian, E. H.; Brews, J. R. *MOS (metal oxide semiconductor) physics and technology*; Wiley New York et al., 1982.

(206) Holmes, P. J. *The electrochemistry of semiconductors / edited by P. J. Holmes*, 1962.

(207) Esaki, L.; Stiles, P. J. New Type of Negative Resistance in Barrier Tunneling. *Phys. Rev. Lett.* **1966,** *16* (24), 1108.

(208) Chang, L.; Stiles, P.; Esaki, L. Electron Tunneling between a Metal and a Semiconductor: Characteristics of Al-Al2O3-SnTe and– GeTe Junctions. *J. Appl. Phys.* **1967,** *38* (11), 4440.

(209) Yaffe, O.; Qi, Y. B.; Scheres, L.; Puniredd, S. R.; Segev, L.; Ely, T.; Haick, H.; Zuilhof, H.; Vilan, A.; Kronik, L.et al. Charge transport across metal/molecular (alkyl) monolayer-Si junctions is dominated by the LUMO level. *Phys. Rev. B* **2012,** *85* (4), 045433.

(210) Dahlke, W. E.; Sze, S. M. Tunneling in metal-oxide-silicon structures. *Solid-State Electron.* **1967,** *10*, 865

(211) Vilan, A.; Cahen, D.; Kraisler, E. Rethinking Transition Voltage Spectroscopy within a Generic Taylor Expansion View. *Acs Nano* **2013,** *7* (1), 695.

(212) Yu, L. H.; Gergel-Hackett, N.; Zangmeister, C. D.; Hacker, C. A.; Richter, C. A.; Kushmerick, J. G. Molecule-induced interface states dominate charge transport in Si–alkyl–metal junctions. *J. Phys.: Condens. Matter* **2008,** *20* 374114.

(213) Ricoeur, G.; Lenfant, S.; Guerin, D.; Vuillaume, D. Molecule/Electrode Interface Energetics in Molecular Junction: A "Transition Voltage Spectroscopy" Study. *J. Phys. Chem. C* **2012,** *116*, 20722.

(214) All current-voltage equation are written for an n-type semiconductor; for a p-type one, J and V need to be multiplied by -1.

(215) Akkerman, H. B.; Boer, B. d. Electrical conduction through single molecules and self-assembled monolayers. *J. Phys.: Condens. Matter* **2008,** *20*, 013001.

(216) Vilan, A. Insulator charging limits direct current across tunneling metal-insulator-semiconductor junctions. *J. Appl. Phys.* **2016,** *119* (1), 014504.

(217) Tung, R. T.; Sullivan, J. P.; Schrey, F. On the inhomogeneity of Schottky barriers. *Mater. Sci. Eng., B* **1992,** *14* (3), 266.

(218) Jiang, L.; Sangeeth, C. S. S.; Wan, A.; Vilan, A.; Nijhuis, C. A. Defect Scaling with Contact Area in EGaIn-Based Junctions: Impact on Quality, Joule Heating, and Apparent Injection Current. *J. Phys. Chem. C* **2015,** *119* (2), 960.

(219) Junay, A.; Guézo, S.; Turban, P.; Delhaye, G.; Lépine, B.; Tricot, S.; Ababou-Girard, S.; Solal, F. Spatially resolved band alignments at Au-hexadecanethiol monolayer-GaAs(001) interfaces by ballistic electron emission microscopy. *J. Appl. Phys.* **2015,** *118* (8), 085310.

(220) Junay, A.; Guézo, S.; Turban, P.; Tricot, S.; Le Pottier, A.; Avila, J.; Ababou-Girard, S.; Schieffer, P.; Solal, F. Effective Metal Top Contact on the Organic Layer via Buffer-Layer-Assisted Growth: A Multiscale Characterization of Au/Hexadecanethiol/n-GaAs(100) Junctions. *J. Phys. Chem. C* **2016,** *120* (42), 24056.

(221) Yu, X.; Lovrincic, R.; Sepunaru, L.; Li, W.; Vilan, A.; Pecht, I.; Sheves, M.; Cahen, D. Insights into Solid-State Electron Transport through Proteins from Inelastic Tunneling Spectroscopy: The Case of Azurin. *ACS Nano* **2015,** *9* (10), 9955.

(222) Simeone, F. C.; Yoon, H. J.; Thuo, M. M.; Barber, J. R.; Smith, B.; Whitesides, G. M. Defining the Value of Injection Current and Effective Electrical Contact Area for EGaIn-Based Molecular Tunneling Junctions. *J. Am. Chem. Soc.* **2013,** *135* (48), 18131.

(223) Afanas' ev, V. V. *Internal Photoemission Spectroscopy: Principles and Applications*; Elsevier, 2010.

(224) Ahktari-Zavareh, A.; Li, W.; Kavanagh, K. L.; Trionfi, A. J.; Jones, J. C.; Reno, J. L.; Hsu, J. W. P.; Talin, A. A. Au∕Ag and Au∕Pd molecular contacts to GaAs. *J. Vac. Sci. Technol., B* **2008,** *26* (4), 1597.

(225) Buzio, R.; Gerbi, A.; Marré, D.; Barra, M.; Cassinese, A. Ballistic electron and photocurrent transport in Au/organic/Si(001) diodes with PDI8-CN2 interlayers. *J. Vac. Sci. Technol., B* **2016,** *34* (4), 041212.

(226) Troadec, C.; Kunardi, L.; Chandrasekhar, N. Ballistic emission spectroscopy and imaging of a buried metal∕organic interface. *Appl. Phys. Lett.* **2005,** *86* (7), 072101.

(227)  Troadec, C.; Kunardi, L.; Gosvami, N. N.; Knoll, W.; Chandrasekhar, N. Charge transport across metal molecule interfaces probed by BEEM. *J. Phys.: Conf. Ser.* **2007,** *61* (1), 647.

(228)  Fadjie-Djomkam, A. B.; Ababou-Girard, S.; Hiremath, R.; Herrier, C.; Fabre, B.; Solal, F.; Godet, C. Temperature dependence of current density and admittance in metal-insulator-semiconductor junctions with molecular insulator. *J. Appl. Phys.* **2011,** *110* (8), 083708.

(230)  Werner, J.; Ploog, K.; Queisser, H. J. Interface-State Measurements at Schottky Contacts: A New Admittance Technique. *Phys. Rev. Lett.* **1986,** *57* (8), 1080.

(231)  Barsoukov, E.; Macdonald, J. R. *Impedance spectroscopy: theory, experiment, and applications*; John Wiley & Sons, 2005.

(232)  Kar, S. Study of silicon–organic interfaces by admittance spectroscopy. *Appl. Surf. Sci.* **2006,** *252* (11), 3961.

(233)  Peng, W.; DeBenedetti, W. J. I.; Kim, S.; Hines, M. A.; Chabal, Y. J. Lowering the density of electronic defects on organic-functionalized Si(100) surfaces. *Appl. Phys. Lett.* **2014,** *104* (24), 241601.

(234)  Godet, C.; Fadjie-Djomkam, A.-B.; Ababou-Girard, S.; Tricot, S.; Turban, P.; Li, Y.; Pujari, S. P.; Scheres, L.; Zuilhof, H.; Fabre, B. Dynamics of Substituted Alkyl Monolayers Covalently Bonded to Silicon: A Broadband Admittance Spectroscopy Study. *J. Phys. Chem. C* **2014,** *118* (13), 6773.

(235)  Godet, C. Entropy effects in the collective dynamic behavior of alkyl monolayers tethered to Si(111). *Beilstein J. Nanotechnol.* **2015,** *6*, 583.

(236)  Godet, C. Dielectric relaxation properties of carboxylic acid-terminated n -alkyl monolayers tethered to Si(1 1 1): dynamics of dipoles and gauche defects. *J. Phys.: Condens. Matter* **2016,** *28* (9), 094012.

(237)  Werner, J.; Levi, A. F. J.; Tung, R. T.; Anzlowar, M.; Pinto, M. Origin of the Excess Capacitance at Intimate Schottky Contacts. *Phys. Rev. Lett.* **1988,** *60* (1), 53.

(238)  Werner, J. H.; Güttler, H. H. Barrier inhomogeneities at Schottky contacts. *J. Appl. Phys.* **1991,** *69* (3), 1522.

(239)  Seitz, O.; Vilan, A.; Cohen, H.; Hwang, J.; Haeming, M.; Schoell, A.; Umbach, E.; Kahn, A.; Cahen, D. Doping molecular monolayers: Effects on electrical transport through alkyl chains on silicon. *Adv. Funct. Mater.* **2008,** *18* (14), 2102.

(240)  Stroscio, J. A.; Feenstra, R. M.; Fein, A. P. Electronic Structure of the Si(111)2×1 Surface by Scanning-Tunneling Microscopy. *Phys. Rev. Lett.* **1986,** *57* (20), 2579.

(241)  Godet, C.; Fadjie-Djomkam, A.-B.; Ababou-Girard, S.; Solal, F. Tunnel barrier parameters derivation from normalized differential conductance in Hg/organic monomolecular layer-Si junctions. *Appl. Phys. Lett.* **2010,** *97* (13), 2105.

(242)  The exact values for the $NDC_{MX}/V_{FB}$ ratio and ξ were slightly adjusted from their nominal values to convey the message; we stress that this method is not exact.

(243)  Cheung, S. K.; Cheung, N. W. Extraction of Schottky diode parameters from forward current-voltage characteristics. *Appl. Phys. Lett.* **1986,** *49* (2), 85.

(244)  Werner, J. H. Schottky barrier and pn-junctionl/V plots—Small signal evaluation. *Appl. Phys. A: Solids Surf.* **1988,** *47* (3), 291.

(245)  Haj-Yahia, A.-E. M.Sc. thesis, Weizmann Institute of Science, 2011.